\documentclass{article}
\usepackage[english]{babel}
\usepackage{amsmath}
\usepackage[left]{lineno}
\usepackage{caption, subcaption, setspace}

\usepackage[letterpaper,top=2cm,bottom=2cm,left=3cm,right=3cm,marginparwidth=1.75cm]{geometry}

\usepackage{graphicx}
\usepackage{siunitx}
\usepackage{fancyhdr,graphicx,amsmath,amssymb}
\usepackage[ruled,vlined]{algorithm2e}
\usepackage[utf8]{inputenc}

\usepackage{natbib,hyperref}

\usepackage{mathtools}
\usepackage{booktabs}
\usepackage{amsmath}
\usepackage[font=small,labelfont=bf]{caption}
\usepackage{ulem}
\usepackage{float}
\usepackage{authblk}

\usepackage{xcolor}

\begin{document}

	\setstretch{1.5}

	\author[1,*]{Alejandro F. Visentini}
	\author[1]{Juan J. Hidalgo}
	\author[1]{Marco Dentz}

	\affil[1]{Institute of Environmental
		Assessment and Water Research, Barcelona, Spain}

	\title{Compressible gas flow through porous media under forced pressure and fluxes: compressibility number and pressure curvature}

	\date{}
	\maketitle

	\renewcommand\Authands{ and }

	\noindent\textbf{Corresponding author:} *Alejandro F. Visentini (afvash@cid.csic.es; afvisentini@gmail.com)

	\begin{abstract}

 Gas compressibility induces a nonlinear relationship between boundary forcing and pressure evolution in porous media. This matters in operations and experiments where inlet pressure, volumetric flux or mass flux is controlled, as each forcing produces a distinct transient approach to steady state. We analyze this systematically for isothermal gas flow, using the porous medium equation (PME), nondimensionalized in terms of a Peclet-type number $\Pi$ that compares the time scales of gas flow and pressure diffusion and measures the steady flow compression. At steady state, increasing $\Pi$ leads to a boundary layer near the outlet, of width $\sim\Pi^{-1}$, where the velocity and flow expansion scale as $\Pi$ and $\Pi^2$, respectively. Since $\Pi$ fixes the steady compression for all three inlet conditions, their transients can be compared directly. We derive a semi-analytical perturbation series in $\beta=2\Pi/(1+2\Pi)<1$, valid for the three inlet conditions, which remains convergent at large $\Pi$. Numerical solutions for $10^{-2}\leq\Pi\leq10^3$ show that the effective pressure-squared diffusivity plateaus at boundary-condition-dependent values for $\Pi\ll1$, and scales as $\Pi^{1/2}$ for $\Pi\gg1$, explained by the slowest eigenmode of the linearized PME about its steady state. To resolve local flow compression, we introduce the pressure curvature, $\kappa=\nabla^2 p$, and interpret its evolution through a nonlinear advection-diffusion-reaction equation.
 Curvature fields reveal localized compression pulses that migrate downstream and dissipate as the system relaxes, while curvature moments highlight a $\Pi$-dependent crossover from compression- to expansion-dominated transients. Applications to gas chromatography and underground hydrogen storage are illustrated.

	\end{abstract}


	\section{Introduction}
	\label{sec:introduction}

	Compressible gas flow through porous media occurs in a wide range of natural and industrial settings, including volcanic degassing \cite[e.g.,][]{massol1999generation, oppenheimer2013volcanic}, production and storage of natural gas \cite[e.g.,][]{tek1989underground}, hydrogen \cite[e.g.,][]{heinemann2021enabling} and carbon dioxide \cite[e.g.,][]{bachu2008co2} and gas chromatography \cite[e.g.,][]{giddings2017dynamics}, among many other applications \cite[e.g.,][]{hunt1988nonaqueous, hu2010physical,scandella2016ephemerality,lee2019compressible}. In most such settings, the flow behavior is strongly affected by the compressibility of the gas, which induces a nonlinear relationship between boundary forcing and pressure evolution. 
	In applications involving gas injection or pumping, such as gas storage operations or gas chromatography, gas flow is driven through different prescribed boundary conditions, namely, pressure, mass flux or volumetric flux, leading to qualitatively distinct pressure evolution paths toward steady state. 

	Gas flow in porous media is generally described by combining mass and momentum (e.g., Darcy law) conservation equations, and an equation of state \cite[e.g.,][]{liepmann2001elements}, the resulting system being highly nonlinear and admitting analytical treatments only for few special cases \cite[e.g.,][]{bear2013dynamics}. This nonlinearity originates in the pressure dependence of many gas flow parameters, namely, gas density, viscosity, temperature and compressibility factor, the latter measuring departure of the gas from ideal behavior \cite[e.g.,][]{chierici1994principles}. However, for many practical cases, the pressure dependence of the gas density dominates over the other parameters, which can often be assumed constant. For such a case, the governing system reduces to a porous-medium equation (PME) for the pressure squared with a pressure-dependent diffusivity \cite[e.g.,][]{vazquez2007porous}. The PME is used to describe compressible gas flow across hydrogeology \cite[e.g.,][]{baehr1995updated}, low-permeability (tight) media where the gas slippage (or Klinkenberg) effect \citep{klinkenberge1941permeability} is accounted for through a pressure-dependent apparent permeability \cite[e.g.,][]{wu1998gas,monteiro2012mathematical}, and reservoir engineering \cite[e.g.,][]{muskat1934flow,al1966flow, chierici1994principles,chaudhry2003gas}. The domain of applicability of the PME extends far beyond gas flow and concerns any problem involving state-dependent (e.g., pressure, heat, solute concentration) diffusivity, such as temperature-dependent heat conduction \cite[e.g.,][]{zeldovich1950theory}, groundwater infiltration described through the Boussinesq equation \cite[e.g.,][]{polubarinova1962theory}, and population growth in biology \cite[e.g.,][]{murray2002biological}. See \citet{vazquez2007porous} for a review.

	Notable advances in describing compressible gas flow come from petroleum engineering literature, which considered the PME early on \cite[e.g.][]{muskat1934flow,muskat1937use}, in combination with transformations such as the pseudo-pressure or pseudo-time to model the flows of real gases \cite[e.g.,][]{al1966flow,aanonsen1985application}.
	In these approaches, the PME is typically linearized at some representative value of gas diffusivity (e.g., at mean pressure), enabling analytical treatment of well test data \cite[e.g.,][]{chaudhry2003gas} and average pressure forecasting, among other applications \cite[e.g.,][]{matthews1967pressure, tek1989underground, chierici1994principles, patzek2013gas}. However, these approaches focus primarily on aggregated reservoir behavior and therefore do not resolve the spatial structure of the evolving pressure, velocity and derived fields where compression effects can be characterized. 

	In parallel, analytical solutions of the PME were famously derived by \citet{zeldovich1950theory} and \citet{barenblatt1952some}, for an instantaneous point source of heat and of gas, respectively, within an infinite domain, and later by \citet{kidder1957unsteady} for a step decrease in pressure applied from the inlet of a semi-infinite domain, all based on the self-similar structure of the PME solution. Interestingly, 
	in the context of groundwater infiltration (i.e., the Boussinesq equation), \citet{polubarinova1948nonlinear} had previously obtained a solution applicable to both a step decrease and increase in inlet pressure  \citep[see][]{polubarinova1962theory}. Both Kidder's and Polubarinova-Kochina's solutions are perturbation series in different expansion parameters. The PME is continuously investigated within the mathematical theory of nonlinear diffusion, with classical
	results concerning finite-speed propagation \cite[e.g.,][]{martinson1976finite}, self-similar solutions \cite[e.g.,][]{aronson1993selfsimilar,barenblatt2003scaling},
	geometrical properties of solutions of the PME \cite[e.g.,][]{lee2003geometrical}, and bounds on the pressure Laplacian, or curvature, known as Aronson--Bénilan estimates \citep{aronson1979regularite,vazquez2007porous,lu2009local}.

	 Despite this extensive work, there is, to our knowledge, no systematic analysis of how imposed pressure, mass flux and volumetric flux differ in their relaxation toward the same steady state in a bounded domain. Investigations have typically been developed around a single inlet condition, for example imposed pressure by \citet{kidder1957unsteady} or, in more recent studies of gas displacing liquid, imposed mass flux \citep{castellucci2026dynamics} or volumetric rate \citep[e.g.,][]{cuttle2023dynamics,morrow2023gas}, the latter studies each defining its own compressibility number. Further, it has not been established how flow compression is distributed during this relaxation.
	Here, we address this gap using the PME. Nondimensionalizing it through a flow time scale yields a Peclet-type number, $\Pi$, comparing the time scales of pressure diffusion and gas flow, which can be expressed equivalently in terms of the steady inlet pressure, mass flux or volumetric flux. We further show that $\Pi$ is directly related to the isothermal compression of the gas. Thus, for a given $\Pi$, the three inlet conditions lead to the same steady state and their transients can be compared directly. We solve the transients numerically and semi-analytically, through a $\Pi$-based perturbation series that by construction applies to the three inlet conditions and 
	converges for all tested $\Pi$. We then   
	explain the resulting relaxation times through the eigenmodes of the PME linearized about the steady state. To locate where and when the gas is compressed during these transients, we use the pressure curvature, which is proportional to the local compression rate and we show it obeys a nonlinear advection--diffusion--reaction equation.  

 	The article is organized as follows. Section~\ref{sec:gov_equations} presents the PME, introduces $\Pi$ and the evolution equation for the pressure curvature, and sets up the problem. Sections~\ref{subsec:steady_state_analysis} and~\ref{subsec:transient_state_analysis} analyze the steady and transient states, respectively, as functions of $\Pi$ and the inlet condition. Section~\ref{sec:applications} illustrates the results in physical units for gas chromatography and underground hydrogen storage, and Section~\ref{sec:conclusions} discusses the results and concludes the paper. The appendices derive the PME, show that $\Pi$ can also be interpreted as a compressibility number and establish the $\Pi$-range over which the PME applies for $\text{H}_2$, $\text{CH}_4$ and $\text{CO}_2$ (Appendix~\ref{subsec:appendix_A}), describe the numerical setup (Appendix~\ref{subsec:appendix_B}), and present the perturbation series (Appendix~\ref{subsec:appendix_C}) and the eigenmodes of the PME linearized about the steady state (Appendix~\ref{app:eigenvalue_problem_degenerate_diffusion}).

	\section{Gas flow equations} \label{sec:gov_equations}

	\subsection{Porous medium equation and compressibility number $\Pi$} \label{subsec:governing_equations}

	The porous medium equation (PME) on the pressure squared  $\psi \equiv p^2$ describes isothermal flow of a gas with constant compressibility factor in porous media \cite[e.g.,][]{al1966flow, chierici1994principles, vazquez2007porous} (Appendix~\ref{subsec:appendix_A}). It is given by
	\begin{equation} \label{eq:PME}
		\frac{\partial \psi}{\partial t} = \frac{k\sqrt{\psi}}{\phi \mu} \nabla^2\psi,
	\end{equation}
	where $k$ is permeability, $\mu$ dynamic viscosity and $\phi$ porosity. When referring to pressure, we will use interchangeably the variables $p$ and $\sqrt{\psi}$. In order to non-dimensionalize Equation~\ref{eq:PME}, we rescale lengths by the horizontal reservoir length $L$, pressure-squared by the outlet pressure-squared $\psi_L = p^2_L$ and times by the flow time-scale $\tau_u=L/u_s$, with $u_s$ a characteristic gas flow speed given by Darcy's law (see Appendix~\ref{subsec:appendix_A}) and defined below. In summary, 
	\begin{align} \label{eq:nondimensional_variables}
		\hat{t}=t/\tau_u,  
		\qquad 
		\hat{x}=x/L, 
		\qquad \hat{\psi}=\psi/\psi_L,
	\end{align}
	and Equation~\ref{eq:PME} becomes
	\begin{align}
		\frac{\partial \hat{\psi}}{\partial \hat{t}} = \frac{\sqrt{\hat{\psi}}}{\Pi} \hat{\nabla}^2 \hat{\psi}, 
		\label{eq:non_dimensional_PME}
	\end{align}
	where 
	\begin{align} \label{eq:Pi_Peclet_number}
		\Pi =  \frac{\tau_D}{\tau_u} = \frac{u_s L}{D_L} 
	\end{align}
	\noindent is a dimensionless number comparing time-scales associated with steady flow speed and pressure diffusion of the gas, with 
	\begin{align}
		\tau_D = \frac{L^2}{D_L},
	\end{align} 
	\noindent and $D_L = k p_L/ \phi \mu$ the $\psi$ diffusivity evaluated at the outlet pressure $\sqrt{\psi_L}$. As such, $\Pi$ has the form of a Peclet number. If $\tau_D$ instead of $\tau_u$ is used to scale time, as is typically done \cite[e.g.,][]{vazquez2007porous}, then, in terms of the diffusive time $\theta \equiv t/\tau_D = \hat{t}/\Pi$, $\Pi$ drops out of Equation~\ref{eq:non_dimensional_PME} and remains only in the boundary data. 
	Defining the velocity scale $u_s$ through the steady mass flux and outlet fluid density, respectively, $q_\text{s}^\mathrm{m}$  (in kg~m$^{-2}$~s$^{-1}$)  and $\rho_L$ (in kg~m$^{-3}$),
	\begin{align}
		u_s = \frac{q_\text{s}^\mathrm{m}}{\phi \rho_L},
		\label{eq:velocity_scale}
	\end{align}
	$\Pi$ can be expressed as
	\begin{align} \label{eq:Pi_mass_flux}
		\Pi =  \frac{q_\text{s}^\mathrm{m} RT \mu L}{kW \psi_L },
	\end{align}
	\noindent thereby linking $\Pi$ with $q_\text{s}^\mathrm{m}$, which is useful when mass flux at the inlet is imposed. Alternatively, using Equation~\ref{eq:Pi_mass_flux} one can also obtain,
	\begin{align} \label{eq:Pi_pressure_ratio}
		\Pi &= \frac{\psi_0-\psi_L}{2\psi_L} \\
		&= \frac{1}{2}\left(\hat{\psi}_0 - 1\right), 
	\end{align}
	\noindent which is useful when the inlet pressure $\sqrt{\psi_0}$ is imposed. From Equation~\ref{eq:Pi_pressure_ratio} one can also write 
	\begin{align} \label{eq:Pi_compressibility_number}
		\Pi
		=
		\frac{1}{2}
		\left[
		\exp(2C_s)-1
		\right],
	\end{align}
	\noindent with $C_s = \ln \left(\rho_0/\rho_L \right)$ the isothermal compression of an ideal gas associated with a pressure change $\Delta p = p_0 - p_L$. See Appendix~\ref{sec:appA_Pi_expressions} for a short derivation of expressions~\ref{eq:Pi_mass_flux}-\ref{eq:Pi_compressibility_number}. Consequently, the parameter $\Pi$ provides a measure of flow compression across the domain. Further, for weak compression, $C_s\ll1$, we have $\Pi \simeq C_s$.  Finally, Equation~\ref{eq:Pi_pressure_ratio} also relates $\Pi$ with the classical expansion parameters of \citet{kidder1957unsteady} and \citet{polubarinova1962theory} as we will see in Section~\ref{subsec:transient_state_analysis}. For simplicity of notation, in the following we drop the hats marking nondimensional quantities, except in Subsection~\ref{subsec:problem_statement} and Section~\ref{sec:applications}, where dimensional and nondimensional quantities, such as initial and boundary data, are compared.

	\subsection{Evolution equation for the pressure curvature $ \kappa$} \label{subsec:curvature_PDE}

	We now consider the pressure curvature $\kappa(x,t)$,
	\begin{align}
		\kappa(x,t) = \nabla^2 p(x,t),
	\end{align}
	\noindent as a measure of local flow compression rate.
	Indeed, regions where $\kappa(x,t) <0$, $\kappa(x,t) >0$ and $\kappa(x,t) = 0$ correspond to expansion, compression and locally incompressible flow, respectively. Thus $\kappa$ provides structural information on how flow compression distributes over space and time. This follows from Darcy's law, which implies $\kappa(x,t) \propto -\nabla \cdot \mathbf{q}$, with $\mathbf{q}$ the Darcy velocity. Note that fluid and flow compression are equivalent concepts when the fluid density depends only on pressure, as assumed here.

	To better interpret the behavior of the local flow compression $\kappa(x,t)$, we derive its evolution equation from the governing PME (Eq.~\ref{eq:non_dimensional_PME} ). Using that $\partial_t\psi = 2p\,\partial_t p$ and $\nabla^2\psi = 2p\nabla^2p + 2|\nabla p|^2$, and dividing the PME by $2p$,  we have
	\begin{align}
		\frac{\partial p}{\partial t}
		=
		\frac{1}{\Pi}
		\left(
		p \kappa + |\nabla p|^2
		\right).
		\label{eq:p_equation_nd}
	\end{align}
	Applying $\nabla^2$ to Equation~\eqref{eq:p_equation_nd} gives 
	\begin{align}
		\frac{\partial \kappa}{\partial t}
		=
		\frac{1}{\Pi}
		\nabla^2
		\left(
		p\kappa + |\nabla p|^2
		\right),
		\label{eq:kappa_compact}
	\end{align}
	\noindent which, after expanding the right-hand side, reads in non-conservative form as
	\begin{align}
		\frac{\partial \kappa}{\partial t}
		=
		\frac{1}{\Pi}
		\left[
		p\nabla^2\kappa
		+
		4\nabla p\cdot\nabla\kappa
		+
		\kappa^2
		+
		2|\nabla\nabla p|^2
		\right],
		\label{eq:kappa_general_dimension}
	\end{align}
	where $|\nabla\nabla p|^2=\sum_{i,j}(\partial_{ij}p)^2$ denotes the squared norm of the Hessian of $p$, $\nabla \nabla p$. Equation~\ref{eq:kappa_general_dimension} is a nonlinear advection diffusion reaction (ADR) equation, with curvature advection velocity $u_{\kappa}$ given as
	\begin{align}
		\label{eq:kappa_adv_velocity}
		u_{\kappa} = -\frac{4 \nabla p}{\Pi},
	\end{align}
	\noindent curvature diffusivity $D_{\kappa}$ given as
	\begin{align}
		D_{\kappa} =\frac{ p}{\Pi},
		\label{eq:kappa_diff_coeff}
	\end{align}
	\noindent that is, the same as for $\psi = p^2$ (Eq.~\ref{eq:non_dimensional_PME}), and the nonlinear reaction term given as
	\begin{align}
		\label{eq:kappa_reaction}
		R_{\kappa} =  \kappa^2
		+
		2|\nabla\nabla p|^2.
	\end{align}

	Thus, the pressure curvature field $\kappa$ evolves as a signed transported scalar, following its own transport Equation~\ref{eq:kappa_general_dimension}. As such, despite $\kappa$ being a derived field, it resolves the distribution of flow compression that otherwise appears integrated in the pressure and flow fields.   
	This appears analogous to flow vorticity, which is a derived field of the flow velocity vector, and follows its own evolution equation, the vorticity transport equation in the case of incompressible flow, characterizing rotational flow
	\cite[e.g.,][]{acheson1990elementary,saffman1992vortex,salmon1998lectures,burgers2013nonlinear}.

	 Note we have smooth $ p( x,t) >0$ throughout and, consequently, spatial derivatives up to fourth order of $p( x,t)$ are well defined, permitting formal derivation of Equation~\eqref{eq:kappa_general_dimension}. 
	For the one-dimensional problem considered here, $\nabla \nabla p =p_{xx} = \kappa$ and
	$|\nabla\nabla p|^2=p_{xx}^2=\kappa^2$, so Equation~\eqref{eq:kappa_general_dimension}
	reduces to
	\begin{align}
		\frac{\partial \kappa}{\partial t}
		=
		\frac{1}{\Pi}
		\left[
		p\frac{\partial^2\kappa}{\partial x^2}
		+
		4\frac{\partial p}{\partial x}
		\frac{\partial \kappa}{\partial x}
		+
		3\kappa^2
		\right],
		\label{eq:kappa_1D}
	\end{align}
	\noindent and so the reaction term simplifies to $R_\kappa = 3 \kappa^2$. The advection and diffusion terms keep their form in any dimension $d$, whereas for $d>1$ the non-negative reaction term depends on the full Hessian of $p$ and not only on its trace $\kappa$. Equation~\ref{eq:kappa_reaction} implies that anisotropic second derivatives of $p$ can make the reaction term positive while $\kappa =0$, for example at a saddle point of $p$ (e.g.,~$\partial_{xx}p=-\partial_{yy}p$). Thus, anisotropic pressure fields can generate compression in regions where the flow is locally incompressible. Pressure fields with anisotropic Hessian can arise in radial geometry around wells. In general, the Cauchy--Schwarz inequality bounding the trace of the Hessian $\kappa$ by its norm,  $|\nabla\nabla p|^2\geq\kappa^2/d$ \citep[e.g.,][p.~8]{lu2009local} leads to the the non-negative reaction term bounded as 
			\begin{align}
				R_{\kappa} \geq (1 + 2/d)\kappa^2,
				\label{eq:R_bound}
			\end{align} 
			\noindent bound which is attained exactly in 1-D. 

			Equation~\ref{eq:kappa_general_dimension} is known in the PME literature for the Laplacian of the pressure variable \citep[e.g.,][Prop.~9.4]{vazquez2007porous} where, together with the bound of Equation~\eqref{eq:R_bound}, it leads to the Aronson--Bénilan estimate \citep{aronson1979regularite,lu2009local}, which bounds the curvature from below in the absence of boundaries and, in our variables, reads
	\begin{align}
		\kappa\geq-\frac{d}{d+2}\frac{\Pi}{t},
		\label{eq:aronson_benilan}
	\end{align}
	\noindent that is, $\kappa\geq-\Pi/(3t)$ in one dimension. In Section~\ref{subsec:transient_state_analysis} we use Equation~\ref{eq:kappa_1D} to interpret transient evolution under different boundary conditions.

	 Finally, note that, since Equation~\ref{eq:kappa_compact} is in conservative form, its split into advection, diffusion and reaction terms (Eq.~\ref{eq:kappa_general_dimension}) is not unique. Indeed, in the conservative one-dimensional form, $\partial_t\kappa= (1/\Pi)\partial_x(p\,\partial_x\kappa+3\,\partial_xp\,\kappa)$, the flux consists of a diffusive part, $p\,\partial_x\kappa$, and an advective part, $3\,\partial_xp\,\kappa$, with advection velocity $-3\partial_xp/\Pi$, and the reaction term $R_\kappa$ is absorbed. We use the non-conservative form, which keeps $R_\kappa$ explicit and whose lower bound (Eq.~\ref{eq:R_bound}) leads to the Aronson--B\'enilan estimate.

	\subsubsection{Pressure curvature diagnostics}

	We focus on the evolution of the mass, or zeroth-order moments  \cite[e.g.,][]{taylor1950instability,aris1956dispersion}, of the positive and negative parts of $\kappa(x,t)$, respectively, $ \kappa^+( x, t) \coloneq \max(\kappa,0) $  and $ \kappa^-( x, t) \coloneq \max(-\kappa,0)$, given as 
	\begin{equation} 
		M^{\kappa\pm}(t) =\int^1_0    \kappa^{\pm}(x,t) \, dx,\label{eq:curvature_mass}
	\end{equation}
	\noindent to delineate compression or expansion regimes across the system. To track the transient compression front migrating away from the inlet once the flow is started, we further consider the normalized first moment, or centroid, of $\kappa^{+}(x,t)$,
	\begin{equation}
		X^{\kappa+}(t) = \frac{1}{M^{\kappa+}(t)}\int_0^1 x\,\kappa^{+}(x,t)\,dx,
		\label{eq:curvature_centroid}
	\end{equation}
	\noindent and the curvature amplitude at the centroid, $\kappa^{+}(X^{\kappa+}(t),t)$. 

	\subsection{Problem setup} \label{subsec:problem_statement} 

	We consider one-dimensional flow through a homogeneous porous medium, $x \in [0,L]$, initially at the outlet pressure $p_L$, kept fixed at $x=L$ throughout, while at $t=0$ an imposed pressure $p_0$, mass flux $q^\mathrm{m}_0$ or volumetric flux $q_0$ is switched on at the inlet, $x=0$. Using Darcy's law and the equation of state (Eqs.~\ref{eq:darcys_law}--\ref{eq:eos}) to express the fluxes, the PME (Eq.~\ref{eq:PME}) is solved subject to
	\begin{align}
		\psi(x,0)=\psi_L, \qquad \psi(L,t)=\psi_L, \qquad
		\begin{cases}
			\psi(0,t)=\psi_0 & \text{imposed pressure},\\[4pt]
			-\dfrac{Wk}{2RT\mu}\,\partial_x\psi(0,t)=q^\mathrm{m}_0 & \text{imposed mass flux},\\[8pt]
			-\dfrac{k}{\mu}\,\partial_x\sqrt{\psi}\,(0,t)=q_0 & \text{imposed volumetric flux}.
		\end{cases}
		\label{eq:problem_bcs_dimensional}
	\end{align}

	\noindent Using the nondimensional variables of Equation~\ref{eq:nondimensional_variables} the nondimensional initial and boundary data in terms of $\Pi$ read: 
	\begin{align}
		\hat\psi(\hat x,0)=1, \qquad \hat\psi(1,\hat t)=1, \qquad
		\begin{cases}
			\hat\psi(0,\hat t)=1+2\Pi & \text{imposed pressure},\\[4pt]
			\partial_{\hat x}\hat\psi(0,\hat t)=-2\Pi & \text{imposed mass flux},\\[4pt]
			\partial_{\hat x}\sqrt{\hat\psi}\,(0,\hat t)=-\dfrac{\Pi}{\sqrt{1+2\Pi}} & \text{imposed volumetric flux},
		\end{cases}
		\label{eq:problem_bcs_nondimensional}
	\end{align}
	\noindent which follow directly from the expressions of $\Pi$ in Subsection~\ref{subsec:governing_equations}. That is, the imposed pressure from Equation~\ref{eq:Pi_pressure_ratio}, the imposed mass flux from Equation~\ref{eq:Pi_mass_flux}, and the imposed volumetric flux from both Equation~\ref{eq:Pi_pressure_ratio} and~\ref{eq:Pi_mass_flux}, since $p_0/p_L=\sqrt{1+2\Pi}$, which gives $\hat q_0=\Pi/\sqrt{1+2\Pi}$ with $\hat q=q\mu	 L/(kp_L)$, while the pressure curvature scales as $\hat\kappa=\kappa L^2/p_L$. For a given $\Pi$, Equations~\ref{eq:problem_bcs_nondimensional} provide three different inlet conditions whose transients relax toward the same steady state, and hence toward the same compression across the domain, which allows us to compare them in Section~\ref{subsec:transient_state_analysis}. That is, a given compression can be achieved through any of the three inlet conditions. For example, a targeted mass flux, often the relevant inlet condition in gas reservoir operations, could be achieved through an imposed pressure or volumetric flux, which are the typically realizable controls, through mapping the mass flux via $\Pi$ to the other conditions.  

	The one-dimensional homogeneous permeability setup isolates the effect of compression from those of radial geometry and permeability heterogeneity, both relevant to large-scale reservoirs. Note that a fixed outlet pressure allows studying both semi-infinite dynamics at early times and boundary-modulated dynamics at later times and represents, for example, the outlet of gas permeameters \cite[e.g.,][]{klinkenberge1941permeability,tanikawa2009comparison}, gas-chromatography columns \citep{giddings2017dynamics} (Section~\ref{subsec:app_GC}), 
	or a far boundary at hydrostatic brine pressure in aquifer gas storage \citep{sainz2017assessment,castellucci2026dynamics}  (Section~\ref{subsec:app_UHS}), among other examples such as air inlet wells in soil vapor extraction \citep{massmann1989applying,ross1994efficiency,neeper2003harmonic}.
	Finally, the magnitudes of the outlet (lowest) pressure and of the permeability set the Knudsen number, $\mathrm{Kn}$, and hence the importance of gas slippage . In the examples of Section~\ref{sec:applications}, they give $\mathrm{Kn}\approx10^{-3}$, so that slippage changes the permeability by less than $1\%$ (Appendix~\ref{subsec:slippage}). For larger Knudsen numbers, up to $\mathrm{Kn}\approx10^{-1}$, slippage can be represented by a constant Klinkenberg coefficient $b$, in which case the shifted pressure $p+b$ satisfies the PME \citep{wu1998gas}.

	\section{Steady-state analysis}
	\label{subsec:steady_state_analysis}

	We now analyze the effect of flow compression at steady state, denoting steady distributions using the subscript "$\text{s}$". Setting the time derivative equal to zero in Equation~\ref{eq:non_dimensional_PME}, the governing Laplace equation is:
	\begin{align} \label{eq:steady_state_nondimensional_PME}
		\frac{d^2 \psi_\text{s}}{d x^2}(x)= 0, 
	\end{align}
	\noindent subject to the boundary data of Equation~\ref{eq:problem_bcs_nondimensional}, where, for a given $\Pi$, the three inlet conditions give the same steady state. Integrating Equation~\ref{eq:steady_state_nondimensional_PME} once we have (see Eq.~\ref{eq:app_Pi_nablapsquared}):
	\begin{align}
		\label{eq:Pi_gradphatsquared}
		\frac{d\psi_\text{s}}{dx} = -2\Pi. 
	\end{align}
	\noindent  Integrating again, the steady solution reads:
	\begin{align}
		\label{eq:general_solution_pressure}
		\psi_\text{s}(x) = 1 + 2\Pi(1-x).
	\end{align}
	\noindent Differentiating Equation~\ref{eq:general_solution_pressure} once, the non-dimensional flow velocity reads:  
	\begin{align}
		\label{eq:general_solution_velocity}
		q_\text{s}(x) &= \frac{\Pi}{\sqrt{\psi_\text{s}(x)}}.
	\end{align}
	\noindent Differentiating again, the pressure curvature $\kappa_\text{s} ( x)$ reads:
	\begin{align}
		\label{eq:pressure_curvature}
		\kappa_\text{s} ( x) =  -\frac{\Pi^{2}}{\psi^{3/2}_\text{s}(x)}.
	\end{align}

\begin{figure}[H] 
		\includegraphics[width=0.6\textwidth]{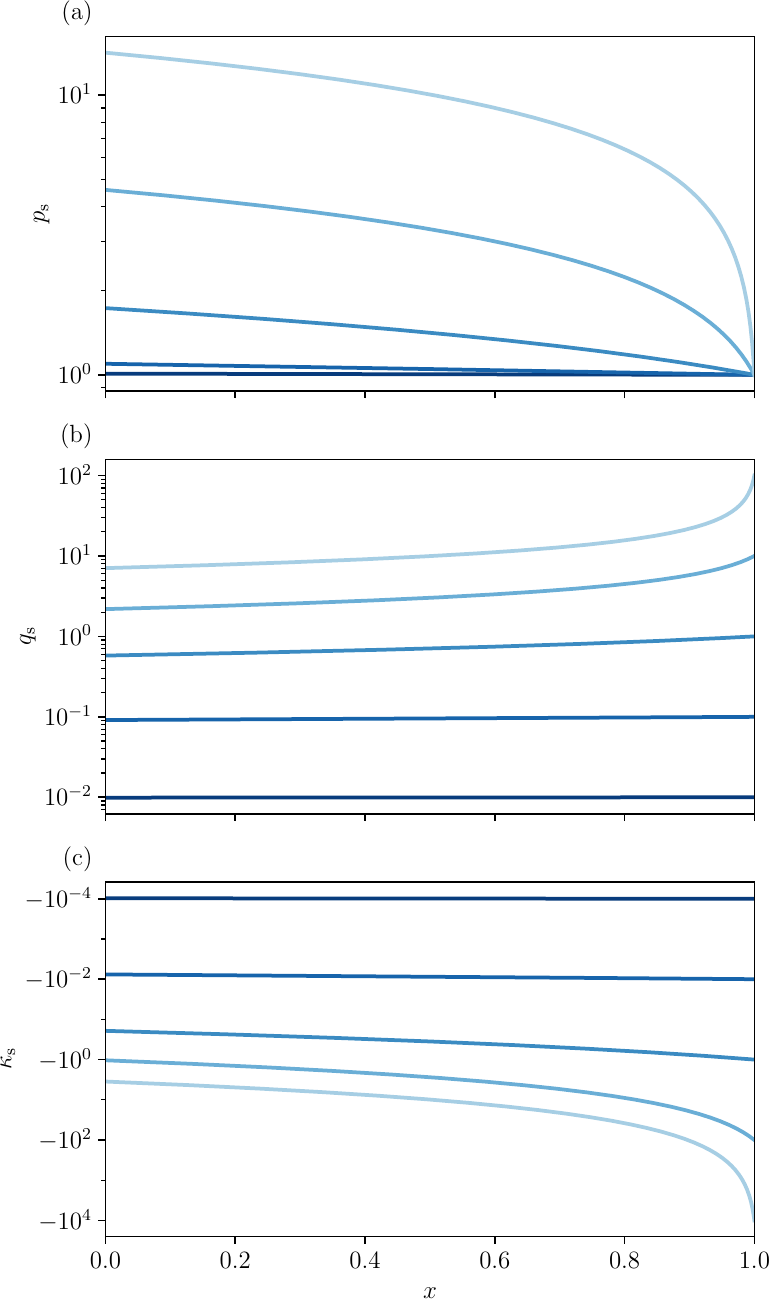}
		\centering
		\caption{Profiles of steady (a) pressure $ p_\text{s}$, (b) flow velocity $ q_\text{s}$ and (c)  pressure curvature $ \kappa_\text{s}$ for $\Pi$-values of (dark to light blue) $(0.01, 0.1, 1, 10, 100)$. }
		\label{fig:figure_1}
	\end{figure}

	Figure~\ref{fig:figure_1} illustrates profiles of steady non-dimensional pressure $\sqrt{\psi}$, flow velocity $q_\text{s}(x)$ and pressure curvature $\kappa_\text{s}(x)$ for $\Pi \in \{0.01, 0.1,1,10,100\}$. 

	The pressure profiles show the well-known concave shape of compressible flows, which accentuates with increasing compression, quantified by $\Pi$ (Fig.~\ref{fig:figure_1}a). The steeper pressure gradient, or higher-amplitude flow  $ q_\text{s}$, toward the outlet (Fig.~\ref{fig:figure_1}b) is reflected in the curvature $\kappa_\text{s}$ as increasing negative amplitude (Fig.~\ref{fig:figure_1}c) toward the same region. Physically, it reflects that the flow undergoes an increasing volumetric expansion rate as it goes downstream, to conserve the steady mass flux $q_\text{s}^{\text{m}} = \rho q_\text{s}$ while 
	satisfying the imposed pressure at the outlet. 

	To characterize how strong flow compression is expressed in the steady pressure $p_\text{s}(x)$, velocity $q_\text{s}(x)$ and curvature $\kappa_\text{s}(x)$, we interpret the steady solution (Eq.~\ref{eq:general_solution_pressure}) asymptotically in the limit $\Pi\gg1$, using the expansion parameter
	\begin{align}
		\varepsilon=\Pi^{-1}\ll1.
	\end{align}
	\noindent In the outer region away from the outlet, where $1-x=\mathcal{O}(1)$, expanding the steady solution (Eq.~\ref{eq:general_solution_pressure}) in $\varepsilon$ gives 
	\begin{align}
	p_\text{s}(x)=\left[2\Pi(1-x)\right]^{1/2}\left[1+\frac{\varepsilon}{4(1-x)}+\mathcal{O}(\varepsilon^2)\right],
	\label{eq:outer_approximation}
	\end{align}
	\noindent so that, to leading order,
	\begin{align}
		p_\text{s}(x) \sim \left[ 2 \Pi ( 1 - x ) \right]^{1/2},
	\end{align}
	\noindent and, using Equations~\ref{eq:general_solution_velocity}-~\ref{eq:pressure_curvature} the velocity scales as
	\begin{align}
		q_\text{s}(x)
		\sim
		\Pi^{1/2}
		\left[
		2(1-x)
		\right]^{-1/2},
	\end{align}
	\noindent and the curvature as
	\begin{align}
		\kappa_\text{s}(x)
		\sim
		-
		\Pi^{1/2}
		\left[
		2(1-x)
		\right]^{-3/2}.
	\end{align}
	That is, $p_\text{s}(x)$, $q_\text{s}(x)$ and $|\kappa_\text{s}(x)|$ scale as $\mathcal{O}(\Pi^{1/2})$, up to spatial factors depending on the distance to the outlet. This outer approximation satisfies the inlet condition, $p_\text{s}(0)=\sqrt{1+2\Pi}\sim(2\Pi)^{1/2}$, but not the outlet one $p_\text{s}(1)=1$, and it breaks down where the correction terms become comparable to the leading term, that is, where $1-x=\mathcal{O}(\varepsilon)$ (Eq.~\ref{eq:outer_approximation}). This defines an inner outlet region, described by the stretched distance from the outlet
	\begin{align*}
	\xi=\frac{1-x}{\varepsilon}=\Pi(1-x),
	\end{align*}
	\noindent with $\xi=\mathcal{O}(1)$ in the inner region and $\xi\gg1$ in the outer region. In the inner outlet region, the steady solution reads, exactly,
	\begin{align}
		p_\text{s}(\xi) = (1 + 2 \xi)^{1/2}.
		\label{eq:inner_solution}
	\end{align}
	\noindent The corresponding velocity and curvature are
	\begin{align}
		q_\text{s}(\xi) = \Pi(1 + 2\xi)^{-1/2},
	\end{align}
	\noindent and,
	\begin{align}
		\kappa_\text{s}(\xi) = -\Pi^2 (1 + 2 \xi)^{-3/2}.
	\end{align}
	\noindent  Therefore,  within the inner outlet region,
	\begin{align}
		p_\text{s} = \mathcal{O}(1), \qquad q_\text{s} = \mathcal{O}(\Pi), \qquad | \kappa_\text{s} | = \mathcal{O}(\Pi^2).
	\end{align}

	A diagnostic width for this inner region can be defined by requiring the nondimensional pressure at the inner edge of the layer to remain $\mathcal{O}(1)$,
	\begin{align}
		p_\text{s}(1-\delta_\text{BL})=c,
	\end{align}
	with $c>1$ an $\mathcal{O}(1)$ constant. Substitution into the steady solution gives
	\begin{align}
		c^2
		=
		1+2\Pi\delta_\text{BL},
	\end{align}
	and hence
	\begin{align} \label{eq:boundary_layer_width}
		\delta_\text{BL}
		=
		\frac{c^2-1}{2\Pi}.
	\end{align}
    This corresponds to $\xi=(c^2-1)/2=\mathcal{O}(1)$, consistent with the inner scaling. 
	Thus, increasing flow compression produces an outlet boundary layer of width $\delta_\text{BL} \sim \Pi^{-1}$, within which velocity and curvature amplitudes scale as $\mathcal{O}(\Pi)$ and $\mathcal{O}(\Pi^2)$, respectively. The pressure-squared field $\psi_\text{s}$ is linear and therefore does not exhibit a boundary-layer structure. For example, choosing $c = \sqrt{3}$ in Equation~\ref{eq:boundary_layer_width}, gives $\delta_\text{BL} \in [100, 10, 1, 0.1, 0.01]$ for the
	chosen $\Pi$-values. When $\delta_\text{BL} > 1$, the diagnostic layer exceeds the domain size, indicating that curvature is weak and distributed over the entire domain, consistent with an almost linear pressure profile.  

	Finally, note that since the steady solution is known, the inner and outer approximations need not be matched to construct it. Indeed, the inner solution (Eq.~\ref{eq:inner_solution}) is exact, since $\psi_\text{s}=1+2\xi$, and its expansion for $\xi\to\infty$, 
	\begin{align}
		p_\text{s}=(2\xi)^{1/2}\left[1+\frac{1}{4\xi}+\mathcal{O}(\xi^{-2})\right],
	\end{align}
	recovers the outer expansion (Eq.~\ref{eq:outer_approximation}) when written with $\xi=\Pi(1-x)$. %

	\section{Transient analysis} \label{subsec:transient_state_analysis}

	 We now examine transients of pressure, velocity and curvature for $\Pi\in[10^{-2},10^2]$ and the three inlet conditions, with the PME solved numerically as detailed in Appendix~\ref{subsec:appendix_B}. We first present the perturbation series solution of the PME and compare it with the numerical solution and with classical expansions (Subsec.~\ref{subsec:analytical_comparison}). We then characterize the relaxation through the effective pressure-squared diffusivity (Subsec.~\ref{subsec:effective_pressure_diffusivity}), and finally use the pressure curvature and its moments to locate flow compression during the transients (Subsec.~\ref{subsec:transient_curvature}). 

	As a reference, we will consider the linear diffusion model obtained by evaluating the pressure-dependent diffusivity in the PME (Eq.~\ref{eq:PME}) at a constant reference pressure $p_\text{ref}$,
	\begin{align}
		\frac{\partial \psi}{\partial t} = \frac{p_\text{ref}}{\Pi}\,\frac{\partial^2 \psi}{\partial x^2},
		\label{eq:linearized_diffusion_model}
	\end{align}
	\noindent with $p_\text{ref}=(1+p_0)/2 =\left[1+(1+2\Pi)^{1/2}\right]/2$, the arithmetic mean of the steady inlet and outlet pressures, classically used to interpret steady gas permeability measurements \citep{klinkenberge1941permeability,chierici1994principles,tanikawa2009comparison} and giving for the constant diffusivity $p_\text{ref}/\Pi$ the harmonic mean of the steady diffusivity. %
 	Equation~\ref{eq:linearized_diffusion_model} is solved with the inlet conditions of Equation~\ref{eq:problem_bcs_nondimensional}.
 	Thus, for imposed fluxes, the inlet pressure of the linear model also evolves in time toward its steady value.

	Also, to compare the transients, we define the relaxation time $t_\text{rel}$ as the time at which the normalized $\ell_2$ distance between $\psi(x,t)$ and the steady state $\psi_\text{s}(x)$ (Eq.~\ref{eq:general_solution_pressure}) falls below a tolerance $r_\text{tol}$,
	\begin{align}
		t_\text{rel} \coloneq\inf\big\{ t \mid r(t)\le r_\text{tol} \big\},
		\label{eq:observed_diffusion_timescale}
	\end{align}
	\noindent where
	\begin{align}
		r(t)=\frac{\| \psi(x,t)-\psi_\text{s}(x)\|}{\|\psi(x,0)-\psi_\text{s}(x)\|},
		\label{eq:normalized_l2_residual}
	\end{align}
	\noindent with $\| \cdot \|$ the $\ell_2$ distance. We choose $r_\text{tol}=0.001$, while other choices shift $t_\text{rel}$ by an amount proportional to $\ln(1/r_\text{tol})$, since the residual decay at late times is exponential (see Subsec.~\ref{subsec:effective_pressure_diffusivity}). For example, choosing $r_\text{tol} = 0.0001$ increases $t_\text{rel}$ between $25\%$ and $35\%$, depending on $\Pi$ and inlet condition, but the trends discussed below do not change.

	\subsection{Perturbation series solution of the PME} \label{subsec:analytical_comparison}

	We now present the semi-analytical solution of the PME in our bounded domain, valid for imposed pressure, mass flux and volumetric flux. A detailed presentation is given in Appendix~\ref{subsec:appendix_C}, while here we outline the approach. We start by  writing the PME in terms of the diffusive time $\theta$ (Subsec.~\ref{subsec:governing_equations}),
	\begin{align}
		\frac{\partial \psi}{\partial \theta} -\sqrt{\psi}\,
		\frac{\partial^2 \psi}{\partial x^2}  = 0,
		\label{eq:psi_pme_draft}
	\end{align}
	\noindent 
	and writing $\psi(x,\theta)$ as a perturbation around the steady-state inlet pressure-squared, $\psi_0 = p^2_0 = 1 + 2\Pi$ (Eq.~\ref{eq:problem_bcs_nondimensional}),
	\begin{align}
		\psi(x,\theta)
		=
		\psi_0
		\left[
		1+\beta V(x,\theta)
		\right],
		\label{eq:general_ansatz}
	\end{align}
	\noindent with expansion parameter $\beta$, given by the relative difference between the steady inlet value $\psi_0$ and the initial and outlet value $1$,
	\begin{align}
		\beta = \frac{\psi_0 - 1}{\psi_0} = \frac{2 \Pi}{1 + 2 \Pi}.
		\label{eq:beta_definition}
	\end{align}
	\noindent As such, $\beta \in [0,1)$ remains bounded for any $\Pi$. $V(x,\theta)$ is a dimensionless perturbation field such that $-1 \leq V \leq 0$. Note that expressing $\beta$ in terms of $\Pi$ makes perturbation expansions around reference mass or volumetric fluxes direct, since the three inlet conditions with the same $\Pi$ share the steady inlet value $\psi_0=1+2\Pi$ (Eq.~\ref{eq:problem_bcs_nondimensional}), and hence $\beta$.

	Substituting Equation~\ref{eq:general_ansatz} into Equation~\ref{eq:psi_pme_draft} and further scaling time through the diffusivity at the nondimensional inlet pressure, that is, 
	\begin{align}
		\theta' =\sqrt{\psi_0}\,\theta,
		\label{eq:scaling_inlet_diffusivity}
	\end{align}
	\noindent gives the PDE for $V(x,\theta')$ as
	\begin{align}
		\frac{\partial V}{\partial \theta'}
		-
		\sqrt{1+\beta V}
		\frac{\partial^2 V}{\partial x^2} = 0.
		\label{eq:general_V_equation}
	\end{align}

	Writing $V(x,\theta')$ in powers of $\beta$,
	\begin{align}
		V(x,\theta')
		=
		\sum_{i=0}^{\infty}\beta^i V_i(x,\theta'),
		\label{eq:power_series_V}
	\end{align} 
	\noindent and substituting it, together with the Taylor expansion of $\sqrt{1+\beta V}$ about $\beta V=0$, into Equation~\ref{eq:general_V_equation} and collecting powers of $\beta$ \citep{bender2013advanced} gives a recursion of linear diffusion equations for the coefficients $V_i(x,\theta')$:
	\begin{align}
		\frac{\partial V_i}{\partial \theta'}
		-\frac{\partial^2 V_i}{\partial x^2}&=S_i(x,\theta';V_0,\cdots,V_{i-1}), \qquad i=0,1,\ldots 
		\label{eq:perturbation_recursion}
	\end{align}
	\noindent where the source term $S_i(x,\theta';V_0,\cdots,V_{i-1})$ is $S_0 \equiv 0$ for $i=0$, 
	and 
	\begin{align}
		S_i=
		\sum_{k=1}^{i}\binom{1/2}{k}
		\sum_{j_0+\cdots+j_k=i-k}
		\left(\prod_{r=1}^{k}V_{j_r}\right)
		\frac{\partial^2V_{j_0}}{\partial x^2},
		\label{eq:source_term_analytical}
	\end{align}
	\noindent for $i \geq 1$.
	 For imposed pressure and mass flux, the initial and boundary data written for $V$ are independent of $\beta$, so they are carried entirely by $V_0$, and the $V_i$, $i\geq1$, have zero initial data and homogeneous boundary data. For imposed volumetric flux, instead, the inlet condition depends on $\beta$ and gives non-homogeneous inlet data at all orders (Appendix~\ref{subsec:app_vol_flux_condition}). The sources $S_i$ are generated by the pressure dependence of the diffusivity $\sqrt{1+\beta V}=p/p_0$ in Equation~\ref{eq:general_V_equation}. 
	 For example, $\sqrt{1+\beta V}\approx1+\beta V_0/2$ gives $S_1=\tfrac12V_0\,\partial_x^2V_0$. 	

	The solutions $V_i$ of Equations~\ref{eq:perturbation_recursion}  are expressed as an expansion in the eigenfunctions of the diffusion operator $-\partial_x^2$ on the bounded domain. 
	At zeroth order the problem is unforced ($S_0=0$), and $V_0$ relaxes from its initial value $-1$ to $V_\mathrm{s}(x)=-x$, the steady value of $V$ corresponding to $\psi_\mathrm{s}=1+2\Pi(1-x)$ (Eq.~\ref{eq:general_solution_pressure}). At higher orders, $V_i$ is driven by the source $S_i$, through a convolution (Duhamel) integral over time  \cite[e.g.,][]{carslaw1959conduction}. The corrections $V_i$, $i\geq1$, vanish at steady state, since the steady state is captured exactly at zeroth order.

	After computing $V_i$, the $N$-th order perturbation approximation of $\psi(x,\theta)$ follows as
	\begin{align}
		\tilde\psi(x,\theta) = \psi_0\Big[1+\beta\sum_{i=0}^{N}\beta^i V_i(x,\theta')\Big].
		\label{eq:series_reconstruction}
	\end{align}
	\noindent Since the fields $V_i$ are independent of $\Pi$, they are computed only once for each inlet condition, and Equation~\ref{eq:series_reconstruction} then gives the solution for any $\Pi$.

	Since the number of terms in Equation~\ref{eq:source_term_analytical} grows as $2^i-1$, analytical evaluation of the sources becomes impractical beyond $i\approx5$, a known difficulty of perturbation approaches \citep{polubarinova1962theory, bender2013advanced}. We instead compute $S_i$ through a nested recursion for the power-series coefficients of $\sqrt{1+\beta V}$ \citep[][\S 4.7]{knuth1997art}, which builds $S_i$ from $\mathcal{O}(i)$ products of lower-order fields and also provides the inlet data of each order for imposed volumetric flux.
	 Details of the construction, the nested recursion and the convergence with the truncation order $N$ are given in Appendix~\ref{subsec:appendix_C}.

	\subsection{Connection to classical Polubarinova-Kochina's and Kidder's expansions for imposed pressure}	

	Expanding around $\psi_0$
	follows  \citet{polubarinova1948nonlinear,polubarinova1962theory}, who used a perturbation approach to solve the Boussinesq equation governing groundwater level. Their expansion parameter, $\alpha_\text{P} = (1 - \psi_0)/\psi_0$ is the same as ours up to the sign:
	\begin{align}
		\alpha_\text{P} = -\beta = - \frac{2 \Pi}{1 + 2 \Pi},
	\end{align}
	\noindent If one would expand instead around the initial and outlet value $\psi = 1$, the parameter used by \citet{kidder1957unsteady} would be obtained, $\alpha_\text{K} = 1 - \psi_0$, which relates to $\Pi$, and our parameter $\beta$, as
	\begin{align}
		\alpha_\text{K}= -2\Pi = -\frac{\beta}{1-\beta}.
		\label{eq:kidder_parameter}
	\end{align} 
	\noindent Since \citet{kidder1957unsteady} consider a depressurizing setting, whereby the applied inlet pressure is smaller than the initial pressure ($\psi_0 <1$), they have bounded $\left| \alpha_K \right| <1$, by construction. However, in our setting of pressurization or injected flux, using $\alpha_K$ implies restricting $\Pi \lesssim 1/2$, or $p_0 < \sqrt{2}$, to guarantee a convergent series, whereas $\beta <1$ for any $\Pi$. For $\Pi\ll1$, $\beta=2\Pi+\mathcal{O}(\Pi^2)$, and thus $\alpha_\text{K}\approx-\beta$. 

\begin{figure}[H]
		\includegraphics[width=\textwidth]{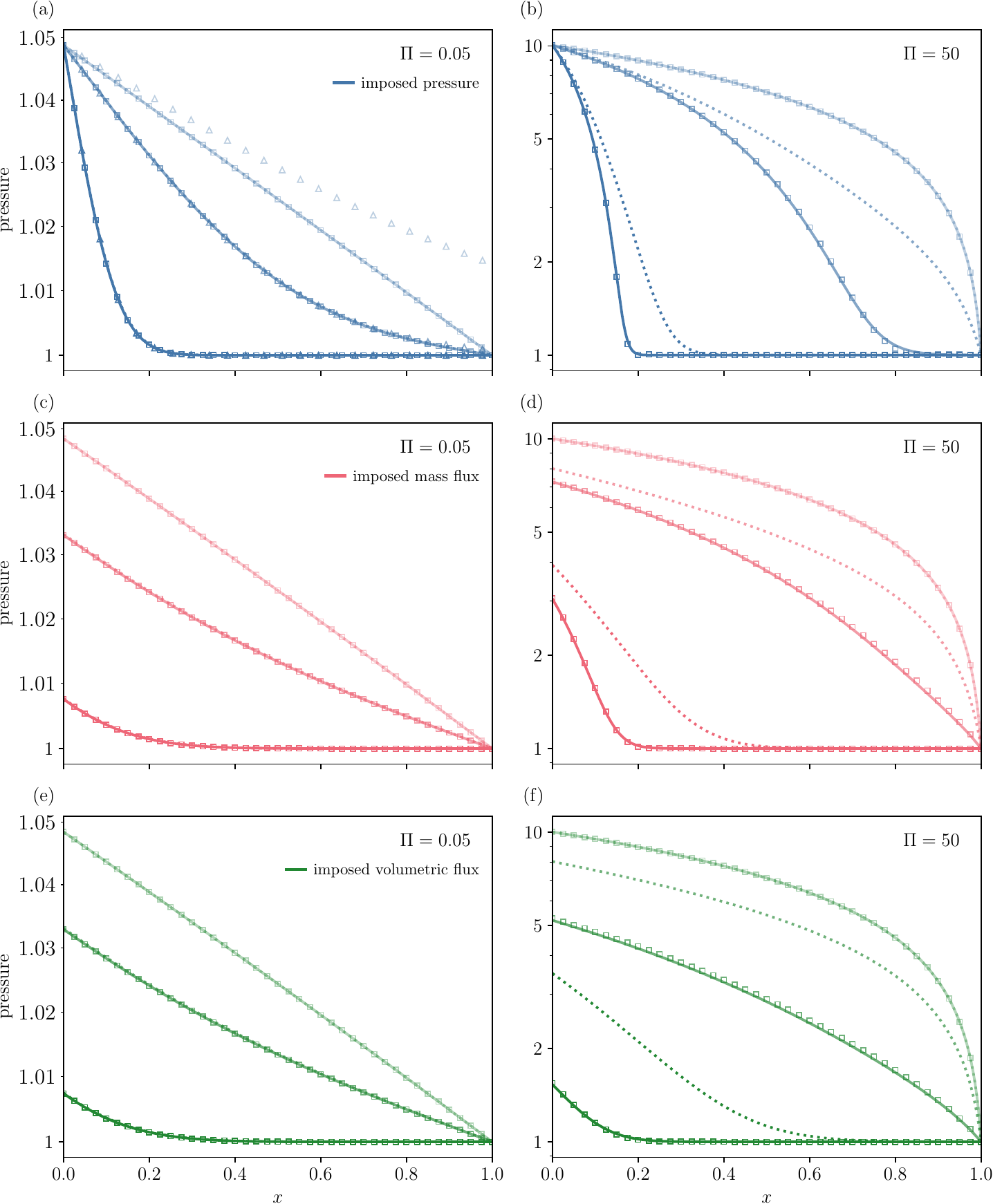}
		\centering
		\caption{Pressure as a function of nondimensional distance $x$ for imposed inlet (a)-(b) pressure, (c)-(d) mass flux and (e)-(f) volumetric flux at compressibility number values (left) $\Pi = 0.05$ and (right) $\Pi = 50$. Shown are solutions of the PME (Eq.~\ref{eq:non_dimensional_PME}) calculated (solid lines) numerically, (squares) semi-analytically through the perturbation series in $\beta = 2\Pi/(1+2\Pi)$, truncated at order 100 and using recursively-computed source coefficients (Appendix~\ref{subsec:appendix_C}), and (triangles) analytically through Kidder's third-order perturbation series in $\alpha_\text{K} = -2\Pi$ for the semi-infinite domain \citep{kidder1957unsteady}, valid before the pressure front reaches the outlet, shown only for imposed pressure at $\Pi = 0.05$ since its convergence requires $\Pi\lesssim1/2$ (Section~\ref{subsec:transient_state_analysis}); (dotted lines) the eigenfunction-expansion solution of the linear diffusion model (Eq.~\ref{eq:linearized_diffusion_model}) with the arithmetic mean of the steady inlet and outlet pressures as $p_\text{ref}$. The selected times are (dark to light) $0.01t_\text{rel}$, $0.2t_\text{rel}$ and $t_\text{rel}$, where $t_\text{rel}$ is the estimated $\psi$-diffusion time (Eq.~\ref{eq:observed_diffusion_timescale})} \label{fig:profiles_beta_series}
	\end{figure}

	Figure~\ref{fig:profiles_beta_series} compares time snapshots of pressure corresponding to the numerical solution of the PME, the perturbation series truncated at order 100, Kidder's semi-infinite solution and the linearized PME  (Eq.~\ref{eq:linearized_diffusion_model}). The time snapshots are obtained relative to the relaxation time $t_\text{rel}$  (Eq.~\ref{eq:observed_diffusion_timescale}).

	For $\Pi = 0.05$ (left column, Fig.~\ref{fig:profiles_beta_series}), the numerical solution of the PME is practically indistinguishable from the linearized model at the arithmetic mean pressure (Eq.~\ref{eq:linearized_diffusion_model})%
	, since the imposed inlet values are small (Eq.~\ref{eq:problem_bcs_nondimensional}) and hence the diffusivity varies only mildly. Kidder's semi-infinite solution \citep{kidder1957unsteady} agrees with the PME at early times, but departs from it once the pressure front reaches the outlet (light triangles, Fig.~\ref{fig:profiles_beta_series}a). Note for imposed fluxes the inlet pressure rises gradually toward its steady value (Figs.~\ref{fig:profiles_beta_series}c,e).

	For $\Pi = 50$ (right column, Fig.~\ref{fig:profiles_beta_series}), the imposed inlet values lead to strong pressure gradients, and the nonlinear PME departs from the linearized model, which runs ahead of the PME at early and intermediate times, since its constant diffusivity, $p_\text{ref}/\Pi$, exceeds that of the PME, $p/\Pi$, near the pressure front, where $p\approx1$. The perturbation series follows the PME at all times, with maximum relative residuals at order truncation order $N=100$, $\epsilon_{100}$, of $0.4\%$, $2\%$ and $4\%$ for imposed pressure, mass flux and volumetric flux, respectively. In general, for the three inlet conditions, $\epsilon_N$ decreases and increases monotonically with $N$ and $\Pi$, respectively, and is well described by the product of geometric, $\beta^N$, and algebraic, $N^{-a}$, factors, 
	\begin{align}
		\epsilon_N \sim \beta^N N^{-a},
	\end{align}
	\noindent with $a \approx 0.8-1.2$ across the different imposed inlet data. Thus, the residual decays faster than that of a geometric series, $\beta^N$, which decreases slowly as $\beta \to 1$ (i.e., $\Pi$ grows). For example, at $\Pi = 50$ and $N=100$ a purely geometric decay would give $\beta^{100} \approx 0.37$, that is, a $37\%$ error. At small $\Pi$, $\beta^N$ decays fast and controls the convergence, whereas at large $\Pi$, the algebraic factor $N^{-a}$ does. As example, imposing the limit $N=100$ while requiring $\epsilon_{100}< 1\%$ restricts $\Pi$ to $\Pi<100$, $<30$ and $<20$ for imposed pressure, mass flux and volumetric flux, respectively. See Appendix~\ref{subsec:convergence_analysis} for further details.

	\subsection{Effective pressure-squared diffusivity} \label{subsec:effective_pressure_diffusivity}

	The effective diffusion coefficient of $p^2 = \psi$, normalized by the nondimensional diffusivity at the outlet, $D_1 = 1/\Pi$, reads
	\begin{align} \label{eq:observed_effective_diffusivity}
		\frac{D_\text{eff}}{D_1} &= \frac{1}{\theta_\text{rel}},
	\end{align}
	\noindent  with $\theta_\text{rel} = t_\text{rel}/\Pi$ the relaxation time (Eq.~\ref{eq:observed_diffusion_timescale}) 
	written in units of the diffusive time (Subsec.~\ref{subsec:governing_equations}).

\begin{figure}[H] 
		\includegraphics[scale = 0.6]{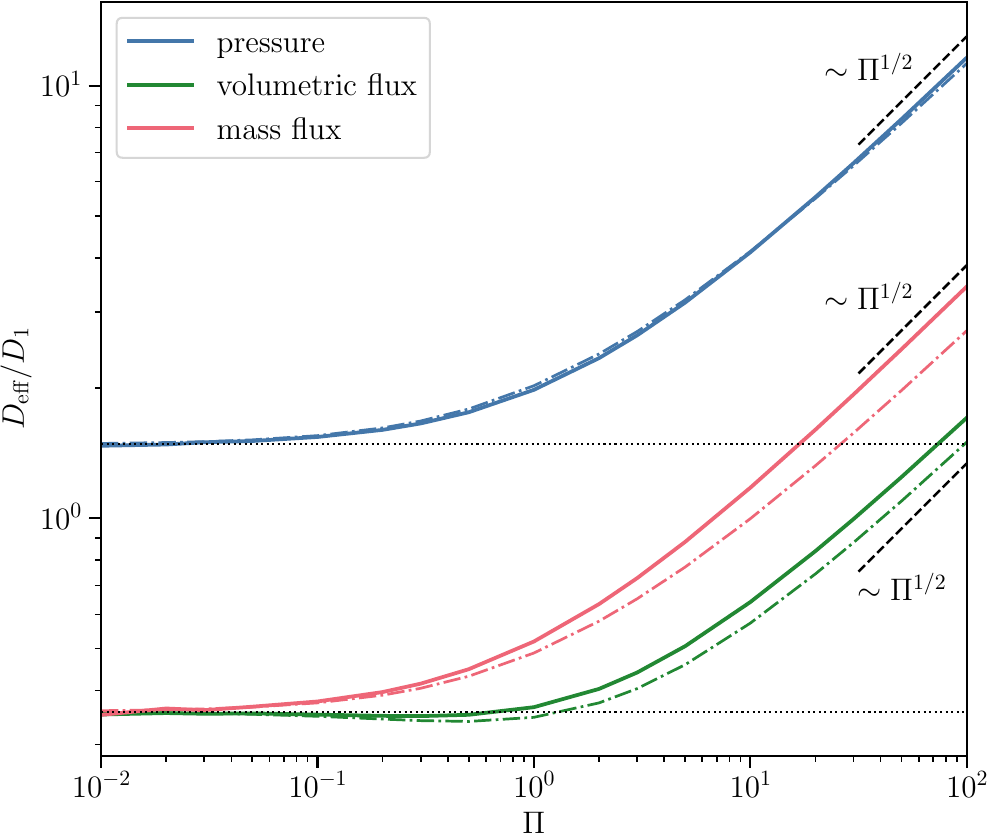}
		\centering
		\caption{Normalized effective $\psi$-diffusion coefficient $D_\text{eff}/D_1$ as a function of $\Pi$ (Eq.~\ref{eq:observed_effective_diffusivity}) obtained from (solid lines) the numerical solutions of the PME for imposed inlet (blue) pressure, (red) mass flux and (green) volumetric flux, and $D_1=1/\Pi$ the diffusivity at the outlet pressure. The dash-dot lines correspond to the linearized PME model (Eq.~\ref{eq:linearized_diffusion_model}) using as $p_\text{ref}$ the arithmetic mean of the steady inlet and outlet pressures. The thin dotted black lines indicate the small-$\Pi$ limits and the dashed segments the $\Pi^{1/2}$ scaling. At small $\Pi$, the curves for imposed volumetric and mass flux overlap, since both inlet conditions coincide when $p\approx1$.}\label{fig:tau_diffusion}
	\end{figure}

	Figure~\ref{fig:tau_diffusion} reports $D_\text{eff}/D_1$ as a function of $\Pi$ for the three inlet conditions.

	For small $\Pi$, $D_\text{eff}/D_1$ plateaus at $\approx 1.48$ for imposed pressure and at $\approx 0.36$ for both imposed fluxes (Fig.~\ref{fig:tau_diffusion}). As $\Pi$ increases above $0.1$, $D_\text{eff}/D_1$ increases for imposed pressure and mass flux, whereas for imposed volumetric flux it remains nearly constant up to $\Pi \approx 1$. For large $\Pi$, $D_\text{eff}/D_1$ grows as $\sim (2\Pi)^{1/2}/c$ for the three inlet conditions (dashed segments), with 
	$c\approx 1.27$, $4.24$ and $9.0$ for imposed pressure, mass flux and volumetric flux, respectively. 
	At the considered tolerance $r_\text{tol} = 10^{-3}$, the linearized model (dash-dot lines) follows the PME within $3\%$ for imposed pressure but underestimates it by up to $21\%$ and $18\%$ at $\Pi=100$ for imposed mass and volumetric flux, respectively (Fig.~\ref{fig:tau_diffusion}).

	To interpret these results, we linearize the PME about the steady state $\psi_\text{s} = 1+2 \Pi(1-x)$ (Eq.~\ref{eq:general_solution_pressure}). Writing $\psi = \psi_\text{s} + \delta \psi$, with $\delta \psi$ a small deviation at late times, substituting into the PME, dropping terms $\mathcal{O}(\left[\delta \psi\right]^2)$ and scaling time with the inlet diffusivity, $\theta'=\sqrt{\psi_0}\,\theta$ (Eq.~\ref{eq:scaling_inlet_diffusivity}), the equation for $\delta \psi$ reads:
	\begin{align}
		\frac{\partial \delta\psi}{\partial \theta'} = \sqrt{1-\beta x} \,\frac{\partial^2\delta\psi}{\partial x^2},
		\label{eq:linearized_PME_around_ss}
	\end{align}
	\noindent with $\sqrt{1 - \beta x} = \sqrt{\psi_\text{s}/\psi_0}$ and $\beta=2\Pi/(1+2\Pi)$ (Eq.~\ref{eq:beta_definition}), so that is the steady diffusivity relative to its inlet value. At the outlet we have $\delta \psi = 0$ and, at the inlet $\delta \psi = 0$, $\partial_x \delta \psi = 0$ and $\partial_x \delta \psi = -(\beta/2)\,\delta\psi$, for imposed pressure, mass flux and volumetric flux, respectively, with the latter condition linearized to first order in $\delta\psi$. The solution of Equation~\ref{eq:linearized_PME_around_ss} is a sum of eigenmodes decaying exponentially in $\theta'$,
	\begin{align}
		\delta \psi(x,\theta')= \sum_n b_n \Phi_n(x) e^{-\mu'_n \theta'},
	\end{align}
	\noindent with modal coefficients $b_n$, modes $\Phi_n(x)$ given in terms of Bessel functions and eigenvalues $\mu'_n$ given analytically by the corresponding roots, depending on $\beta$ and the inlet condition (Appendix~\ref{app:eigenvalue_problem_degenerate_diffusion}). The late-time modal coefficients $b_n$ depend on the preceding nonlinear transient of the PME and can be estimated from its numerical or semi-analytical solution (Subsection~\ref{subsec:analytical_comparison}). Consequently, at late times, the residual is $r(t)=\| \delta\psi(\cdot,t)\| / \| \delta\psi(\cdot,0)\|$ and dominated by the slowest-decaying eigenmode \cite[e.g.,][]{carslaw1959conduction,crank1975mathematics},
	\begin{align}
		r(\theta')\approx a_1\,e^{-\mu'_1\theta'},
		\label{eq:slowest-decaying_mode}
	\end{align}
	\noindent with $a_1=|b_1|\|\Phi_1\|/\|\delta\psi(\cdot,0)\|$ the normalized amplitude and $\mu'_1$ the smallest eigenvalue setting the decay rates. Evaluating Equation~\ref{eq:slowest-decaying_mode} at $r=r_\text{tol}$ gives the relaxation time as 
	\begin{align}
		\theta'_\text{rel}\approx \frac{\ln{ \left(a_1/r_\text{tol} \right) }}{\mu'_1},
		\label{eq:theta_rel_split}
	\end{align}
	\noindent and, correspondingly,
	\begin{align}
		\frac{D_\text{eff}}{D_1}=\frac{\sqrt{\psi_0}}{\theta'_\text{rel}}\approx\frac{\sqrt{1 + 2\Pi}\mu'_1}{\ln(a_1/r_\text{tol})}.
		\label{eq:deff_general}	
	\end{align}
	\noindent %

Equation~\ref{eq:deff_general}, reproduces the observed $D_\text{eff}/D_1$ within $2\%$ for all $\Pi$. For example, as $\Pi \to 0$ the problem is linear and the slowest-decaying mode is the Fourier mode with coefficients $a_1=(2/\pi)\sqrt{3/2}$ and $a_1=(8/\pi^2)\sqrt{3/2}$ for imposed inlet pressure and fluxes, respectively, and eigenvalues $\mu'_1=\pi^2 $ and $\mu'_1=(\pi/2)^2$ for the same inlet conditions, which give the plateaus $1.48$ and $0.36$. For large $\Pi$, $a_1$, estimated from the simulations, are $\approx3$, $4$ and $16$, and the Bessel roots give $\mu'_1\approx6.41$, $1.96$ and $1.10$ (Appendix~\ref{app:eigenvalue_problem_degenerate_diffusion}) for imposed pressure, mass flux and volumetric flux, respectively, so that $D_\text{eff}/D_1\approx\sqrt{2\Pi}/c$, with $c=\ln(a_1/r_\text{tol})/\mu'_1$.

For imposed mass flux, the inlet pressure is free and the perturbation can only relax through the outlet, as in a domain twice as long, so $\mu'_1$ and hence $D_\text{eff}/D_1$ are smaller than for imposed pressure, by a factor $4$ at small $\Pi$ and $\approx3.3$ at large $\Pi$. For imposed volumetric flux, $D_\text{eff}/D_1$ is smaller than for imposed mass flux because the injected mass flux remains below its steady value during the transient, further delaying the relaxation. Accordingly, the ratio of mass to volumetric flux eigenvalues is $\approx1.8$ at large $\Pi$.

To interpret how the linearized model at arithmetic mean pressure, often used in gas permeameter measurements, compares to the PME, we consider the ratio of the effective diffusivity of the linearized model, $D^\text{lin}_\text{eff}/D_1\approx p_\text{ref}\,\mu^\text{lin}_1/\ln(a^\text{lin}_1/r_\text{tol})$, to $D_\text{eff}/D_1$ (Eq.~\ref{eq:deff_general}),
\begin{align}
\frac{D_\text{eff}^\text{lin}}{D_\text{eff}}\approx\frac{p_\text{ref}\,\mu^\text{lin}_1}{\sqrt{1+2\Pi}\,\mu'_1}\,\frac{\ln(a_1/r_\text{tol})}{\ln(a^\text{lin}_1/r_\text{tol})} 
\label{eq:deff_ratio_linearized}
\end{align}
\noindent where $a^\text{lin}_1\approx0.8$, $1$ and $1.6$, and $\mu^\text{lin}_1=\pi^2$, $(\pi/2)^2$ and $\nu^2$, with $\nu\cot\nu=\beta/2$, correspond to the linearized model for imposed pressure, mass flux and volumetric flux, respectively. 
As $r_\text{tol} \to 0$, we have $\ln(a_1/r_\text{tol})/ \ln(a^\text{lin}_1/r_\text{tol}) \to 1$. Further, as $\Pi \to \infty$, $p_\text{ref}/\sqrt{1+2\Pi} \to1/2$ while the analytical eigenvalues give $\mu^\text{lin}_1/\mu'_1\approx 1.54$, $1.26$ and $1.24$ for imposed pressure, mass flux and volumetric flux, respectively. Consequently,  $D^{\text{lin}}_\text{eff} / D_\text{eff} \to 0.77$, $0.63$ and $0.62$. For example, at $r_\text{tol}=10^{-3}$ and $\Pi=100$, the ratios are $0.97$, $0.79$ and $0.82$ (Fig.~\ref{fig:tau_diffusion}).

	\subsection{Characterization of flow compression through pressure curvature $\kappa(x,t)$} \label{subsec:transient_curvature}

\begin{figure}[H] 
		\includegraphics[width=\textwidth]{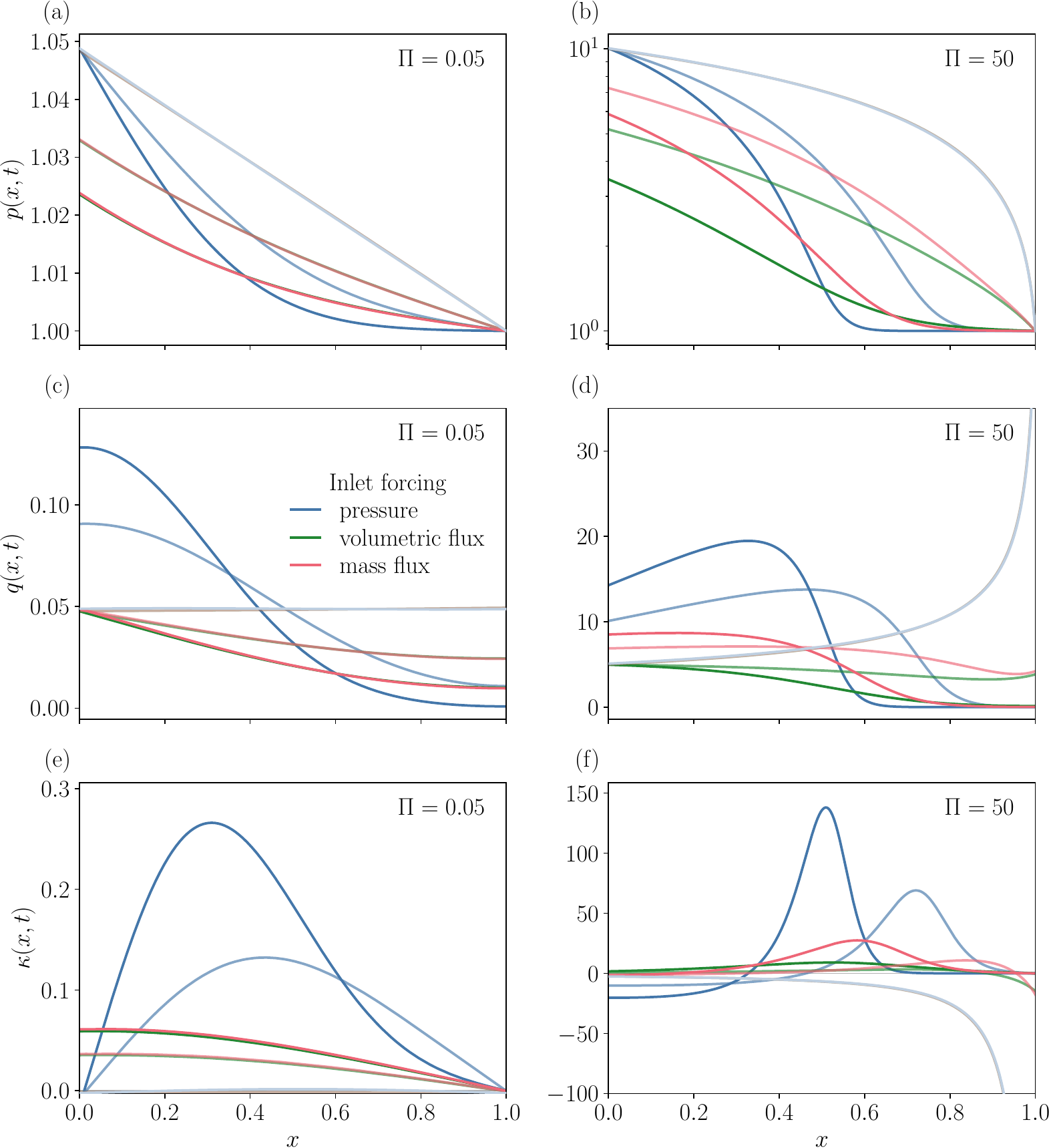}
		\centering
		\caption{Profiles of (a)-(b) pressure, (c)-(d) volumetric flux and (e)-(f) pressure curvature as a function of nondimensional distance $x$ at times (dark to light) $0.1 \times t_\text{rel}$, $0.2 \times t_\text{rel}$ and $ 1\times t_\text{rel}$ with $t_\text{rel}$ the relaxation time-scale (Eq.~\ref{eq:observed_diffusion_timescale}) for (left column) $\Pi = 0.05$ and (right column) $\Pi = 50$. Blue, green, and red lines correspond to imposed inlet pressure, volumetric flux and mass flux, respectively. For $\Pi=0.05$, the curves for imposed volumetric and mass flux overlap.}
		\label{fig:figure_4}
	\end{figure}

	We now use the pressure-curvature field $\kappa(x,t)=\nabla^2p(x,t)$ to characterize how transient compression and expansion organize inside the domain. Figure~\ref{fig:figure_4} shows snapshots of $\kappa(x,t)$ for $\Pi = 0.05$ and $\Pi = 50$ under the three inlet conditions. For reference comparison, profiles of pressure $p(x,t)$ and volumetric flux $q(x,t)$ are also shown.

	For $\Pi = 0.05$, both $q(x,t)$ and $\kappa(x,t)$ remain smooth during the transient. The $\kappa(x,t)$-field exhibits a broad positive peak of small amplitude that progressively relaxes toward the steady-state. Imposing pressure (blue) leads to the larger $\kappa(x,t)$ amplitudes, reflecting stronger compression for this case compared to imposed fluxes.
 	Since $\kappa=-\partial_x q$, compression is set by the decrease of $q$ from the inlet toward the front, which is larger for imposed pressure with inlet flux exceeding its steady value during the transient. In contrast, for imposed fluxes, the inlet flux remains close to its steady value (Fig.~\ref{fig:figure_4}c). For $\Pi = 50$ the flow $q(x,t)$ develops a sharp advancing front, more and less pronounced for imposed pressure and fluxes, respectively (Fig.~\ref{fig:figure_4}d). Correspondingly, $\kappa(x,t)$ (Fig.~\ref{fig:figure_4}f) forms localized positive pulses at the front, representing high flow compression, which are transported downstream, consistent with the positive curvature advection velocity $u_\kappa$ (Eq.~\ref{eq:kappa_adv_velocity}), and smoothed while the system relaxes toward steady-state. Behind the pulse, $\kappa<0$ for imposed pressure, indicating flow expansion near the inlet, while at late times strong expansion develops near the outlet as $\kappa$ approaches the steady boundary-layer distribution (Section~\ref{subsec:steady_state_analysis}).  The field $\kappa(x,t)$ provides a direct visualization of the emergence and subsequent distribution of flow compression. In the absence of boundaries, negative curvature (i.e., gas expansion) is bounded from below by the Aronson--Bénilan estimate, $\kappa\geq-\Pi/(3t)$ in one dimension (Eq.~\ref{eq:aronson_benilan}). Our solutions satisfy this bound until the outlet is felt. For imposed pressure at $\Pi=50$ we have for example $\min_x \kappa \approx -0.5 \Pi/(3t)$, meaning the strongest expansion reaches about half the bound.

\begin{figure}[H] 
		\includegraphics[width=\textwidth]{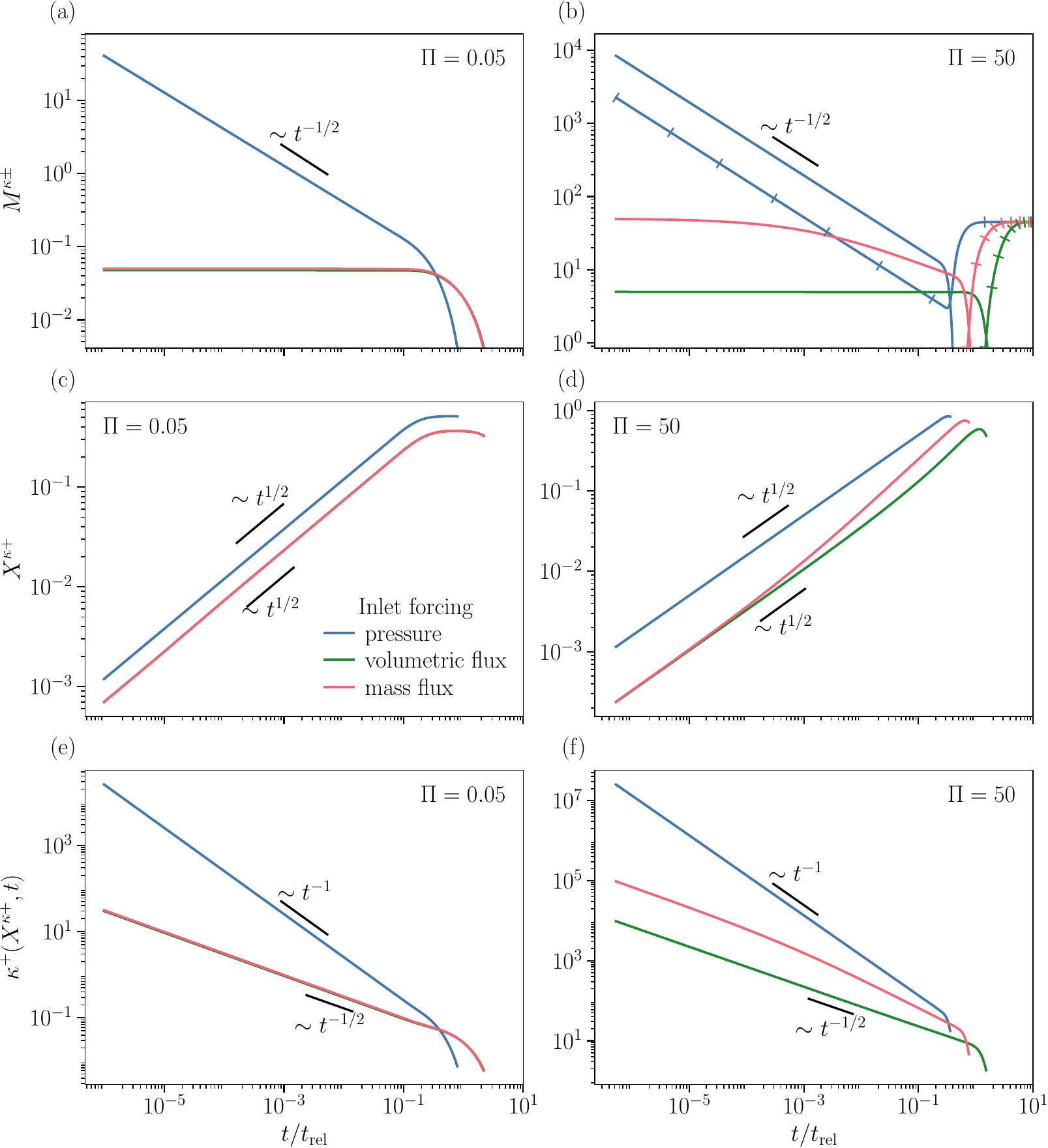}
		\centering
		\caption{Time series of (a)-(b) mass of positive and negative pressure curvature, $M^{\kappa\pm}$ (Eq.~\ref{eq:curvature_mass}), (c)-(d) centroid of positive curvature, $X^{\kappa+}$ (Eq.~\ref{eq:curvature_centroid}), and (e)-(f) positive curvature at the centroid, $\kappa^{+}(X^{\kappa+},t)$, as a function of time normalized by the relaxation time $t_\text{rel}$ for imposed inlet (blue) pressure, (green) volumetric flux and (red) mass flux, at (left) $\Pi = 0.05$ and (right) $\Pi = 50$. In (b), $M^{\kappa-}$ is indicated by lines with tick marks. Panels (c)-(f) are shown up to the crossover time $t_c$ at which $M^{\kappa+} = M^{\kappa-}$. Black segments indicate power-law scalings. For $\Pi=0.05$, the curves for imposed volumetric and mass flux overlap.} \label{fig:curvature_spatial_moments}
	\end{figure}

	Figure~\ref{fig:curvature_spatial_moments} reports time-series of mass of positive and negative curvature, respectively, $M^{\kappa+}(t)$ and $M^{\kappa-}(t)$, the centroid of positive curvature, $X^{\kappa+}(t)$, and the curvature at the centroid, $\kappa^{+}(X^{\kappa+},t)$ (Eqs.~\ref{eq:curvature_mass}--\ref{eq:curvature_centroid}), as a function of time normalized by $t_\text{rel}$ (Eq.~\ref{eq:observed_diffusion_timescale}), for $\Pi = 0.05$ and $\Pi = 50$ for the three inlet conditions.

	At early times, before the outlet boundary is felt, the positive curvature mass $M^{\kappa+}$ (Fig.~\ref{fig:curvature_spatial_moments}a-b) decays as $t^{-1/2}$ for imposed pressure, remains constant for imposed volumetric flux and decreases for imposed mass flux (for $\Pi=50$).
	 The negative curvature mass is negligible for $\Pi=0.05$ and for $\Pi=50$ it decays as $t^{-1/2}$ as well for imposed pressure, consistent with the expansion region behind the compression pulse near the inlet (Fig.~\ref{fig:figure_4}f). These trends reflect the inlet forcing, since integrating $\kappa=-\partial_x q$ over the domain gives $M^{\kappa+}-M^{\kappa-}=q(0,t)-q(1,t)$ and, before the outlet is felt, it is $q(1,t) \approx 0$ and thus $M^{\kappa+} -M^{\kappa-} \approx q(0,t)$. Further, for small $M^{\kappa-}$, occurring for $\Pi=0.05$ or imposed fluxes, we have $M^{\kappa+}\approx q(0,t)$.  Therefore, the early-time diffusive decay of the curvature mass for imposed pressure reflects the diffusive decay of the inlet flux $q(0,t)$, occurring because the pressure profile is self-similar in $x/t^{1/2}$ \cite[e.g.,][]{kidder1957unsteady}, so that the inlet pressure gradient decays as $t^{-1/2}$. For imposed volumetric flux, instead, the inlet flux is fixed at $q_0=\Pi/p_0$, whereas for imposed mass flux, $p(0,t)\,q(0,t)=\Pi$ is fixed, so that the inlet flux $q(0,t)=\Pi/p(0,t)$ decreases from $\Pi$ to $\Pi/p_0$ as the inlet pressure rises, consistent with Figure~\ref{fig:curvature_spatial_moments}b.

The centroid $X^{\kappa+}$ (Fig.~\ref{fig:curvature_spatial_moments}c-d) advances as $t^{1/2}$ for all inlet conditions, consistent with the diffusive propagation of the compression front, and saturates as the front reaches the outlet. At $\Pi=50$, the centroid advances faster than $t^{1/2}$ at intermediate times for imposed fluxes, since the inlet pressure, and thus the diffusivity behind the front, increases with time, whereas for imposed pressure it is fixed from the start. The curvature at the centroid $\kappa^{+}(X^{\kappa+},t)$ (Fig.~\ref{fig:curvature_spatial_moments}e-f) decays as $t^{-1}$ for imposed pressure and as $t^{-1/2}$ for imposed fluxes. These decays are consistent with 
\begin{align}
\kappa^{+}(X^{\kappa+},t)\sim M^{\kappa+}/\sigma^{\kappa+}		
\end{align}
\noindent with $\sigma^{\kappa+}$ the spread of positive curvature (second central moment), which grows as $t^{1/2}$ for all inlet conditions before the outlet is felt (not shown), that is, the compression pulse preserves its shape while it spreads. For imposed mass flux at $\Pi=50$, the decay is faster, since $M^{\kappa+}$ decreases while the centroid accelerates. 

\begin{figure}[H] 
		\includegraphics[width=0.7\textwidth]{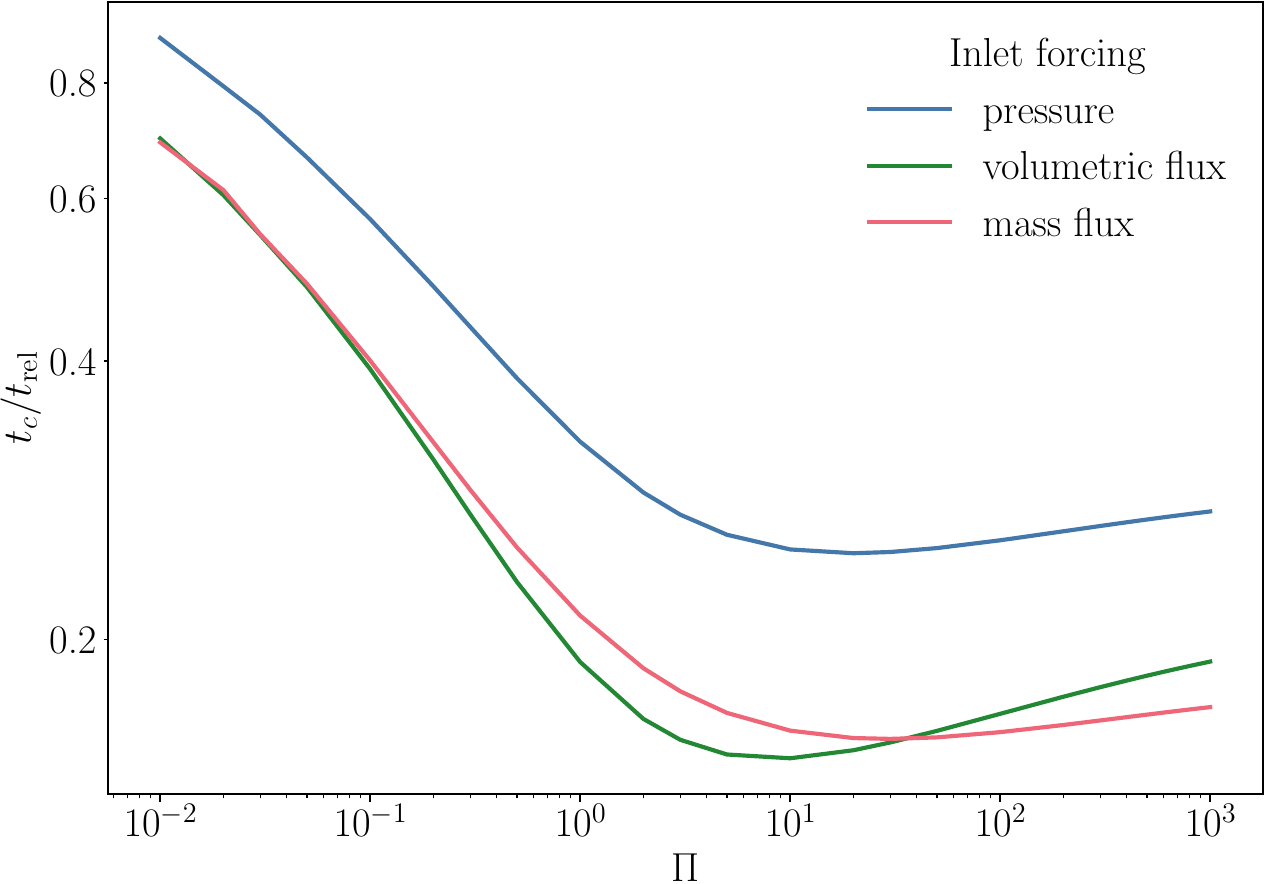}
		\centering
		\caption{Crossover time $t_c$, at which $M^{\kappa+}=M^{\kappa-}$ and the flow transitions from compression- to expansion-dominated, normalized by the relaxation time $t_\text{rel}$ of each inlet condition, as a function of $\Pi$, for imposed inlet (blue) pressure, (green) volumetric flux and (red) mass flux (logarithmic axes). At small $\Pi$, the curves for imposed volumetric and mass flux overlap.} \label{fig:figure_6}
	\end{figure}

	Figure~\ref{fig:figure_6} illustrates the crossover time $t_c$ for which $M^{\kappa+} = M^{\kappa-}$, normalized by $t_\text{rel}$ of each inlet condition, as a function of $\Pi$. Such time marks the transition from compression- to expansion-dominated flow which is also when the outlet volumetric flux equals the inlet one, since $\int^1_0 \kappa \, dx = M^{\kappa+} -  M^{\kappa-} = q(0,t) - q(1,t)$.

	For small $\Pi$, the crossover from  $M^{\kappa+}$ to $M^{\kappa-}$   (Fig.~\ref{fig:figure_6}) occurs at $t_c/t_\text{rel} \approx0.7\text{--}0.9$, indicating that the flow remains compression-dominated during most of the relaxation. As $\Pi$ increases, $t_c/t_\text{rel}$ decreases to $\approx0.25$ for imposed pressure and $\approx0.15$ for imposed fluxes at $\Pi\approx10$, indicating that flow expansion dominates earlier during the transient. Since the crossover occurs when the outlet volumetric flux equals the inlet one, its timing follows from the steady fluxes, $q_\text{s}=\Pi/p_\text{s}$ (Eq.~\ref{eq:general_solution_velocity}), that is, $\Pi/p_0$ at the inlet and $\Pi$ at the outlet, which differ by the steady expansion, $M^{\kappa-}_\text{s}=\Pi(1-1/p_0)$. During the transient, the outlet flux rises from zero toward $\Pi$, while the inlet flux decreases toward $\Pi/p_0$, or equals it for imposed volumetric flux. The crossover therefore requires the outlet flux to reach at least a fraction $1/p_0$ of its steady value. For small $\Pi$, inlet and outlet fluxes are nearly equal and
	$1/p_0\approx1-\Pi \approx 1$ so that fraction is nearly the steady value, which occurs only late in the relaxation. 
	For large $\Pi$, instead, the steady outlet-to-inlet flux ratio is large, $p_0\approx(2\Pi)^{1/2}$, because the expansion in the outlet boundary layer, with curvature $\sim\Pi^2$ over a width $\sim\Pi^{-1}$, raises the outlet flux to $\Pi$, whereas the inlet flux is only $\Pi/p_0\approx(\Pi/2)^{1/2}$. Then $1/p_0\ll1$, and the outlet flux equals the inlet flux already when it reaches a small fraction of its steady value, well before full relaxation.

	\section{Two application examples from gas chromatography and underground hydrogen storage} \label{sec:applications}

We now consider application examples from
gas-chromatography (GC) and underground hydrogen storage (UHS). In both settings, gas flows from an inlet, where pressure or flux is controlled, toward an outlet at fixed pressure. 
Standard operating conditions give $\Pi \approx 1$-$10$ for GC
and $\Pi \approx 0.1$-$7$ in UHS, within the nonlinear regime of the PME.  

\subsection{Gas chromatography (GC)} \label{subsec:app_GC}

Gas chromatography (GC) is an analytical technique used to separate and characterize mixtures of volatile components carried by a gas flowing through an elongated tube, either empty or packed with a porous medium. The mixture components are separated due to their different interactions with the stationary phase, and are eluted at the outlet at different retention times defined as $t_R=(1+k_R)\,t^m$, where $k_R>0$ is the retention constant of each component and $t^m$ the transit time unaffected by the stationary phase \citep{guiochon1988fundamentals,giddings2017dynamics}. 

GC studies usually assume steady flow so that transit times are computed as $t^{m}_{\text{steady}}=\int_0^L \mathrm{d}x/u_\text{s}(x)$, with $u_s(x)$ the steady velocity. Our parametrization gives $t^{m}_{\text{steady}}=\tau_D\,[(1+2\Pi)^{3/2}-1]/(3\Pi^2)$ (Eq.~\ref{eq:Pi_pressure_ratio}). However, when component particles are injected at the same time at which the flow is started-up, or for flow-programmed GC, whereby the carrier-gas flow is increased 
during the analysis,
to shorten the elution of strongly retained components \cite[e.g.,][]{guiochon1988fundamentals}, the steady flow is only an approximation.

\begin{figure}[H]
	\centering
	\includegraphics[width=0.7\textwidth]{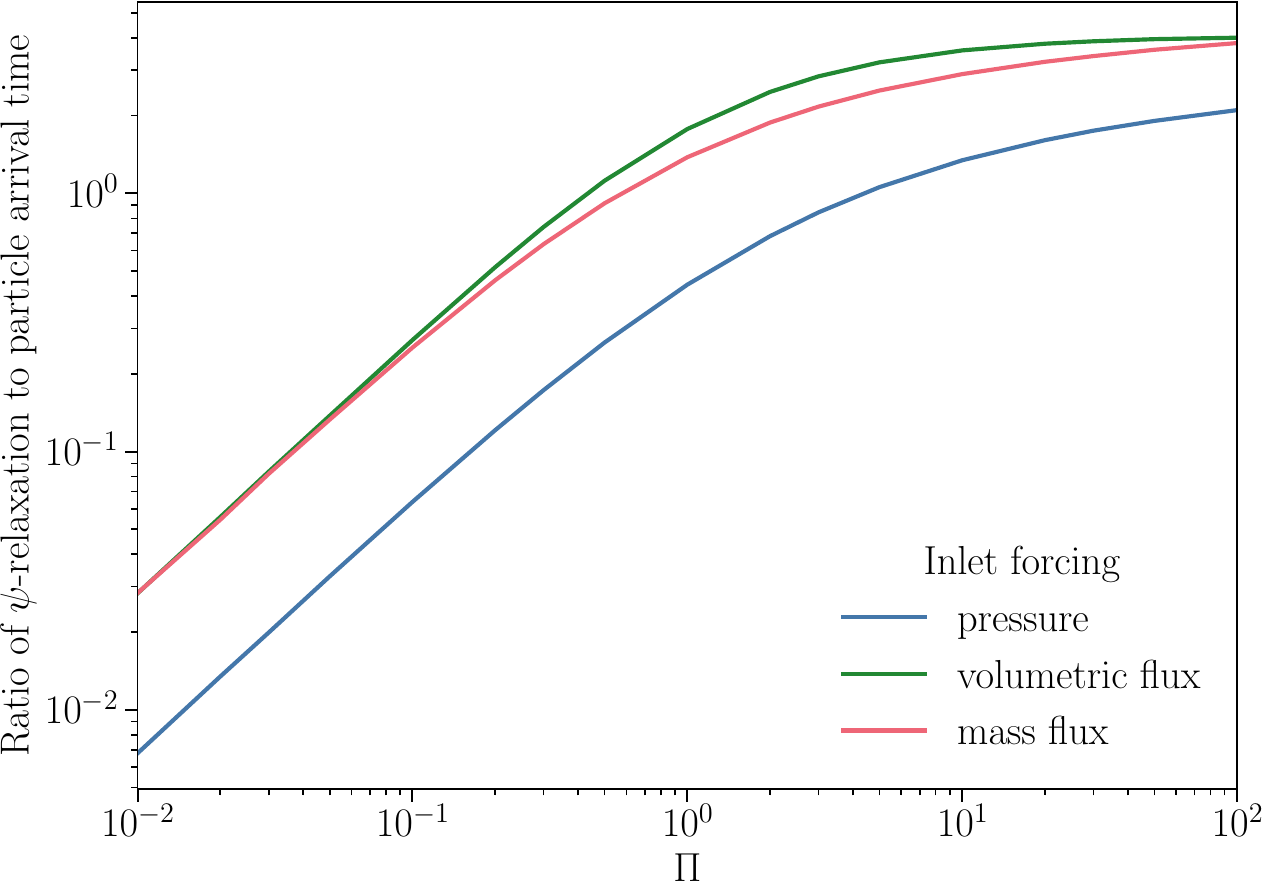}
	\caption{Relaxation time $t_\text{rel}$ of $\psi = p^2$ divided by the arrival time of a particle transported passively by the gas as a function of $\Pi$, for imposed inlet (blue) pressure, (green) volumetric flux and (red) mass flux (logarithmic axes).} \label{fig:GC_trel_tm}
\end{figure}

Figure~\ref{fig:GC_trel_tm} shows the ratio of the relaxation time $t_\text{rel}$ to the arrival time $t_\text{arr}$ of a passive particle starting at the inlet when the flow is started from rest. This ratio
depends only on $\Pi$ and inlet condition, independent of the used gas or column.

A threshold such as $t_\text{rel}/t_\text{arr} \leq 0.1$ gives allowable $\Pi\lesssim 0.16$ and $\Pi\lesssim0.037$ for imposed pressure and fluxes, respectively, while most settings typically have $\Pi \approx 1$--$10$.
For example, a standard setting can be an empty tubular column of radius $r=125\,\mu$m and length $L=60$~m operated with helium and whose outlet is kept at atmospheric pressure (Table~\ref{tab:app_GC}). The porosity is $\phi=1$ and Poiseuille flow gives permeability $k=r^2/8$, yielding $\tau_D\approx364$~s. Imposing an inlet pressure of  $p_0\approx3.3$~atm, a mass flux of $0.13$~kg~m$^{-2}$~s$^{-1}$ 
or a volumetric flux of $0.25$~m~s$^{-1}$ gives $\Pi=5$, with $t^m_\text{steady}\approx172$~s, relaxation times $t_\text{rel}\approx 116$, $413$ and $718$~s, and arrival times $t_\text{arr}\approx109$, $165$ and $223$~s, that is, $t_\text{rel}/t_\text{arr}\approx1.1$, $2.5$ and $3.2$, for the same inlet conditions.

\begin{table}[H]
	\centering
	\caption{Parameters and characteristic times for a GC capillary column at $\Pi=5$, with a pressure controller ($p$), a mass-flow controller ($q^\mathrm{m}$) and imposed volumetric flux ($q$).}
	\label{tab:app_GC}
	\begin{tabular}{ll}
		\hline
		Permeability $k$ & $r^2/8=1.95\times10^{-9}$ m$^2$ ($r=125\,\mu$m) \\
		Porosity $\phi$ & $1$ \\
		Viscosity $\mu$ (He) & $2.0\times10^{-5}$ Pa s \\
		Length $L$ & $60$ m \\
		Temperature $T$ & $25\,^\circ$C \\
		Outlet and inlet pressures $p_L$, $p_0$ & $1$ atm, $3.32$ atm \\
		Inlet mass flux $q^\mathrm{m}$ / volumetric flux $q$ & $0.13$ kg m$^{-2}$ s$^{-1}$ / $0.25$ m s$^{-1}$ \\
		Diffusive time $\tau_D$ at outlet pressure & $364$ s \\
		Steady transit time $t^m_\text{steady}$ & $172$ s \\
		Relaxation time $t_\text{rel}$ ($p$ / $q^\mathrm{m}$ / $q$) & $116$ / $413$ / $718$ s \\
		Arrival time $t_\text{arr}$ ($p$ / $q^\mathrm{m}$ / $q$) & $109$ / $165$ / $223$ s \\
		Ratio $t_\text{rel}/t_\text{arr}$ ($p$ / $q^\mathrm{m}$ / $q$) & $1.1$ / $2.5$ / $3.2$ \\
		Crossover time $t_c$ ($p$ / $q^\mathrm{m}$ / $q$) & $30$ / $69$ / $108$ s \\
		\hline
	\end{tabular}
\end{table}

The range of $\Pi$ values achieved in GC is $0.6\lesssim\Pi\lesssim5\times10^3$, since steady inlet-to-outlet pressure ratios from about $1.5$ up to $100$ are considered, the largest being required for long columns, used to separate components of similar $k_R$ \citep{guiochon1988fundamentals}.

\subsection{Underground hydrogen storage (UHS)} \label{subsec:app_UHS}

In UHS, hydrogen is injected into a porous formation through wells operated either at constant bottomhole pressure or at constant flow rate at standard conditions, which correspond to imposed inlet pressure and mass flux \citep[e.g.,][]{paterson1983implications,hagemann2024model}. We consider an idealized linear geometry, in which the injected gas flows over a length $L$ toward a boundary saturated with stationary brine and hence held at fixed (hydrostatic) pressure \citep[e.g.,][]{sainz2017assessment,castellucci2026dynamics}.

The ratio $t_\text{rel}/t_\text{arr}$ of Figure~\ref{fig:GC_trel_tm} is also relevant in UHS, whereby $t_\text{arr}$ represents the arrival time of injected hydrogen gas to the outlet, which is undesirable, or compounds transported passively by the gas, such as sulfur, which even at ultra-low concentrations can render hydrogen unsuitable for fuel cells \cite[e.g.,][]{shi2007hydrogen,xu2022accurate,bacquart2022measuring}. Thus, since $t_\text{rel}/t_\text{arr} \sim \mathcal{O}(1)$ except for $\Pi \lesssim 10^{-1}$, passive transport by hydrogen occurs over unsteady flow in most cases, an observation to be taken into account in future studies characterizing such transport. Since the single-gas model does not resolve the displacement of a different resident gas \cite[e.g.,][]{visentini2025gas} or brine \cite[e.g.,][]{castellucci2026dynamics}, these arrival times are estimates for such cases. Indeed, a more viscous resident fluid delays the arrival of hydrogen under pressure control, aside from viscous fingering effects that cause uncontrollable breakthrough \citep{paterson1983implications, homsy1987viscous}.

The pressure curvature $\kappa$ might find applications in reservoir operations.  
For instance, the crossover time $t_c$ marks when most of the compression has passed and
can be measured from the inlet and outlet volumetric fluxes, for example at a monitoring well close to the fixed outlet pressure, or in the laboratory, at the outlet of a column. Such measurements may help interpret gas-invasion dynamics, shown to be strongly affected by gas compression
that delays pressure build-up and leads to burst like behavior \citep{lee2019compressible,cuttle2023dynamics}, and delays the onset of viscous fingering \citep{morrow2023gas}. Further, in well testing \citep{chaudhry2003gas}, the distance first reached by a pressure perturbation is typically estimated by the linearized radius of investigation $r_\text{inv}=2 \sqrt{D_\text{ref}t}$, with the diffusivity linearized at some reference pressure. Since the curvature centroid $X^{\kappa+}$ tracks the position of this traveling compression front,
which advances with a pressure-dependent diffusivity and, for imposed fluxes, accelerates as the inlet pressure rises (Section~\ref{subsec:transient_curvature}, Fig.~\ref{fig:curvature_spatial_moments}c-d), it could help to better interpret the radius of investigation. %

As an example in physical units, we consider a sandstone with $k=100$~mD ($9.9\times10^{-14}$~m$^2$) %
, porosity $0.2$, $L=1$~km and $p_L=50$~bar, for which $\tau_D\approx42$~days (Table~\ref{tab:app_UHS}). Injection at $p_0\approx87$~bar gives $\Pi=1$, relaxation times $t_\text{rel}\approx21$, $82$ and $115$~days, crossover times $t_c\approx7$, $17$ and $22$~days and arrival times $t_\text{arr}\approx48$, $59$ and $65$~days for imposed pressure, mass flux and volumetric flux, respectively.

\begin{table}[H]
	\centering
	\caption{Parameters and characteristic times for UHS in a sandstone formation at $\Pi=1$, for imposed inlet pressure ($p$), mass rate ($q^\mathrm{m}$) and volumetric rate ($q$).}
	\label{tab:app_UHS}
	\begin{tabular}{ll}
		\hline
		Permeability $k$ & $100$ mD ($9.9\times10^{-14}$ m$^2$) \\
		Porosity $\phi$ & $0.2$ \\
		Viscosity $\mu$ (H$_2$) & $9\times10^{-6}$ Pa s \\
		Length $L$ & $1$ km \\
		Outlet and inlet pressures $p_L$, $p_0$ & $50$ bar, $87$ bar \\
		Diffusive time $\tau_D$ at outlet pressure & $42$ days \\
		Steady transit time $t^m_\text{steady}$ & $59$ days \\
		Relaxation time $t_\text{rel}$ ($p$ / $q^\mathrm{m}$ / $q$) & $21$ / $82$ / $115$ days \\
		Crossover time $t_c$ ($p$ / $q^\mathrm{m}$ / $q$) & $7$ / $17$ / $22$ days \\
		Arrival time $t_\text{arr}$ ($p$ / $q^\mathrm{m}$ / $q$) & $48$ / $59$ / $65$ days \\
		\hline
	\end{tabular}
\end{table} 

Regarding the range of $\Pi$, the injection pressure $p_0$ is bounded from above by the fracture pressure of the formation, itself bounded by the overburden pressure that generally increases with the depth of the reservoir \cite[e.g.,][]{chierici1994principles}. At the same time, the initial fluid pressure $p_L$ is close to hydrostatic, unless the reservoir has been depleted. Fracture pressure and initial fluid pressure set the attainable working pressure ratio $p_0/p_L$, and hence $\Pi$, such that mostly depleted reservoirs at higher depths typically allow the largest working pressure ratio, and hence $\Pi$ values, and vice versa. For example, \citet{lubon2020numerical} consider a $\approx1.2$~km deep aquifer initially at hydrostatic pressure of $\approx120$~bar, with fracture pressures of $200$--$255$~bar, giving $p_0/p_L\lesssim1.7$--$2$ or $\Pi\lesssim1$--$1.6$. \citet{okoroafor2022toward} consider reservoirs depleted to $32$~bar and bottom-hole pressures up to $120$~bar, giving $p_0/p_L\approx3.75$, that is, $\Pi\sim6.5$. Operating conditions may also lie below, for example, \citet{pfeiffer2017hydrogen} consider injection limited to $65$~bar in an aquifer initially at $\approx50$~bar, giving $\Pi\approx0.4$, whereas \citet{hagemann2024model} consider injection at $90$--$102$~bar into a gas field at $81.6$~barat $\approx1.2$~km depth, giving $\Pi\approx0.1$--$0.3$. Nonlinear effects become noticeable already for $\Pi\gtrsim0.1$  (e.g., Fig.~\ref{fig:tau_diffusion}). 

	\section{Discussion and conclusions} \label{sec:conclusions}

We have nondimensionalized the porous medium equation (PME) for the pressure-squared, $\psi = p^2$, through a Peclet-type number, denoted $\Pi$, that compares time scales of pressure diffusion and gas flow, and quantifies isothermal flow compression across the domain (i.e., a compressibility number, Eq.~\ref{eq:app_Pi_Cs}). Since a given $\Pi$ can be equivalently expressed in terms of the steady inlet pressure, volumetric flux and mass flux, the three inlet conditions lead to the same steady state, and hence flow compression, and their transients can be compared directly.

At steady
state, analytical expressions for pressure $p$, volumetric flux $q = -\nabla p$ and pressure-curvature $\kappa=\nabla^2p$ depend only on $\Pi$ and were used to show the presence of an outlet boundary layer of width $\delta_\text{BL} \sim \Pi^{-1}$, within which the velocity grows as $\sim \Pi$ and the amplitude of negative curvature, associated with flow expansion, grows as $\Pi^2$. For the transients, we obtained a semi-analytical solution of the PME as a perturbation series in the parameter $\beta=2\Pi/(1+2\Pi)$, by expanding around the inlet value of $\psi$, as done by \citet{polubarinova1962theory} for the Boussinesq equation. This gives $| \beta | <1$ for any $\Pi$, and the series converged for all tested $\Pi$ up to 100 (Appendix~\ref{subsec:appendix_C}). This is different from previous formulations such as that of \citet{kidder1957unsteady}, who used a different expansion parameter, $\alpha = -2\Pi$, which makes the perturbation series convergent only for $\Pi \lesssim \mathcal{O}(1)$ where the linearized PME is still accurate anyway. Instead, our solution was demonstrated for $\Pi = 50$, well within the nonlinear regime of the PME, 
which required truncation at order 100, illustrating the slow convergence of the series as $\Pi$ grows. Since such orders lie beyond what is tractable through closed-form expressions, 
we provided a recursive formula whereby the forcing terms at each order are computed from the lower orders, requiring only a numerical quadrature. The expansion in terms of $\beta(\Pi)$ applies to the three inlet conditions, by construction.

The transients were then analyzed from two complementary perspectives. First, we examined the domain-scale effective pressure-squared diffusivity, $D_\text{eff}$, which determines the relaxation time required for $\psi$ to reach its steady state. For $\Pi \ll 1$, $D_\text{eff}$ approaches constant plateau values, and for $\Pi \gg 1$ it scales as $\Pi^{1/2}$, with prefactors depending on the inlet condition. Both limits are explained by the slowest eigenmode of the PME linearized about the steady state. In particular, relaxation is slower for imposed volumetric flux than for imposed mass flux, because the reduced injection of gas mass during the transient makes the domain effectively longer. Further, linearizing the PME at the arithmetic mean of the steady inlet and outlet pressures, classically done to interpret gas permeability measurements \cite[e.g.,][]{klinkenberge1941permeability,chierici1994principles,tanikawa2009comparison}, gives, as $\Pi$ increases, effective diffusivities that approach $0.77$, $0.63$ and $0.62$ times the PME diffusivity $D_\text{eff}$, for imposed pressure, mass flux and volumetric flux, respectively. 
Second, we used the pressure curvature $\kappa$, which is proportional to the flow compression rate and vanishes for incompressible flows, to examine how flow compression distributes during the transients. We interpreted the evolution of $\kappa$ through a nonlinear advection--diffusion--reaction equation, implying that local compression and expansion evolve as a transported scalar field with sign. Pulses of positive $\kappa$ emerge near the inlet, migrate downstream and dissipate diffusively while the pressure approaches steady state, and, before the outlet is felt, $\kappa$ satisfies the Aronson--B\'enilan lower bound \citep{aronson1979regularite,lu2009local}, which is expressed in terms of $\Pi$ as $\kappa \geqslant -\Pi/(3t)$. The masses of positive and negative $\kappa$ show a crossover from compression- to expansion-dominated flow, when the outlet volumetric flux equals the inlet one, which occurs late in the relaxation for $\Pi \ll 1$ and early for $\Pi \gg 1$, depending also on the inlet condition. 

We finally considered two examples where our results could be useful, namely, gas chromatography (GC) and underground
hydrogen storage (UHS). In GC, typical operating conditions lie in the range $\Pi \approx 1$-$10$, with $\Pi$ up to $\approx5\times10^3$ being attainable for long chromatography columns designed to
separate components of similar retention times. In UHS, the fracture pressure limits $\Pi\lesssim2$ for reservoirs at hydrostatic pressure, with values up to $\Pi \approx 7$ being possible for depleted
reservoirs at high depths, and most typical operating conditions lying at $\Pi \approx 0.1$-$1.5$. Since gas flow nonlinearity becomes noticeable already for $\Pi \gtrsim 0.1$, all such settings are affected by these effects. For example, the effective $\psi$-diffusivities for imposed mass flux and volumetric flux differ by $\approx7\%$ at $\Pi=0.1$ and by $\approx42\%$ at $\Pi=1$ (Fig.~\ref{fig:tau_diffusion}). Further, computed ratios between the $\psi$ relaxation time and the arrival time of a particle transported passively by the flow show that for most cases in GC and UHS the flow transient lasts a substantial fraction of, or longer than, the transport time scale.

The present analysis is based on the PME, which makes a series of assumptions (see Appendix~\ref{sec:PME_region_validty}) including pressure-independent compressibility factor $Z$ and viscosity $\mu$, and negligible inertia through Darcy's law. These assumptions bound the inlet-outlet working pressure ratio set by $\Pi$ where the PME is applicable (Table~\ref{tab:PME_validity_Pi}).  Inertia does not bound $\Pi$ in the explored range because our estimations for hydrogen gas ($\text{H}_2$) at $320$~K give Reynolds numbers $\mathrm{Re}\ll1$ for $\Pi\leq10^3$. On the other hand, the pressure dependence of $Z$ and $\mu$ bounds $\Pi$ for $\text{H}_2$ as $\Pi\lesssim10^2$ at reference outlet pressure $p_L=10$~bar, while at $p_L=100$~bar no restriction appears up to inlet pressures of $200$~bar (Table~\ref{tab:PME_validity_Pi}). For methane, $\text{CH}_4$, the bounds are tighter, $\Pi\lesssim8$ and $\Pi\lesssim0.5$ at $p_L=10$ and $100$~bar, and for carbon dioxide, $\text{CO}_2$, even more, owing to its proximity to the critical point. 

This work introduces a framework to examine the nonlinear relationship between inlet forcing and gas pressure evolution through a Peclet-type number $\Pi$ and the pressure curvature. Besides relaxing the above simplifications and considering radial geometry around wells for more direct application to gas reservoirs, where the steady $\psi$ profile becomes logarithmic in the distance to the well, three natural extensions follow from the present formulation. First, spatial variability in the permeability, $k(\mathbf{x})$, leads to a more general PME \citep[][\S2.1.1]{vazquez2007porous}, given in Appendix~\ref{subsec:appendix_A} (Eq.~\ref{eq:PME_inhomogeneous_appA}). Variability in $k$ affects not only the local diffusion coefficient, $D\propto kp$, but also induces flow compression where the flow has a component along permeability gradients, since $-\mu\nabla \cdot \textbf{q} = k \kappa + \nabla k \cdot \nabla p$, as obtained by differentiating Darcy's law. The present setting could be used as a reference or equivalent 1-D model at given $\Pi$, defined through an effective permeability, to compare against 2-D or 3-D models containing heterogeneous $k(\mathbf{x})$-fields.
 For example, one could assess the impact of heterogeneity in randomly distributed permeability fields \cite[e.g.,][]{dagan1982analysis} on the effective $\psi$-diffusivity for compressible gases. Also, the curvature remains a measure of the compression term $k\kappa$, whose weight relative to the heterogeneity-induced term, $\nabla k\cdot\nabla p$, could be assessed as a function of $k(\mathbf{x})$. Second, in gas--liquid displacements, gas compression is further coupled to capillarity and to the motion of the interface \citep{cuttle2023dynamics,morrow2023gas,castellucci2026dynamics}. 
 For instance, \citet{castellucci2026dynamics} recently presented a vertically integrated (thin-layer) model for an imposed mass flux of gas into a confined, brine-filled porous layer, which reduces to the PME in the absence of liquid. Further, for constant $Z$, their compressibility parameter coincides with $\Pi$, which they consider up to $\Pi=\mathcal{O}(1)$. This suggests that the $\Pi$-based comparison of inlet conditions could be extended to such two-phase settings. Third, oscillatory pressure or flux forcing would introduce the ratio of the forcing period to the diffusive time $\tau_D$ as a second dimensionless number, providing a direct way to study frequency-dependent compression dynamics, including phase lags between boundary forcing, pressure propagation, flux response and curvature production.

	\section{Acknowledgments} \label{sec:acknowledgments}

	The authors acknowledge the financial support through Grant TED2021-129991B-C33 (GreenHUGS) funded by MCIN/AEI/10.13039/501100011033 and the European Union NextGenerationEU/PRTR. We also acknowledge the financial support through Grant 101192337 (HyDRA), co-funded by the European Union and supported by the Clean Hydrogen Partnership and its members. 

	\section{Declaration of interests} \label{sec:declaration_of_interest}

	The authors report no conflict of interests.

	\cleardoublepage
	\phantomsection
	\bibliographystyle{agufull08}
	\bibliography{references}

\begin{thebibliography}{78}
\providecommand{\natexlab}[1]{#1}
\expandafter\ifx\csname urlstyle\endcsname\relax
  \providecommand{\doi}[1]{doi:\discretionary{}{}{}#1}\else
  \providecommand{\doi}{doi:\discretionary{}{}{}\begingroup
  \urlstyle{rm}\Url}\fi

\bibitem[{\textit{Aanonsen}(1985)}]{aanonsen1985application}
Aanonsen, S. (1985), Application of pseudotime to estimate average reservoir
  pressure, in \textit{SPE Annual Technical Conference and Exhibition?}, pp.
  SPE--14,256, SPE.

\bibitem[{\textit{Acheson}(1990)}]{acheson1990elementary}
Acheson, D.~J. (1990), \textit{Elementary {F}luid {D}ynamics}, Oxford
  University Press.

\bibitem[{\textit{Al-Hussainy et~al.}(1966)\textit{Al-Hussainy, Ramey~Jr, and
  Crawford}}]{al1966flow}
Al-Hussainy, R., H.~Ramey~Jr, and P.~Crawford (1966), The flow of real gases
  through porous media, \textit{{J}ournal of {P}etroleum {T}echnology},
  \textit{18}(05), 624--636.

\bibitem[{\textit{Aris}(1956)}]{aris1956dispersion}
Aris, R. (1956), On the dispersion of a solute in a fluid flowing through a
  tube, \textit{Proceedings of the Royal Society of London. Series A.
  Mathematical and Physical Sciences}, \textit{235}(1200), 67--77.

\bibitem[{\textit{Aronson and B{\'e}nilan}(1979)}]{aronson1979regularite}
Aronson, D.~G., and P.~B{\'e}nilan (1979), R{\'e}gularit{\'e} des solutions de
  l'{\'e}quation des milieux poreux dans {$\mathbb{R}^n$}, \textit{Comptes
  Rendus de l'Acad{\'e}mie des Sciences de Paris, S{\'e}rie A--B},
  \textit{288}, A103--A105.

\bibitem[{\textit{Aronson and Graveleau}(1993)}]{aronson1993selfsimilar}
Aronson, D.~G., and J.~Graveleau (1993), A selfsimilar solution to the focusing
  problem for the porous medium equation, \textit{European Journal of Applied
  Mathematics}, \textit{4}(1), 65--81.

\bibitem[{\textit{Bachu}(2008)}]{bachu2008co2}
Bachu, S. (2008), C{O}2 storage in geological media: Role, means, status and
  barriers to deployment, \textit{Progress in {E}nergy and {C}ombustion
  {S}cience}, \textit{34}(2), 254--273.

\bibitem[{\textit{Bacquart et~al.}(2022)\textit{Bacquart, Morris, Bartlett,
  Kjos, Murugan, and Storms}}]{bacquart2022measuring}
Bacquart, T., A.~S.~O. Morris, S.~Bartlett, O.~Kjos, A.~Murugan, and W.~Storms
  (2022), Measuring total sulphur amount fraction at picomol/mol in hydrogen
  fuel: new results from cryo-{GC-SCD} analytical method, \textit{Processes},
  \textit{10}(7), 1393, \doi{10.3390/pr10071393}.

\bibitem[{\textit{Baehr and Joss}(1995)}]{baehr1995updated}
Baehr, A.~L., and C.~J. Joss (1995), An updated model of induced airflow in the
  unsaturated zone, \textit{Water Resources Research}, \textit{31}(2),
  417--421.

\bibitem[{\textit{Barenblatt}(1952)}]{barenblatt1952some}
Barenblatt, G.~I. (1952), On some unsteady motions of a liquid or a gas in a
  porous medium, \textit{Prikladnaya Matematika i Mekhanika}, \textit{16},
  67--78, (in Russian). Zbl 0049.41902, MR0046217.

\bibitem[{\textit{Barenblatt}(2003)}]{barenblatt2003scaling}
Barenblatt, G.~I. (2003), \textit{Scaling}, vol.~34, Cambridge University
  Press.

\bibitem[{\textit{Bear}(1972)}]{bear2013dynamics}
Bear, J. (1972), \textit{Dynamics of {F}luids in {P}orous {M}edia}, Dover
  {P}ublications.

\bibitem[{\textit{Bender and Orszag}(2013)}]{bender2013advanced}
Bender, C.~M., and S.~A. Orszag (2013), \textit{Advanced mathematical methods
  for scientists and engineers I: Asymptotic methods and perturbation theory},
  Springer Science \& Business Media.

\bibitem[{\textit{Burgers}(2013)}]{burgers2013nonlinear}
Burgers, J.~M. (2013), \textit{The nonlinear diffusion equation: asymptotic
  solutions and statistical problems}, Springer Science \& Business Media.

\bibitem[{\textit{Carslaw and Jaeger}(1959)}]{carslaw1959conduction}
Carslaw, H., and J.~Jaeger (1959), \textit{Conduction of {H}eat in {S}olids},
  Clarendon Press, {O}xford.

\bibitem[{\textit{Castellucci et~al.}(2026)\textit{Castellucci, Boya, Ma,
  Chernyavsky, and Jensen}}]{castellucci2026dynamics}
Castellucci, P., R.~Boya, L.~Ma, I.~L. Chernyavsky, and O.~E. Jensen (2026),
  Dynamics of a compressible gas injected into a confined porous layer,
  \textit{Journal of Fluid Mechanics}, \textit{1029}, A28.

\bibitem[{\textit{Chaudhry}(2003)}]{chaudhry2003gas}
Chaudhry, A.~U. (2003), \textit{Gas Well Testing Handbook}, Gulf Professional
  Publishing (Elsevier Science), Burlington, MA.

\bibitem[{\textit{Chierici}(1994)}]{chierici1994principles}
Chierici, G.~L. (1994), \textit{Principles of Petroleum Reservoir Engineering:
  Volume 1}, Springer-Verlag, Berlin, Heidelberg, translated from the Italian
  by P. J. Westaway.

\bibitem[{\textit{Crank}(1975)}]{crank1975mathematics}
Crank, J. (1975), \textit{The {M}athematics of {D}iffusion}, Clarendon Press,
  {O}xford.

\bibitem[{\textit{Cuttle and MacMinn}(2023)}]{cuttle2023dynamics}
Cuttle, C., and C.~W. MacMinn (2023), Dynamics of compression-driven gas-liquid
  displacement in a capillary tube, \textit{Physical Review Letters},
  \textit{130}(11), 114,001.

\bibitem[{\textit{Dagan}(1982)}]{dagan1982analysis}
Dagan, G. (1982), Analysis of flow through heterogeneous random aquifers: 2.
  unsteady flow in confined formations, \textit{Water Resources Research},
  \textit{18}(5), 1571--1585.

\bibitem[{{\relax DLMF}()}]{NIST:DLMF}
{\relax DLMF} (), {\it NIST Digital Library of Mathematical Functions},
  \url{https://dlmf.nist.gov/}, Release 1.2.7 of 2026-06-15, f.~W.~J. Olver,
  A.~B. {Olde Daalhuis}, D.~W. Lozier, B.~I. Schneider, R.~F. Boisvert, C.~W.
  Clark, B.~R. Miller, B.~V. Saunders, H.~S. Cohl, and M.~A. McClain, eds.

\bibitem[{\textit{Giddings}(2017)}]{giddings2017dynamics}
Giddings, J.~C. (2017), \textit{Dynamics of chromatography: principles and
  theory}, CRC Press.

\bibitem[{\textit{Guiochon and Guillemin}(1988)}]{guiochon1988fundamentals}
Guiochon, G., and C.~L. Guillemin (1988), Fundamentals of the chromatographic
  process: flow of gases through chromatographic columns, in
  \textit{Quantitative Gas Chromatography for Laboratory Analyses and On-Line
  Process Control}, \textit{Journal of Chromatography Library}, vol.~42,
  chap.~2, pp. 35--55, Elsevier, \doi{10.1016/S0301-4770(08)70074-5}.

\bibitem[{\textit{Haberman}(2013)}]{haberman2013applied}
Haberman, R. (2013), \textit{Applied Partial Differential Equations with
  Fourier Series and Boundary Value Problems}, 5 ed., Pearson, Boston.

\bibitem[{\textit{Hagemann et~al.}(2024)\textit{Hagemann, Hogeweg, and
  L{\"u}ddeke}}]{hagemann2024model}
Hagemann, B., S.~Hogeweg, and C.~L{\"u}ddeke (2024), Model development for
  microbial growth in subsurface reservoirs, in \textit{ECMOR 2024}, vol. 2024,
  pp. 1--12, European Association of Geoscientists \& Engineers.

\bibitem[{\textit{Heinemann et~al.}(2021)\textit{Heinemann, Alcalde, Miocic,
  Hangx, Kallmeyer, Ostertag-Henning, Hassanpouryouzband, Thaysen, Strobel,
  Schmidt-Hattenberger et~al.}}]{heinemann2021enabling}
Heinemann, N., J.~Alcalde, J.~M. Miocic, S.~J. Hangx, J.~Kallmeyer,
  C.~Ostertag-Henning, A.~Hassanpouryouzband, E.~M. Thaysen, G.~J. Strobel,
  C.~Schmidt-Hattenberger, et~al. (2021), Enabling large-scale hydrogen storage
  in porous media--the scientific challenges, \textit{Energy \& {E}nvironmental
  {S}cience}, \textit{14}(2), 853--864.

\bibitem[{\textit{Homsy}(1987)}]{homsy1987viscous}
Homsy, G.~M. (1987), Viscous fingering in porous media, \textit{Annual {R}eview
  of {F}luid {M}echanics}, \textit{19}(1), 271--311.

\bibitem[{\textit{Hu et~al.}(2010)\textit{Hu, Wu, Liu, Meegoda, and
  Gao}}]{hu2010physical}
Hu, L., X.~Wu, Y.~Liu, J.~N. Meegoda, and S.~Gao (2010), Physical modeling of
  air flow during air sparging remediation, \textit{Environmental {S}cience \&
  {T}echnology}, \textit{44}(10), 3883--3888.

\bibitem[{\textit{Hunt et~al.}(1988)\textit{Hunt, Sitar, and
  Udell}}]{hunt1988nonaqueous}
Hunt, J.~R., N.~Sitar, and K.~S. Udell (1988), Nonaqueous phase liquid
  transport and cleanup: 1. analysis of mechanisms, \textit{Water {R}esources
  {R}esearch}, \textit{24}(8), 1247--1258.

\bibitem[{\textit{Kidder}(1957)}]{kidder1957unsteady}
Kidder, R.~E. (1957), Unsteady flow of gas through a semi-infinite porous
  medium, \textit{Journal of Applied Mechanics}, \textit{24}(3), 329--332,
  \doi{10.1115/1.4011542}.

\bibitem[{\textit{Klinkenberg}(1941)}]{klinkenberge1941permeability}
Klinkenberg, L. (1941), The permeability of porous media to liquids and gases,
  \textit{Drilling and {P}roduction {P}ractice}, pp. 200--213.

\bibitem[{\textit{Knuth}(1997)}]{knuth1997art}
Knuth, D.~E. (1997), \textit{The Art of Computer Programming, Volume 2:
  Seminumerical Algorithms}, 3 ed., Addison-Wesley, Reading, MA.

\bibitem[{\textit{Lasseux et~al.}(2021)\textit{Lasseux, Vald{\'e}s-Parada, and
  Bottaro}}]{lasseux2021upscaled}
Lasseux, D., F.~J. Vald{\'e}s-Parada, and A.~Bottaro (2021), Upscaled model for
  unsteady slip flow in porous media, \textit{Journal of Fluid Mechanics},
  \textit{923}, A37.

\bibitem[{\textit{Lee et~al.}(2019)\textit{Lee, Zhao, Abouatallah, Wang, and
  Bazylak}}]{lee2019compressible}
Lee, C., B.~Zhao, R.~Abouatallah, R.~Wang, and A.~Bazylak (2019),
  Compressible-gas invasion into liquid-saturated porous media: application to
  polymer-electrolyte-membrane electrolyzers, \textit{Physical Review Applied},
  \textit{11}(5), 054,029.

\bibitem[{\textit{Lee and V{\'a}zquez}(2003)}]{lee2003geometrical}
Lee, K.-A., and J.~L. V{\'a}zquez (2003), Geometrical properties of solutions
  of the porous medium equation for large times, \textit{Indiana University
  Mathematics Journal}, pp. 991--1016.

\bibitem[{\textit{Lie}(2019)}]{lie2019introduction}
Lie, K.-A. (2019), \textit{An introduction to reservoir simulation using
  MATLAB/GNU Octave: User guide for the MATLAB Reservoir Simulation Toolbox
  (MRST)}, Cambridge University Press.

\bibitem[{\textit{Lie and M{\o}yner}(2021)}]{lie2021advanced}
Lie, K.-A., and O.~M{\o}yner (2021), \textit{Advanced modelling with the MATLAB
  {R}eservoir {S}imulation {T}oolbox}, Cambridge University Press.

\bibitem[{\textit{Liepmann and Roshko}(2001)}]{liepmann2001elements}
Liepmann, H.~W., and A.~Roshko (2001), \textit{Elements of {G}as {D}ynamics},
  Courier Corporation.

\bibitem[{\textit{Lu et~al.}(2009)\textit{Lu, Ni, V{\'a}zquez, and
  Villani}}]{lu2009local}
Lu, P., L.~Ni, J.-L. V{\'a}zquez, and C.~Villani (2009), Local
  {A}ronson--{B}{\'e}nilan estimates and entropy formulae for porous medium and
  fast diffusion equations on manifolds, \textit{Journal de Math{\'e}matiques
  Pures et Appliqu{\'e}es}, \textit{91}(1), 1--19,
  \doi{10.1016/j.matpur.2008.09.001}.

\bibitem[{\textit{Lubo{\'n} and Tarkowski}(2020)}]{lubon2020numerical}
Lubo{\'n}, K., and R.~Tarkowski (2020), Numerical simulation of hydrogen
  injection and withdrawal to and from a deep aquifer in {NW} poland,
  \textit{International {J}ournal of {H}ydrogen {E}nergy}, \textit{45}(3),
  2068--2083.

\bibitem[{\textit{Martinson}(1976)}]{martinson1976finite}
Martinson, L.~K. (1976), Finite velocity of propagation of thermal
  perturbations in media with constant heat conduction coefficient,
  \textit{Zhurnal Vychislitelnoi Matematiki i Matematicheskoi Fiziki},
  \textit{16}, 1233--1241.

\bibitem[{\textit{Massmann}(1989)}]{massmann1989applying}
Massmann, J.~W. (1989), Applying groundwater flow models in vapor extraction
  system design, \textit{Journal of Environmental Engineering},
  \textit{115}(1), 129--149.

\bibitem[{\textit{Massol and Jaupart}(1999)}]{massol1999generation}
Massol, H., and C.~Jaupart (1999), The generation of gas overpressure in
  volcanic eruptions, \textit{Earth and Planetary Science Letters},
  \textit{166}(1-2), 57--70.

\bibitem[{\textit{Matthews et~al.}(1967)\textit{Matthews, Russell
  et~al.}}]{matthews1967pressure}
Matthews, C.~S., D.~G. Russell, et~al. (1967), \textit{Pressure {B}uildup and
  {F}low {T}ests in {W}ells}, vol.~1, Henry L. Doherty Memorial Fund of AIME
  New York.

\bibitem[{\textit{Mei and Auriault}(1991)}]{mei1991effect}
Mei, C., and J.-L. Auriault (1991), The effect of weak inertia on flow through
  a porous medium, \textit{Journal of Fluid Mechanics}, \textit{222}, 647--663.

\bibitem[{\textit{Monteiro et~al.}(2012)\textit{Monteiro, Rycroft, and
  Barenblatt}}]{monteiro2012mathematical}
Monteiro, P.~J., C.~H. Rycroft, and G.~I. Barenblatt (2012), A mathematical
  model of fluid and gas flow in nanoporous media, \textit{Proceedings of the
  National Academy of Sciences}, \textit{109}(50), 20,309--20,313.

\bibitem[{\textit{Morrow et~al.}(2023)\textit{Morrow, Cuttle, and
  MacMinn}}]{morrow2023gas}
Morrow, L.~C., C.~Cuttle, and C.~W. MacMinn (2023), Gas compression
  systematically delays the onset of viscous fingering, \textit{Physical Review
  Letters}, \textit{131}(22), 224,002, \doi{10.1103/PhysRevLett.131.224002}.

\bibitem[{\textit{Murray}(2002)}]{murray2002biological}
Murray, J.~D. (2002), Biological waves: Single-species models, in
  \textit{Mathematical Biology: I. An Introduction}, pp. 437--483, Springer.

\bibitem[{\textit{Muskat}(1934)}]{muskat1934flow}
Muskat, M. (1934), The flow of compressible fluids through porous media and
  some problems in heat conduction, \textit{Physics}, \textit{5}(3), 71--94.

\bibitem[{\textit{Muskat}(1937)}]{muskat1937use}
Muskat, M. (1937), Use of data oil the build-up of bottom-hole pressures,
  \textit{Transactions of the AIME}, \textit{123}(01), 44--48.

\bibitem[{\textit{Neeper}(2003)}]{neeper2003harmonic}
Neeper, D.~A. (2003), Harmonic analysis of flow in open boreholes due to
  barometric pressure cycles, \textit{Journal of Contaminant Hydrology},
  \textit{60}(3--4), 135--162, \doi{10.1016/S0169-7722(02)00086-4}.

\bibitem[{\textit{Okoroafor et~al.}(2022)\textit{Okoroafor, Saltzer, and
  Kovscek}}]{okoroafor2022toward}
Okoroafor, E.~R., S.~D. Saltzer, and A.~R. Kovscek (2022), Toward underground
  hydrogen storage in porous media: Reservoir engineering insights,
  \textit{International Journal of Hydrogen Energy}, \textit{47}(79),
  33,781--33,802.

\bibitem[{\textit{Oldenburg}(2007)}]{oldenburg2007joule}
Oldenburg, C.~M. (2007), Joule-{T}homson cooling due to {CO}$_2$ injection into
  natural gas reservoirs, \textit{Energy {C}onversion and {M}anagement},
  \textit{48}(6), 1808--1815, \doi{10.1016/j.enconman.2007.01.010}.

\bibitem[{\textit{Oppenheimer et~al.}(2013)\textit{Oppenheimer, Fischer, and
  Scaillet}}]{oppenheimer2013volcanic}
Oppenheimer, C., T.~P. Fischer, and B.~Scaillet (2013), Volcanic degassing:
  Process and impact, \textit{Treatise on Geochemistry: Second Edition}, pp.
  111--179.

\bibitem[{\textit{Paterson}(1983)}]{paterson1983implications}
Paterson, L. (1983), The implications of fingering in underground hydrogen
  storage, \textit{International Journal of Hydrogen Energy}, \textit{8}(1),
  53--59, \doi{10.1016/0360-3199(83)90035-6}.

\bibitem[{\textit{Patzek et~al.}(2013)\textit{Patzek, Male, and
  Marder}}]{patzek2013gas}
Patzek, T.~W., F.~Male, and M.~Marder (2013), Gas production in the barnett
  shale obeys a simple scaling theory, \textit{Proceedings of the National
  Academy of Sciences}, \textit{110}(49), 19,731--19,736.

\bibitem[{\textit{Peng and Robinson}(1976)}]{peng1976new}
Peng, D.-Y., and D.~B. Robinson (1976), A new two-constant equation of state,
  \textit{Industrial \& Engineering Chemistry Fundamentals}, \textit{15}(1),
  59--64.

\bibitem[{\textit{Pfeiffer et~al.}(2017)\textit{Pfeiffer, Beyer, and
  Bauer}}]{pfeiffer2017hydrogen}
Pfeiffer, W.~T., C.~Beyer, and S.~Bauer (2017), Hydrogen storage in a
  heterogeneous sandstone formation: dimensioning and induced hydraulic
  effects, \textit{Petroleum Geoscience}, \textit{23}(3), 315--326,
  \doi{10.1144/petgeo2016-050}.

\bibitem[{\textit{Polubarinova-Kochina}(1948)}]{polubarinova1948nonlinear}
Polubarinova-Kochina, P.~Y. (1948), On a nonlinear differential equation
  encountered in the theory of infiltration, in \textit{Dokl. Akad. Nauk SSSR},
  vol.~63, pp. 623--627.

\bibitem[{\textit{Polubarinova-Kochina}(1962)}]{polubarinova1962theory}
Polubarinova-Kochina, P.~Y. (1962), \textit{Theory of Ground Water Movement},
  Princeton University Press, Princeton, New Jersey.

\bibitem[{\textit{Ross and Lu}(1994)}]{ross1994efficiency}
Ross, B., and N.~Lu (1994), Efficiency of air inlet wells in vapor extraction
  systems, \textit{Water Resources Research}, \textit{30}(2), 581--584,
  \doi{10.1029/93WR02902}.

\bibitem[{\textit{Saffman}(1992)}]{saffman1992vortex}
Saffman, P.~G. (1992), Vortex dynamics, in \textit{Theoretical Approaches to
  Turbulence}, pp. 263--277, Springer.

\bibitem[{\textit{S{\'a}inz-Garc{\'\i}a
  et~al.}(2017)\textit{S{\'a}inz-Garc{\'\i}a, Abarca, Rub{\'\i}, and
  Grandia}}]{sainz2017assessment}
S{\'a}inz-Garc{\'\i}a, A., E.~Abarca, V.~Rub{\'\i}, and F.~Grandia (2017),
  Assessment of feasible strategies for seasonal underground hydrogen storage
  in a saline aquifer, \textit{International {J}ournal of {H}ydrogen {E}nergy},
  \textit{42}(26), 16,657--16,666.

\bibitem[{\textit{Salmon}(1998)}]{salmon1998lectures}
Salmon, R. (1998), \textit{Lectures on geophysical fluid dynamics}, Oxford
  University Press.

\bibitem[{\textit{Scandella et~al.}(2016)\textit{Scandella, Pillsbury, Weber,
  Ruppel, Hemond, and Juanes}}]{scandella2016ephemerality}
Scandella, B.~P., L.~Pillsbury, T.~Weber, C.~Ruppel, H.~F. Hemond, and
  R.~Juanes (2016), Ephemerality of discrete methane vents in lake sediments,
  \textit{Geophysical Research Letters}, \textit{43}(9), 4374--4381.

\bibitem[{\textit{Shi et~al.}(2007)\textit{Shi, Yi, Hou, Jing, and
  Ming}}]{shi2007hydrogen}
Shi, W., B.~Yi, M.~Hou, F.~Jing, and P.~Ming (2007), Hydrogen sulfide poisoning
  and recovery of pemfc pt-anodes, \textit{Journal of power sources},
  \textit{165}(2), 814--818.

\bibitem[{\textit{Tanikawa and Shimamoto}(2009)}]{tanikawa2009comparison}
Tanikawa, W., and T.~Shimamoto (2009), Comparison of {K}linkenberg-corrected
  gas permeability and water permeability in sedimentary rocks,
  \textit{International Journal of Rock Mechanics and Mining Sciences},
  \textit{46}(2), 229--238, \doi{10.1016/j.ijrmms.2008.03.004}.

\bibitem[{\textit{Taylor}(1950)}]{taylor1950instability}
Taylor, G.~I. (1950), The instability of liquid surfaces when accelerated in a
  direction perpendicular to their planes. i, \textit{Proceedings of the Royal
  Society of London. Series A. Mathematical and Physical Sciences},
  \textit{201}(1065), 192--196.

\bibitem[{\textit{Tek}(1989)}]{tek1989underground}
Tek, M.~R. (1989), \textit{Underground {S}torage of {N}atural {G}as: {T}heory
  and {P}ractice}, 171, Springer Science \& Business Media.

\bibitem[{\textit{Thomas}(2013)}]{thomas2013numerical}
Thomas, J.~W. (2013), \textit{Numerical partial differential equations: finite
  difference methods}, vol.~22, Springer Science \& Business Media.

\bibitem[{\textit{V{\'a}zquez}(2007)}]{vazquez2007porous}
V{\'a}zquez, J.~L. (2007), \textit{The {P}orous {M}edium {E}quation:
  {M}athematical {T}heory}, Oxford {U}niversity {P}ress.

\bibitem[{\textit{Visentini et~al.}(2025)\textit{Visentini, Hidalgo,
  Cueto-Felgueroso, and Dentz}}]{visentini2025gas}
Visentini, A.~F., J.~J. Hidalgo, L.~Cueto-Felgueroso, and M.~Dentz (2025), Gas
  spreading and mixing in heterogeneous porous media for underground hydrogen
  storage, \textit{International Journal of Hydrogen Energy}, \textit{146},
  149,753.

\bibitem[{\textit{Wu et~al.}(1998)\textit{Wu, Pruess, and Persoff}}]{wu1998gas}
Wu, Y.-S., K.~Pruess, and P.~Persoff (1998), Gas flow in porous media with
  {K}linkenberg effects, \textit{Transport in {P}orous {M}edia}, \textit{32},
  117--137.

\bibitem[{\textit{Xiao et~al.}(2019)\textit{Xiao, Bernab{\'e}, Evans, Mok,
  Zhao, Ren, and Chen}}]{xiao2019klinkenberg}
Xiao, W., Y.~Bernab{\'e}, B.~Evans, U.~Mok, J.~Zhao, X.~Ren, and M.~Chen
  (2019), Klinkenberg effect and effective pressure for gas permeability of
  tight sandstones, \textit{Journal of Geophysical Research: Solid Earth},
  \textit{124}(2), 1412--1429.

\bibitem[{\textit{Xu et~al.}(2022)\textit{Xu, Zhang, Wang, Dong, Qiu, Xu, Zong,
  and Xie}}]{xu2022accurate}
Xu, C., Y.~Zhang, Y.~Wang, J.~Dong, L.~Qiu, G.~Xu, B.~Zong, and Z.~Xie (2022),
  Accurate quantification of ultra-trace sulfur compounds in hydrogen by
  integrating fill-less trap pre-concentration with gas chromatograph and
  sulfur chemiluminescence detector, \textit{International Journal of Hydrogen
  Energy}, \textit{47}(55), 23,264--23,277,
  \doi{10.1016/j.ijhydene.2022.04.272}.

\bibitem[{\textit{Zel'dovich and Kompanyeets}(1950)}]{zeldovich1950theory}
Zel'dovich, Y.~B., and A.~S. Kompanyeets (1950), Towards a theory of heat
  conduction with thermal conductivity depending on the temperature, in
  \textit{Collection of Papers Dedicated to the 70th Birthday of A. F. Ioffe},
  pp. 61--72, Izd. Akad. Nauk SSSR, Moscow, mR0069381.

\bibitem[{\textit{Ziarani and Aguilera}(2012)}]{ziarani2012knudsen}
Ziarani, A.~S., and R.~Aguilera (2012), Knudsen's permeability correction for
  tight porous media, \textit{Transport in {P}orous {M}edia}, \textit{91}(1),
  239--260, \doi{10.1007/s11242-011-9842-6}.

\end{thebibliography}
	\addcontentsline{toc}{chapter}{Bibliography}

	\appendix

	\section{Governing equations and pressure-squared PME} \label{subsec:appendix_A}
	\renewcommand{\thefigure}{A\arabic{figure}}
	\setcounter{figure}{0}
	\numberwithin{equation}{section}\setcounter{equation}{0}

	The laminar flow of real gases in porous media is described by the following set of equations with adequate initial and boundary conditions:
	\begin{align}
		\frac{\partial \phi \rho}{ \partial t } = -\nabla \cdot (\rho \textbf{q} ), \label{eq:continuity_equation}\\
		\textbf{q}  = - \frac{k}{\mu}  \bigg[ \nabla p + \rho g \nabla z  \bigg], \label{eq:darcys_law}\\
		\rho = \frac{pW}{RTZ} \label{eq:eos}.
	\end{align}

	\noindent Equation~\ref{eq:continuity_equation} expresses fluid mass conservation in the absence of sources or sinks, with $\phi(\textbf{x},t)$ the porosity $\rho(\textbf{x},t)$ the fluid mass density ($\si{kg.m^{-3}}$) and $\textbf{q}(\textbf{x},t)$ the (volumetric) specific discharge ($\si{m.s^{-1}}$). Darcy's law (Eq.~\ref{eq:darcys_law}), is 
	a statement of momentum conservation at the continuum scale assuming laminar flow and absence of inertial effects \cite[e.g.][]{mei1991effect}, with $k(\bold{x})$, $\mu(\bold{x},t)$, $p(\bold{x},t)$ and $g$ denoting, respectively, the intrinsic permeability of the medium (in $\si{m^2}$), the dynamic viscosity of the fluid (in $\si{Pa.s}$), the phase pressure (in $\si{Pa}$) and the constant of gravity acceleration (in $\si{m.s^{-2}}$), and $z$ the elevation. 
	The equation of state~\ref{eq:eos} relates the gas density with its pressure, via the average molecular weight of the gas $W$ ($\text{kg}\,\text{mol}^{-1}$), the universal gas constant $R = 8.314\,\text{J}\,\text{mol}^{-1}\,\text{K}^{-1}$, the temperature $T$ (K) and the compressibility factor $Z(p)$, the latter measuring the deviation of a real gas (or mixture) from ideal behavior. 

Substituting Equations~\ref{eq:darcys_law} and \ref{eq:eos} into \ref{eq:continuity_equation}, with $g=0$ and constant $\phi$ and $T$, gives \cite[e.g.,][]{al1966flow}
	\begin{equation} \label{eq:gas_flow_1_appA}
		\phi\,\frac{\partial}{\partial t}\left(\frac{p}{Z}\right) = \nabla\cdot\left(k\,f_R\,p\,\nabla p\right),
	\end{equation}
	where 
	$f_R=1/\mu Z$ is the real-gas factor. 
		Since $Z=Z(p)$ at constant $T$, the accumulation term on the left-hand side of Equation~\ref{eq:gas_flow_1_appA} can be written as $\partial_t(p/Z)=A_c\,\partial_t p$, with the accumulation coefficient
	\begin{equation} \label{eq:gas_compressibility_appA}
		A_c=\frac{\mathrm{d}}{\mathrm{d}p}\left(\frac{p}{Z}\right)=\frac{p}{Z}\,\gamma_\mathrm{g}, 
	\end{equation}
	where $\gamma_\mathrm{g}$ is the isothermal gas compressibility,
	\begin{align}
	\gamma_\mathrm{g} = \frac{1}{\rho}\frac{\partial \rho}{\partial p}\bigg|_{T} = \frac{1}{p} - \frac{1}{Z}\frac{\partial Z}{\partial p}.
	\end{align}
	\noindent Equation~\ref{eq:gas_flow_1_appA} then reads
	\begin{equation} \label{eq:gas_flow_2_appA}
		\phi\,A_c\,\frac{\partial p}{\partial t} = \nabla\cdot\left(k\,f_R\,p\,\nabla p\right).
	\end{equation}
	Assuming constant $Z$ (not necessarily equal to 1) and $\mu$, we have constant $A_c=1/Z$ and $f_R$, and Equation~\ref{eq:gas_flow_2_appA} becomes
	\begin{equation} \label{eq:PME_inhomogeneous_appA}
		\frac{\partial \psi}{\partial t}  = \frac{\sqrt{\psi}}{\phi\,\mu}\left(k\nabla^2 \psi + \nabla k\cdot\nabla \psi\right),
	\end{equation}
	\noindent where $\psi = p^2$. The permeability $k(\mathbf{x})$ is allowed to vary and we have substituted $\partial_t \sqrt{\psi} = \partial_t \psi/2\sqrt{\psi}$  and $\sqrt{\psi}\nabla \sqrt{\psi}=\nabla\psi/2$. Equation~\ref{eq:PME_inhomogeneous_appA} is equivalent to the non-homogeneous porous medium equation (NHPME) \cite[e.g.,][]{vazquez2007porous}. The validity of the assumptions above is discussed in Appendix~\ref{sec:PME_region_validty}.

\noindent Further considering constant $k$, we obtain the PME for homogeneous media
	\begin{equation} \label{eq:PME_appA}
		\frac{\partial \psi}{\partial t} = D \nabla^2\psi, 
	\end{equation}
	\noindent where the diffusivity $D$ is given by,
	\begin{equation} \label{eq:pressure_diffusivity}
		D = \frac{k\sqrt{\psi}}{\phi \mu}.
	\end{equation}

	\subsection{Nondimensionalization and compressibility number $\Pi$}
	\label{sec:appA_Pi_expressions}

	Substituting the nondimensional variables of Equation~\ref{eq:nondimensional_variables}, that is,
	\begin{align}
		\hat{t}=t/\tau_u,  
		\qquad 
		\hat{x}=x/L, 
		\qquad 
		\hat{\psi}=\psi/\psi_L,
	\end{align}
	with $\psi = p^2$, into the PME (Eq.~\ref{eq:PME_appA}) gives
	\begin{align} \label{eq:nondimensional_PME1_appA}
		\frac{\partial \hat\psi}{\partial \hat t} = \frac{\sqrt{\hat\psi}}{\Pi} \hat \nabla^2 \hat\psi,
	\end{align}
	\noindent where 
	\begin{align}
		\Pi = \frac{u_s L}{D_L} = \frac{\tau_D}{\tau_u}
	\end{align}
	\noindent is a Peclet-type number given by the ratio of pressure-squared diffusion- to flow time-scales, with  
	\begin{align}
		D_L = \frac{k p_L}{\phi \mu}
	\end{align}
	the pressure-squared diffusivity evaluated at the outlet pressure. The parameter $\Pi$ can be interpreted as a nondimensional steady mass flux scale and as a compressibility number. 
	Indeed, writing the steady velocity scale $u_s$ in terms of the spatially-constant steady mass flux $q_\text{s}^\mathrm{m}$ and fluid density evaluated at the outlet $\rho_L$, 
	\begin{align}
		u_s
		=
		\frac{q_\text{s}^\mathrm{m}}{\phi \rho_L},
		\label{eq:velocity_scale_mass_flux_appA}
	\end{align}
	and substituting $\rho_L = p_LW/RT$ (Eq.~\ref{eq:eos}) we have:
	\begin{align} \label{eq:app_Pi_mass_flux_scale}
		\Pi = \frac{q_\text{s}^\mathrm{m} RT \mu L}{kW p^2_L } = \hat{q}^\mathrm{m}_\text{s},
	\end{align}
	\noindent rendering $\Pi$ as the nondimensional steady mass flux $\hat{q}^\mathrm{m}_\text{s}$.
	Further, since the mass flux $q^\mathrm{m}$ can be generally expressed as 
	\begin{align}
		q^\mathrm{m}
		=  -\frac{1}{2}\frac{W}{RT}\frac{k}{\mu}\nabla \psi,
		\label{eq:mass_flux_p_squared_appA}
	\end{align}
	which results from Darcy's law in the absence of gravity (Eq.~\ref{eq:darcys_law}), we have nondimensionally
	\begin{align}
		\hat q^\mathrm{m}
		= -\frac{1}{2}\hat{\nabla}\hat\psi,
	\end{align}
	which, using Equation~\ref{eq:app_Pi_mass_flux_scale}, gives at steady-state 
	\begin{align}
		\hat{\nabla} \hat\psi_\text{s} 
		=
		-2\Pi,
		\label{eq:app_Pi_nablapsquared}
	\end{align}
	\noindent or, writing the constant $ \hat{\nabla} \hat\psi_\text{s} = 1-\hat{\psi}_0$,
	\begin{align}
		\Pi = \frac{1}{2} \left( \hat{\psi_0} - 1 \right).
		\label{eq:app_Pi_compressibility_number}
	\end{align}

		The isothermal compression associated with a pressure change $\Delta p=p_0-p_L$ is 
		\begin{align}
			C_s 
			= 
			\int_{p_L}^{p_0} \gamma_g(p)\, dp 
			= 
			\int_{\rho_L}^{\rho_0} \frac{d\rho}{\rho} 
			= 
			\ln \left(\frac{\rho_0}{\rho_L}\right),
		\end{align}
		and for an ideal gas 
		\begin{align}
			C_s 
			= 
			\ln \left(\frac{p_0}{p_L}\right),
		\end{align}
		\noindent since $\rho/\rho_L=p/p_L$. Using Equation~\eqref{eq:app_Pi_compressibility_number}, this gives
		\begin{align}
			C_s 
			= 
			\frac{1}{2} \ln \left( 1 + 2 \Pi \right).
		\end{align}
		Equivalently,
		\begin{align}
			\Pi
			=
			\frac{1}{2}
			\left[
			\exp(2C_s)-1
			\right].
			\label{eq:app_Pi_Cs}
		\end{align}
		Thus, $\Pi$ is monotonically related to the finite logarithmic compression $C_s$. In the weak-compression limit, $C_s\ll1$ and $\Pi \simeq C_s$.

		\subsection{Validity of the PME assumptions in terms of $\Pi$}
		\label{sec:PME_region_validty}

	The PME (Eq.~\ref{eq:PME_appA}) assumes (i) negligible inertia, (ii) constant accumulation coefficient $A_c=\mathrm{d}(p/Z)/\mathrm{d}p$ and real-gas factor $f_R=1/\mu Z$ (Eq.~\ref{eq:gas_flow_2_appA}), (iii) continuum flow without slippage, (iv) isothermal flow and (v) negligible gravity. We illustrate below where these assumptions hold in terms of $\Pi$. 

		\subsubsection{Negligible inertia and real-gas effects}

\begin{figure}[H]
			\centering
			\includegraphics[width=0.5\textwidth]{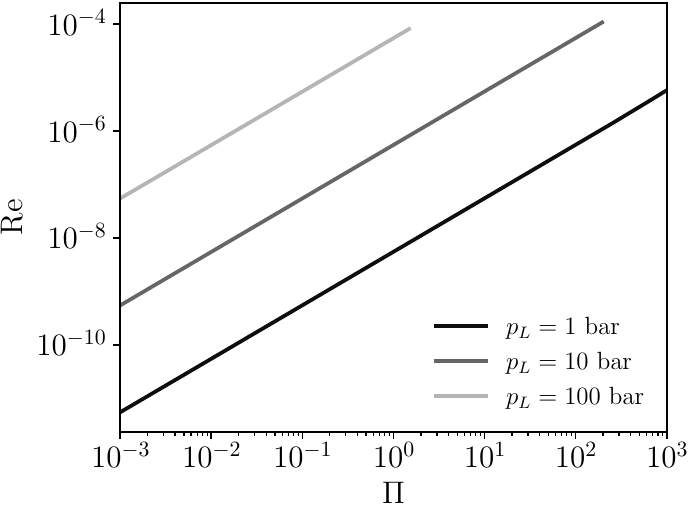}
			\caption{Reynolds number $\text{Re}=\rho u\sqrt{k}/\mu$ of the steady flow of $\text{H}_2$ at $T=320$~K, with $k\approx10^{-13}$~m$^2$ as a function of $\Pi$, for outlet pressures (dark to light) $p_L=1$, $10$ and $100$~bar. The curves are shown for inlet pressures $p_0\leq200$~bar, taken as a typical fracture pressure of storage formations \citep[e.g.,][see also Section~\ref{subsec:app_UHS}]{lubon2020numerical}. This limits the plot $\Pi$-range to $\Pi\leq2\times10^2$ and $\Pi\leq1.5$ for $p_L=10$ and $100$~bar, respectively.} \label{fig:validity_Re_appA}
		\end{figure}

		We start by considering the Reynolds number $\text{Re}$,
		\begin{align}
			\text{Re}= \frac{\rho u \sqrt k}{\mu} 
		\end{align}
		\noindent as a function of $\Pi \in \left[10^{-3},\,10^3  \right]$, which sets the ratio between inlet and outlet pressures, respectively, $p_0$ and $p_L$. We evaluate the characteristic density $\rho$ and viscosity $\mu$ for ideal hydrogen gas, $\text{H}_2$, at temperature $T=320$~K and steady inlet pressure $p_0 =p_L\sqrt{1 + 2\Pi}$ (Eq.~\ref{eq:general_solution_pressure}), with the reference outlet pressure values $p_L = 1, 10, 100\,\text{bar}$. Likewise, the characteristic velocity $u$ is taken as the steady inlet velocity (Eq.~\ref{eq:general_solution_velocity}). The permeability is $k\approx10^{-13}$~m$^2$ ($\approx100$~mD). Across the explored $\Pi$-range, $\mathrm{Re} \ll 1$ (Fig.~\ref{fig:validity_Re_appA}), indicating negligible inertial (non-Darcy) effects for the chosen length scale $\sqrt{k}$.

				To evaluate the real gas factor $f_R=1/\mu Z$ and the accumulation coefficient $A_c = d(p/Z)/dp$ (Eq.~\ref{eq:gas_compressibility_appA}) as a function of pressure we use the flash calculator of the Matlab Reservoir Simulation Toolbox \citep[MRST,][]{lie2019introduction, lie2021advanced}, based on the Peng-Robinson equation of state \citep{peng1976new}, for $\text{H}_2$, carbon dioxide, $\text{CO}_2$, and methane $\text{CH}_4$. We evaluate $f_R$ and $A_c$ at the inlet pressure $p_\text{0}$, and normalize by their values at the outlet pressure $p_L$, so that departures from unity measure maximum variation within the domain. We consider $10\%$ departure of unity as threshold.

		\begin{figure}[H]
			\centering
			\includegraphics[width=\textwidth]{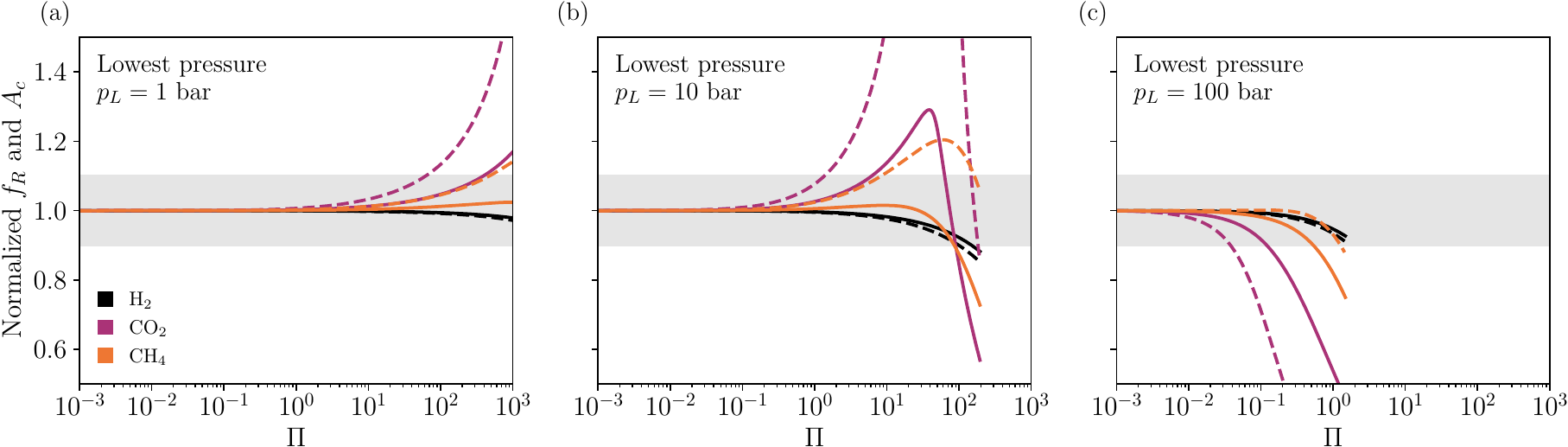}
			\caption{(Solid lines) Real-gas factor $f_R=1/\mu Z$ and (dashed lines) accumulation coefficient $A_c=\mathrm{d}(p/Z)/\mathrm{d}p$, whose constancy is assumed to obtain the PME (Eq.~\ref{eq:gas_flow_2_appA}), both evaluated at the inlet pressure $p_\text{in}$ and normalized by their values at the outlet pressure $p_L$, as a function of $\Pi$, for $\text{H}_2$, $\text{CO}_2$ and $\text{CH}_4$ at $T=320$~K and outlet pressures (a) $p_L=1$, (b) $10$ and (c) $100$~bar, from the Peng--Robinson equation of state. The grey band marks departures within $\pm10\%$. The curves are shown for inlet pressures $p_0\leq200$~bar, taken as a typical fracture pressure of storage formations \citep[e.g.,][see also Section~\ref{subsec:app_UHS}]{lubon2020numerical}. This limits the plot $\Pi$-range to $\Pi\leq2\times10^2$ and $\Pi\leq1.5$ for $p_L=10$ and $100$~bar, respectively.} \label{fig:validity_fR_appA}
		\end{figure}

			For $\text{H}_2$, the normalized $f_R$ remains within $\pm10\%$ from unity for the whole $\Pi$ range at $p_L = 1$~bar, for $\Pi\lesssim10^2$ at $p_L=10$~bar, and up to $\Pi=1.5$, the largest value considered, at $p_L=100$~bar (see caption of Fig.~\ref{fig:validity_fR_appA}). $\text{CH}_4$ and $\text{CO}_2$ depart more strongly because of their non-ideality, which increases with $p_L$. 
			For $\text{CH}_4$, the normalized $f_R$ remains within $\pm10\%$ from unity for the whole $\Pi$ range at $p_L=1$~bar, for $\Pi\lesssim90$ at $p_L=10$~bar and for $\Pi\lesssim0.5$ at $p_L=100$~bar, that is, for inlet pressures below about $130$--$145$~bar, comparable to well testing, where $\mu Z$ is taken as constant for natural gas below about $138$~bar \citep{chaudhry2003gas}. For $\text{CO}_2$, it does so for $\Pi\lesssim4\times10^2$, $6$ and $0.1$ at $p_L=1$, $10$ and $100$~bar, and its non-monotone behavior reflects the proximity to the critical point at $T=320$~K.
			Likewise, for $\text{H}_2$, the normalized accumulation coefficient $A_c$ remains within $\pm10\%$ from unity for the whole $\Pi$ range at $p_L=1$~bar, for $\Pi\lesssim10^2$ at $p_L=10$~bar, and up to $\Pi=1.5$ at $p_L=100$~bar (Fig.~\ref{fig:validity_fR_appA}). For $\text{CH}_4$, it does so for $\Pi\lesssim5\times10^2$, $8$ and $1.3$, and for $\text{CO}_2$ for $\Pi\lesssim60$, $1.4$ and $0.04$, at $p_L=1$, $10$ and $100$~bar, so that $A_c$ is generally more restrictive than $f_R$ (Table~\ref{tab:PME_validity_Pi}). 
			Beyond these values, real-gas effects can be incorporated through the pseudo-pressure \citep[see for example][]{al1966flow,chierici1994principles}, at the cost of a nonlinearity that differs from that of the PME. 

	Table~\ref{tab:PME_validity_Pi} summarizes the resulting upper bounds of $\Pi$, set by the more restrictive of the two conditions. 

\begin{table}[H]
	\centering
	\caption{Upper bound of $\Pi$ for which both the real-gas factor $f_R=1/\mu Z$ and the accumulation coefficient $\mathrm{d}(p/Z)/\mathrm{d}p$ remain within $10\%$ of their outlet values, for three outlet pressures $p_L$ at $T=320$~K. \textsuperscript{a}No departure within the range considered, $\Pi\leq10^3$ and $p_0\leq200$~bar. }
	\label{tab:PME_validity_Pi}
	\begin{tabular}{lccc}
		\hline
		& \multicolumn{3}{c}{Validity $\Pi$-range of the PME} \\
		Gas & $p_L=1$~bar & $p_L=10$~bar & $p_L=100$~bar \\
		\hline
		$\text{H}_2$ & $\Pi\gtrsim10^3$\textsuperscript{a} & $\Pi\lesssim10^2$ & $\Pi\gtrsim1.5$\textsuperscript{a} \\
		$\text{CH}_4$ & $\Pi\lesssim5\times10^2$ & $\Pi\lesssim8$ & $\Pi\lesssim0.5$ \\
		$\text{CO}_2$ & $\Pi\lesssim60$ & $\Pi\lesssim1.4$ & $\Pi\lesssim0.04$ \\
		\hline
	\end{tabular}
\end{table}

 \subsubsection{Continuum flow without slippage} \label{subsec:slippage}

 Darcy's law further assumes continuum flow, that is, a small Knudsen number $\mathrm{Kn}=\lambda/r_p$, with $\lambda\propto1/p$ the molecular mean free path and $r_p$ a characteristic pore radius. Continuum flow holds for $\mathrm{Kn}\lesssim10^{-3}$. For $10^{-3}\lesssim\mathrm{Kn}\lesssim10^{-1}$, gas slippage at the pore walls increases the apparent permeability to $k(1+b/p)$, with $b$ the Klinkenberg coefficient and $b/p\approx4\,\mathrm{Kn}$ \citep{klinkenberge1941permeability}, while transition and free-molecular flow, $\mathrm{Kn}\gtrsim10^{-1}$, require higher-order Knudsen corrections \cite[e.g.,][]{ziarani2012knudsen}. Since the pressure decreases along the domain, $\mathrm{Kn}$ is largest at the outlet, so that its range is set by $p_L$ and the permeability. In the examples of Section~\ref{sec:applications}, $\mathrm{Kn}(p_L)\approx1.5\times10^{-3}$ for helium in the GC column ($\lambda\approx0.19\,\mu$m, $r=125\,\mu$m) and $\approx1\times10^{-3}$ for $\text{H}_2$ in the UHS sandstone ($\lambda\approx2.5$~nm at $50$~bar, $r_p\approx\sqrt{8k/\phi}\approx2\,\mu$m), so that slippage changes the permeability by less than $1\%$. Slippage becomes important in tight and shale formations at low pressure \citep{xiao2019klinkenberg,lasseux2021upscaled}. For constant $b$, however,
 the shifted pressure $P = p + b$ satisfies the PME \citep[see][]{wu1998gas}.

 \subsubsection{Isothermal flow}

Isothermal flow is a standard assumption for gas flow through porous media in most conditions \cite[e.g.,][]{chierici1994principles}, since heat exchange between the gas and the solid matrix (or the column wall in GC) over a pore radius takes a time $r_p^2/\alpha$, with $\alpha$ the thermal diffusivity of the gas, which is $\lesssim10^{-4}$~s in the examples of Section~\ref{sec:applications}, far shorter than the flow time scales considered. In addition, the solid (or wall) represents a large buffer of heat, given its volumetric heat capacity largely exceeding that of the gas, so that temperature changes remain small.
Strongly non-isothermal settings exist, however, such as Joule--Thomson cooling 
during $\text{CO}_2$ injection into depleted low-permeability reservoirs with large pressure drops, as reported by \citet{oldenburg2007joule}. These cases require an additional energy balance accounting for variable temperature $T(x,t)$, rendering the PME inadequate. 

\subsubsection{Negligible gravity}

Neglecting gravity is justified for a horizontal domain of thickness $h$ when the vertical hydrostatic pressure gradient, $\mathrm{d}p/\mathrm{d}z=-\rho g=-p/H$, with $H=RT/(Wg)$ the gas scale height, is established quickly. That is, the vertical pressure diffusion over the time $\tau_h=h^2/D_L$ needs to be short compared with both the diffusive and flow time scales over the domain's length $L$, $\tau_D=L^2/D_L$ and $\tau_u=\tau_D/\Pi$, that is,
\begin{align}
	\frac{\tau_h}{\min(\tau_D,\tau_u)}=\frac{h^2}{L^2}\max(1,\Pi)\ll1,
\end{align}
\noindent  giving $h \ll L $ for $\Pi < 1$ or $h\ll L/\sqrt{\Pi}$ for $\Pi>1$. For example, at $\Pi=100$, requiring $\tau_h/\tau_u\leq0.01$ gives $h/L\leq0.01$, that is, $h\lesssim0.1$, $1$ and $10$~m for $L=10$, $100$ and $1000$~m, respectively. 

Further, once the pressure gradient is established, the pressure, and hence the gas density and the horizontal mass flux, needs to vary little over the thickness $h$. Integrating $\mathrm{d}p/\mathrm{d}z=-p/H$ gives
\begin{equation}
	p(z) = p(0)\,e^{-z/H}\approx p(0)\left(1-\frac{z}{H}\right),\qquad 0\leq z\leq h\ll H.
\end{equation}
Since $H\approx9$~km for air and $\approx130$~km for $\text{H}_2$ at $T=300$~K, this variation is below $1\%$ for $h\lesssim90$~m in air and $h\lesssim1.3$~km in $\text{H}_2$. 

For an inclined domain with elevation difference $ \Delta z$ between inlet and outlet, gravity is further negligible when the hydrostatic pressure difference, $\rho_Lg\Delta z=p_L\Delta z/H$, is small compared with the imposed one, $p_0-p_L$, where $p_0=\sqrt{1+2\Pi}\,p_L$ is the steady inlet pressure (Eq.~\ref{eq:Pi_pressure_ratio}). Taking the ratio of the two differences gives
\begin{align}
	\frac{p_L\Delta z/H}{p_0-p_L}=\frac{\Delta z/H}{\sqrt{1+2\Pi}-1}\ll1,
\end{align}
\noindent or, 
\begin{align}
	\frac{ \Delta z}{H} \ll\sqrt{1+2\Pi}-1.
\end{align}
\noindent Solving for $\Pi$, expanding the square root and keeping terms up to second order yields, 
\begin{align}
	\Pi\gg \frac{\Delta z}{H}+\frac{1}{2}\left(\frac{\Delta z}{H}\right)^2. 
\end{align}
\noindent For example, for $\Delta z=100$~m this gives  $\Pi\gg10^{-3}$ in $\text{H}_2$ and $\Pi\gg10^{-2}$ in air.

		\section{Numerical setup} \label{subsec:appendix_B}
		\renewcommand{\thefigure}{B\arabic{figure}}
		\setcounter{figure}{0}
		\numberwithin{equation}{section}\setcounter{equation}{0}

		We solve Equation~\ref{eq:PME_appA} numerically on a one-dimensional domain of length $L=1800$~m, discretized into $N_x=28\,800$ cells, that is, $\Delta x/L\approx3.5\times10^{-5}$, a resolution selected from the grid-convergence analysis described below. Since the results are presented in nondimensional form and depend only on $\Pi$ and the inlet condition, the choice of $L$ is arbitrary.
		The implemented MATLAB solver approximates spatial derivatives using second-order central finite differences. Time integration is performed using a second-order implicit Crank–Nicolson scheme \cite[e.g.,][]{thomas2013numerical}, 
		which evaluates the diffusion term as the average of its values at the previous time level, where the pressure is known, and at the new time level, where it is sought. Because the diffusion coefficient depends on the pressure, the system is nonlinear. The diffusion coefficient at the previous time level is given by the converged pressure of the previous step. The diffusion coefficient at the new time level is obtained by Picard iterations, whereby at each iteration it is evaluated with the latest iterate of $\psi$, so that each iteration solves a linear problem. The iterations stop when the relative $\ell_2$ change of $\psi$ between successive iterates falls below $10^{-10}$, with at most 50 iterations per step.
	Following the MATLAB Reservoir Simulation Toolbox (MRST) \citep{lie2019introduction, lie2021advanced}, adaptive time stepping was implemented, with $\Delta t$ adjusted depending on the number of Picard iterations and the rate of change of the solution.
	This strategy ensures accuracy during sharp pressure gradients, while allowing larger time steps during smoother pressure profiles. 

\begin{figure}[H]
		\centering
		\includegraphics[width=\textwidth]{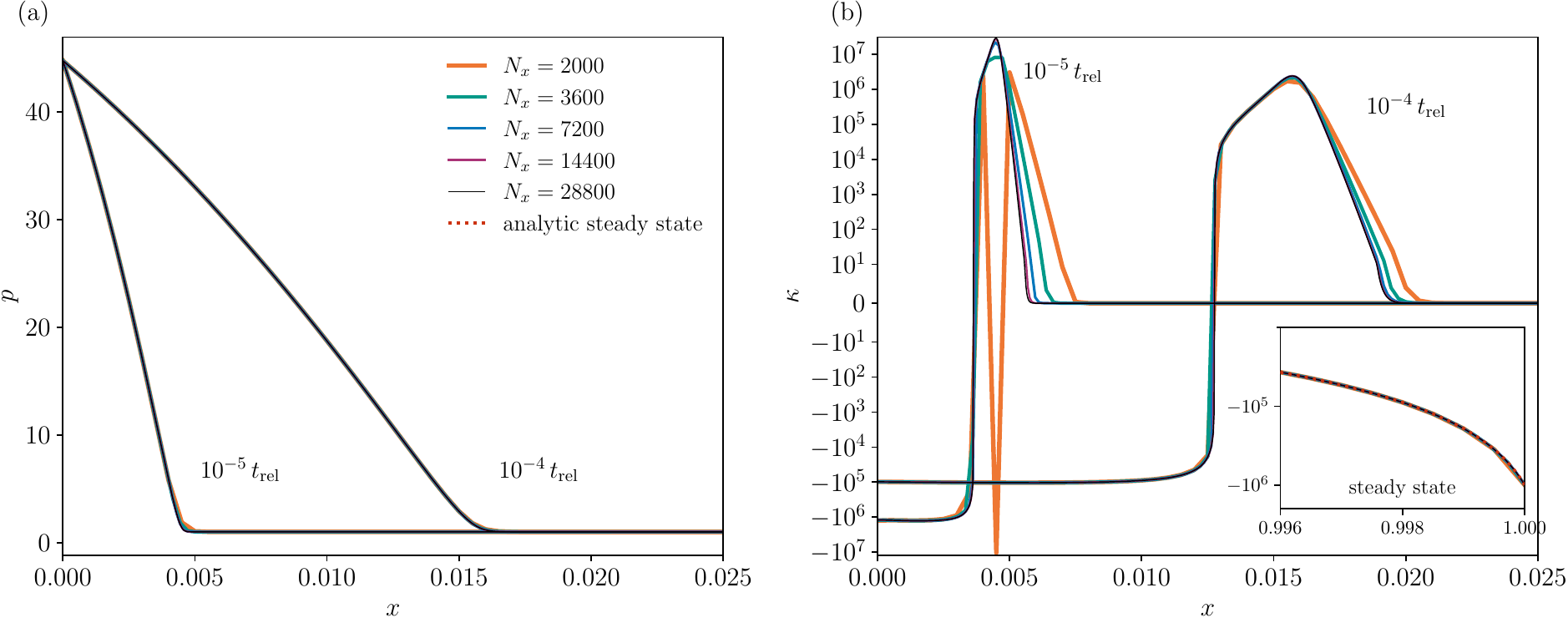}
		\caption{Grid convergence for $\Pi=1000$ and imposed inlet pressure: (a) pressure and (b) pressure curvature (symmetric logarithmic scale) near the inlet (i.e., over the first $2.5\%$ of the domain, $0\leq x\leq0.025$), at $t=10^{-5}$ and $10^{-4}\,t_\text{rel}$, for $N_x=2000$ to $28\,800$ cells. Inset in (b): steady-state curvature at the outlet, $0.996\leq x\leq1$, with the analytical solution (dotted).}
		\label{fig:grid_convergence}
	\end{figure}

	Grid convergence was assessed for $\Pi=1000$ with imposed inlet pressure, the case with the steepest transient fronts and the thinnest outlet boundary layer, using $N_x=2000$, $3600$, $7200$, $14\,400$ and $28\,800$. Figure~\ref{fig:grid_convergence} shows the pressure $p$ and pressure curvature $\kappa$ near the inlet at early times taken as $t=10^{-5}$ and $10^{-4}\,t_\text{rel}$, when the transients are sharpest and hence most sensitive to resolution. The reference relaxation time $t_\text{rel}$ is defined in Equation~\ref{eq:observed_diffusion_timescale}. 

	The pressure is well resolved on all grids, with a relative $\ell_2$ difference to the finest grid below $0.5\%$, whereas the curvature, a second derivative, requires finer grids; at $t=10^{-5}\,t_\text{rel}$, the $\ell_2$ difference is $135\%$, $68\%$, $17\%$ and $4\%$ for, respectively, $N_x=2000$, $3600$, $7200$ and $14\,400$; and at $10^{-4}\,t_\text{rel}$, the differences are $19\%$, $6\%$, $1.5\%$ and $0.3\%$. 
	For curvature, the decrease of the relative $\ell_2$ differences scale as $\sim \Delta x^2$ and, for a given grid, decrease with time as the front widens. At steady state, the grids with $N_x\geq3600$ agree with the analytical solution (Section~\ref{subsec:steady_state_analysis}) within $10^{-4}$ for the curvature, including the outlet boundary layer of width $\sim\Pi^{-1}$, which spans $\sim 30$ cells on the production grid. Thus, the curvature, and hence its integral moments, are resolved from $t\approx10^{-5}\,t_\text{rel}$ onward on the production grid.

	\section{Perturbation series solution}
	\label{subsec:appendix_C}
	\renewcommand{\thefigure}{C\arabic{figure}}
	\setcounter{figure}{0}
	\numberwithin{equation}{section}\setcounter{equation}{0}

	This appendix presents the semi-analytical solution of the PME for our bounded domain, compared in Section~\ref{subsec:transient_state_analysis} against numerical solutions of the PME.

	We start by scaling the nondimensional time as  $\theta = t/\Pi$  (Subsec.~\ref{subsec:governing_equations}), such that the PME (Eq.~\ref{eq:non_dimensional_PME}) reads
	\begin{align}
		\frac{\partial \psi}{\partial \theta} -\sqrt{\psi}\,
		\frac{\partial^2 \psi}{\partial x^2}  = 0,
		\label{eq:app_psi_pme}
	\end{align}

	We first write $\psi(x,\theta)$ as a perturbation around the steady inlet pressure-squared $\psi_0 = p_0^2 = 1 + 2\Pi$ (Eq.~\ref{eq:problem_bcs_nondimensional}) ,
	\begin{align}
		\psi(x,\theta)
		=
		\psi_0
		\left[
		1+\beta V(x,\theta)
		\right],
		\label{eq:app_general_ansatz}
	\end{align}
	\noindent where $\beta$ is used as the expansion parameter and given by the relative difference between the steady inlet value $\psi_0$ and the initial and outlet value $1$,  
	\begin{align}
		\beta = \frac{\psi_0 - 1}{\psi_0}  = \frac{2 \Pi}{1 + 2 \Pi}.
		\label{eq:app_beta_definition}
	\end{align}
	\noindent As such, $\beta \in [0,1)$ remains bounded for any $\Pi$. The field $V(x,\theta)$ is the dimensionless perturbation, with $-1 \leq  V  \leq 0$. Substituting Equation~\eqref{eq:app_general_ansatz} into Equation~\eqref{eq:app_psi_pme} and dividing by $\psi_0\beta$ gives
	\begin{align}
		\frac{\partial V}{\partial \theta}
		-
		\sqrt{\psi_0}
		\sqrt{1+\beta V}
		\frac{\partial^2 V}{\partial x^2}
		=
		0 .
	\end{align}
	Absorbing the constant factor $\sqrt{\psi_0}$ by further scaling time as 
	\begin{align}
		\theta'=\sqrt{\psi_0}\,\theta ,
	\end{align}
	the equation for $V(x,\theta')$ reads
	\begin{align}
		\frac{\partial V}{\partial \theta'}
		-
		\sqrt{1+\beta V}
		\frac{\partial^2 V}{\partial x^2}
		=
		0 .
		\label{eq:app_general_V_equation}
	\end{align}

	\noindent Now, let us write $V(x,\theta')$ in powers of $\beta$,
	\begin{align}
		V(x,\theta')
		=
		\sum_{i=0}^{\infty}\beta^i V_i(x,\theta'),
		\label{eq:app_power_series_V}
	\end{align}
	and expand $\sqrt{1+\beta V}$ in its Taylor series about $\beta V=0$,
	\begin{align}
		\sqrt{1+\beta V}
		&=
		\sum_{k=0}^{\infty}
		\binom{1/2}{k}
		\left(\beta V\right)^k
		\nonumber\\
		&=
		1
		+\frac{\beta V}{2}
		-\frac{\left(\beta V\right)^2}{8}
		+\frac{\left(\beta V\right)^3}{16}
		+\mathcal{O}\!\left( \left[ \beta V\right]^4\right).
		\label{eq:app_general_sqrt_expansion}
	\end{align}
	\noindent Substituting Equation~\ref{eq:app_power_series_V} into Equation~\ref{eq:app_general_sqrt_expansion} and collecting terms with
	equal powers of $\beta$,
	\begin{align}
		\sqrt{1+\beta V}
		&=
		1
		+\frac{1}{2}\beta
		\left(
		V_0+\beta V_1+\beta^2 V_2+\cdots
		\right)
		-\frac{1}{8}\beta^2
		\left(
		V_0+\beta V_1+\beta^2 V_2+\cdots
		\right)^2
		\nonumber\\
		&\quad
		+\frac{1}{16}\beta^3
		\left(
		V_0+\beta V_1+\beta^2 V_2+\cdots
		\right)^3
		+\mathcal{O}(\beta^4),
		\nonumber\\
		&=
		1
		+\frac{V_0}{2}\beta
		+\left(
		\frac{V_1}{2}
		-\frac{V_0^2}{8}
		\right)\beta^2
		+\left(
		\frac{V_2}{2}
		-\frac{V_0V_1}{4}
		+\frac{V_0^3}{16}
		\right)\beta^3
		+\mathcal{O}(\beta^4),
		\label{eq:app_series_square_root}
	\end{align}
	\noindent further substituting Equations~\ref{eq:app_series_square_root} and~\ref{eq:app_power_series_V} into the PDE for $V(x,\theta')$ (Eq.~\ref{eq:app_general_V_equation}),
	and requiring the resulting equality to hold order by order, leads to the following recursion of linear diffusion equations for the perturbation coefficients $V_i(x,\theta')$:
	\begin{align}
		\frac{\partial V_i}{\partial \theta'}
		-\frac{\partial^2 V_i}{\partial x^2}&=S_i(x,\theta';V_0,\cdots,V_{i-1}).
		\label{eq:app_perturbation_recursion}
	\end{align}
	\noindent where $S_0\equiv0$ and, for $i\geq1$,
\begin{align}
	S_i=
	\sum_{k=1}^{i}\binom{1/2}{k}
	\sum_{j_0+\cdots+j_k=i-k}
	\left(\prod_{r=1}^{k}V_{j_r}\right)
	\frac{\partial^2V_{j_0}}{\partial x^2},
	\label{eq:app_source_term_analytical}
\end{align}
\noindent with $ \binom{1/2}{k}
=\frac{(1/2)(1/2-1)\cdots(1/2-k+1)}{k!} $. For $i=1,2,3$ we have for example
\begin{align}
	S_1 &=\frac12V_0V_{0,xx},\\
	S_2 &=
	\frac12\left(V_0V_{1,xx}+V_1V_{0,xx}\right)
	-\frac18V_0^2V_{0,xx},\\
	S_3 &=
	\frac12\left(V_0V_{2,xx}+V_1V_{1,xx}+V_2V_{0,xx}\right)
	-\frac18\left(V_0^2V_{1,xx}+2V_0V_1V_{0,xx}\right)
	+\frac{1}{16}V_0^3V_{0,xx}.
\end{align}

\noindent Through Equation~\ref{eq:app_general_ansatz}, the initial and boundary data of Equation~\ref{eq:problem_bcs_nondimensional} become $V(x,0)=V(1,\theta')=-1$ and, at the inlet, $V(0,\theta')=0$ for imposed pressure or $\partial_xV(0,\theta')=-1$ for imposed mass flux. These data are independent of $\beta$, so they are carried entirely by $V_0$, and the $V_i$, $i\geq1$, have zero initial data and homogeneous boundary data, $V_i(0,\theta')=V_i(1,\theta')=0$ for imposed pressure and $\partial_xV_i(0,\theta')=V_i(1,\theta')=0$ for imposed mass flux. For imposed volumetric flux, the inlet condition $\partial_xp(0,\theta')=-\Pi/\sqrt{\psi_0}$ (Eq.~\ref{eq:problem_bcs_nondimensional}) becomes, with $p=\sqrt{\psi_0}\sqrt{1+\beta V}$ (Eq.~\ref{eq:app_general_ansatz}) and $\psi_0\beta=2\Pi$ (Eq.~\ref{eq:app_beta_definition}),
\begin{align}
	\partial_xV(0,\theta')&=-\sqrt{1+\beta V(0,\theta')}\nonumber\\
		&=-1-\frac{\beta}{2}V_0(0,\theta')-\beta^2\Big[\frac{V_1(0,\theta')}{2}-\frac{V_0^2(0,\theta')}{8}\Big]+\mathcal{O}(\beta^3),
	\label{eq:app_volflux_inlet_V_series}
\end{align}
\noindent which depends on $\beta$ and thus gives non-homogeneous inlet data at all orders (Appendix~\ref{subsec:app_vol_flux_condition}).

The solutions of Equations~\eqref{eq:app_perturbation_recursion} in our bounded interval with fixed outlet pressure can be generally expressed as
\begin{align}
	V_i(x,\theta')
	=
	\delta_{i0}V_{\mathrm{s}}(x)
	+\sum_n c_{i,n}(\theta')X_n(x), \qquad i=0,1,\dots
	\label{eq:app_general_V0_solution}
\end{align}
\noindent where $\delta_{i0}$ is the Kronecker delta (one for $i=0$ and zero otherwise) and $V_{\mathrm{s}}(x)$ is the steady perturbation, obtained from the steady state $\psi_\mathrm{s}(x)=1+2\Pi(1-x)$ (Eq.~\ref{eq:general_solution_pressure}), through Equation~\ref{eq:app_general_ansatz} as
\begin{align}
	V_{\mathrm{s}}(x)
	=
	\frac{\psi_{\mathrm{s}}(x)/\psi_{0}-1}{\beta} = -x.
\end{align}
\noindent Since $V_\mathrm{s}$ does not depend on $\beta$, it is carried entirely by the zeroth-order field, and the fields $V_i$, $i\geq1$, have no steady part. The second term of Equation~\eqref{eq:app_general_V0_solution}  is the transient part, described through the expansion in the eigenfunctions $X_n(x)$ of the negative Laplacian, $\mathcal L=-\partial_x^2$, that is, $-X_n''=\lambda_nX_n$ with $X_n(1)=0$ and $X_n(0)=0$ for imposed pressure or $X_n'(0)=0$ for imposed fluxes. Projecting Equation~\ref{eq:app_perturbation_recursion} onto $X_n$,
\begin{align}
	\int_0^1\left(\frac{\partial V_i}{\partial\theta'}-\frac{\partial^2V_i}{\partial x^2}\right)X_n\,dx=\int_0^1S_iX_n\,dx,
	\label{eq:app_projection}
\end{align}
\noindent each modal coefficient obeys the linear ordinary differential equation
\begin{align}
	\frac{\mathrm{d}c_{i,n}}{\mathrm{d}\theta'}=-\lambda_nc_{i,n}+s_{i,n}(\theta'),
	\qquad
	s_{i,n}=\frac{1}{\|X_n\|^2}\int_0^1S_iX_n\,dx,
	\label{eq:app_modal_ode}
\end{align}
\noindent with $s_{i,n}$ the source projected onto mode $n$ (see \citealt[][\S 8.3, Eq. 8.3.10]{haberman2013applied} for the derivation). Its solution, obtained with an integrating factor, is the decay of the initial value plus a time-convolution (Duhamel) integral of the source,
\begin{align}
	c_{i,n}(\theta')
	&=
	c_{i,n}(0)e^{-\lambda_n\theta'}
	+\frac{1}{\|X_n\|^2}
	\int_0^{\theta'}
	e^{-\lambda_n(\theta'-s)}
	\left[\int_0^1 S_i(x,s)X_n(x)\,dx\right]ds, \qquad i=0,1,\dots
	\label{eq:app_general_duhamel}
\end{align}
\noindent  with
\begin{align}
	c_{i,n}(0)
	&=
	\frac{1}{\|X_n\|^2} \displaystyle\int_0^1
	[V_i(x,0)-\delta_{i0} V_{\mathrm{s}}(x)]X_n(x)\,dx,
\end{align}
\noindent and $\|X_n\|^2=\int_0^1X_n^2(x)\,dx$. For imposed pressure (i.e., Dirichlet-Dirichlet) it is
\begin{align}
	X_n(x)=\sin(n\pi x),
	\qquad
	\lambda_n=n^2\pi^2,
	\qquad n=1,2,\ldots,
	\label{eq:app_general_DD_basis}
\end{align}
\noindent while for imposed mass and volumetric flux (i.e., Neumann-Dirichlet) it is
\begin{align}
	X_n(x)=\cos\left[(n+1/2)\pi x\right],
	\qquad
	\lambda_n=(n+1/2)^2\pi^2,
	\qquad n=0,1,\ldots  
	\label{eq:app_general_ND_basis}
\end{align}
\noindent 

At zeroth order ($S_0=0$), only the first term of Equation~\ref{eq:app_general_duhamel} contributes, and $V_0$ decays from $V_0(x,0)=-1$ toward $V_\mathrm{s}(x)=-x$ as a linear diffusion process with the diffusivity of the inlet pressure. At higher orders, only the Duhamel integral contributes, with sources $S_i$ generated by the lower-order fields $V_0,\ldots,V_{i-1}$. For imposed volumetric flux, the inlet data of $V_i$ contribute an additional source (Eq.~\ref{eq:volflux_inlet_Vi}).

The $N$-th order approximation for any $\Pi$ then follows as
\begin{align}
	\tilde\psi(x,\theta) = \psi_0\Big[1+\beta\sum_{i=0}^{N}\beta^i V_i(x,\theta')\Big].
	\label{eq:app_series_reconstruction}
\end{align}

\subsection{Semi-analytical implementation and convergence analysis}

Although $\beta<1$ for any $\Pi$, the convergence of the series in Equation~\ref{eq:app_series_reconstruction} slows down as $\Pi$ grows and $\beta \to 1$, requiring higher order $N$ to achieve a given accuracy. Since the source of each order is explicit in the lower-order fields, any order is in principle accessible analytically. 
However, because the number $2^i -1$ of terms grows combinatorially to calculate the sources $S_i$ in Equation~\ref{eq:app_source_term_analytical},  and the size of each term grows as well, analytical computation becomes impractical past $i=4$ or $i=5$. This is a well-known challenge in perturbation approaches \cite[e.g.,][]{polubarinova1962theory, bender2013advanced}.

Instead, to evaluate higher orders, we compute each source term $S_i$ from a nested recursion constructed as follows. We first note that the coefficient of $\beta^i$ in the expansion of $\sqrt{1+\beta V}\,\partial_x^2V$ is equal to $S_i+\partial_x^2V_i$, since Equation~\eqref{eq:app_general_V_equation} must hold order by order in $\beta$. Let us re-write $\sqrt{1 + \beta V}$ as its own power series in $\beta$,
\begin{align}
	\sqrt{1+\beta V} = \sum_{m=0}^{\infty} \beta^m U_m(x,\theta'). 
	\label{eq:app_Um_series}
\end{align}

Differentiating Equation~\eqref{eq:app_power_series_V} twice in $x$ term by term gives $\partial_x^2V=\sum_j\beta^j V_{j,xx}$ which, multiplied by Equation~\eqref{eq:app_Um_series} term by term and arranging products gives
\begin{align}
	\sqrt{1+\beta V}\,\partial_x^2V
	=
	\left(\sum_{m=0}^{\infty}\beta^m U_m\right)\left(\sum_{j=0}^{\infty}\beta^j V_{j,xx}\right)
	=
	\sum_{m=0}^{\infty}\sum_{j=0}^{\infty}\beta^{m+j}\,U_m\,V_{j,xx}.
\end{align}
Collecting, for each power $i=m+j$, all pairs $(m,j)$ with $j=i-m$ (i.e., $m=0,\ldots,i$) yields
\begin{align}
	\sqrt{1+\beta V}\,\partial_x^2V = \sum_{i=0}^{\infty}\beta^i\sum_{m=0}^{i}U_m\,(V_{i-m})_{xx}. 
	\label{eq:app_implementation_expansion_1}
\end{align}
Equating the right-hand side of Equation~\ref{eq:app_implementation_expansion_1} with $S_i+\partial_x^2V_i$ we obtain: 
\begin{align}
	S_i = \sum_{m=1}^{i} U_m\,(V_{i-m})_{xx}
	\label{eq:app_source_term_Um}
\end{align}
\noindent   for $i \geq 1$, with $S_0 = 0$. Finally, to close the approach  we calculate the fields $U_m$ as follows. We square Equation~\eqref{eq:app_Um_series} and insert the series for $V_i(x,\theta')$ in its left hand side, which gives
\begin{align}
	1+\beta V = 1+\beta\sum^{\infty}_{i=0}V_i\beta^i = \sum^{\infty}_{m=0}\beta^m\sum_{k=0}^mU_kU_{m-k}.
\end{align}
\noindent Matching powers of $ \beta$ yields $U_0=1$ for $m=0$, while for $m\geq1$, we have 
\begin{align}
	2U_m+\sum_{k=1}^{m-1}U_kU_{m-k}=V_{m-1},
\end{align}
\noindent or,
\begin{align}
	U_m
	=
	\frac{1}{2}
	\left[
	V_{m-1}
	-
	\sum_{k=1}^{m-1}U_kU_{m-k}
	\right].
	\label{eq:app_Um_recursion}
\end{align}
The recursion of Equation~\eqref{eq:app_Um_recursion}, together with Equation~\eqref{eq:app_source_term_Um}, replaces the direct evaluation of Equation~\eqref{eq:app_source_term_analytical}, building $S_i$ from $\mathcal{O}(i)$ products of previously computed fields, instead of a number of terms that grows combinatorially with $i$, which makes high orders accessible. 
Our approach is a standard recursive computation of powers of a power series \citep[][\S 4.7]{knuth1997art}.  

\subsubsection{Volumetric flux inlet condition} \label{subsec:app_vol_flux_condition}
Since the right-hand side of the inlet condition (Eq.~\ref{eq:app_volflux_inlet_V_series}) is just the series of Equation~\eqref{eq:app_Um_series} evaluated at $x=0$, matching powers of $\beta$ yields
\begin{align}
	\partial_x V_i(0,\theta') = -U_i(0,\theta'), \qquad i\geq0,
	\label{eq:volflux_inlet_Vi}
\end{align}
with $V_i(1,\theta')=0$ for $i\geq1$ and $U_i(0,\theta')$ depending only on $V_0,\ldots,V_{i-1}$ (Eq.~\ref{eq:app_Um_recursion}). However, since the eigenfunctions satisfy $dX_n/dx(0)=0$, differentiating the expansion of $V_i$ (Eq.~\ref{eq:app_general_V0_solution}) term by term would give $\partial_xV_i(0,\theta')=0$, inconsistent with Equation~\ref{eq:volflux_inlet_Vi}. Following \citet[][\S 8.4]{haberman2013applied}, we instead project $\partial_x^2V_i$ onto $X_n$ and use Green's second identity to obtain
\begin{align}
	\int_0^1X_n\,\frac{\partial^2V_i}{\partial x^2}\,dx &= \Big[X_n\frac{\partial V_i}{\partial x}-V_i\frac{dX_n}{dx}\Big]_0^1-\lambda_n\int_0^1V_iX_n\,dx \nonumber\\
	&= U_i(0,\theta')-\lambda_n\|X_n\|^2c_{i,n},
\end{align}
\noindent where the inlet flux enters through the boundary term, $-X_n(0)\,\partial_xV_i(0,\theta')=U_i(0,\theta')$, since $X_n(0)=1$. Dividing by $\|X_n\|^2=1/2$, the modal coefficients of $\partial_x^2V_i$ become $-\lambda_nc_{i,n}+2U_i(0,\theta')$ and the inlet flux thus acts as a source $2U_i(0,\theta')$ in the modal equation (Eq.~\ref{eq:app_modal_ode}), added to the integrand of Equation~\ref{eq:app_general_duhamel}. Consequently, the second derivative used to build the sources of the next orders is evaluated as \citep[][Eqs.~8.4.12--8.4.13]{haberman2013applied}
\begin{align}
	\frac{\partial^2V_i}{\partial x^2}=-\sum_n\lambda_n\Big[c_{i,n}-\frac{2U_i(0,\theta')}{\lambda_n}\Big]X_n(x).
	\label{eq:app_volflux_d2V}
\end{align}

The Duhamel integrals (Eq.~\ref{eq:app_general_duhamel}) are evaluated by numerical quadrature on a time grid of $N_t=20000$ points, refined at early times as $\theta'_j=\theta'_{\max}\,[j/(N_t-1)]^4$ to ensure resolving the early-time behaviour near the inlet, and the fields $V_i$ are computed on a uniform grid of $N_x=2801$ points, with 4000 eigenmodes for $V_0$ and 1200 for $V_i$, $i\geq1$, up to order $N=100$. 

\subsubsection{Convergence analysis} \label{subsec:convergence_analysis}

\begin{figure}[H]
	\centering
	\includegraphics[width=0.6\textwidth]{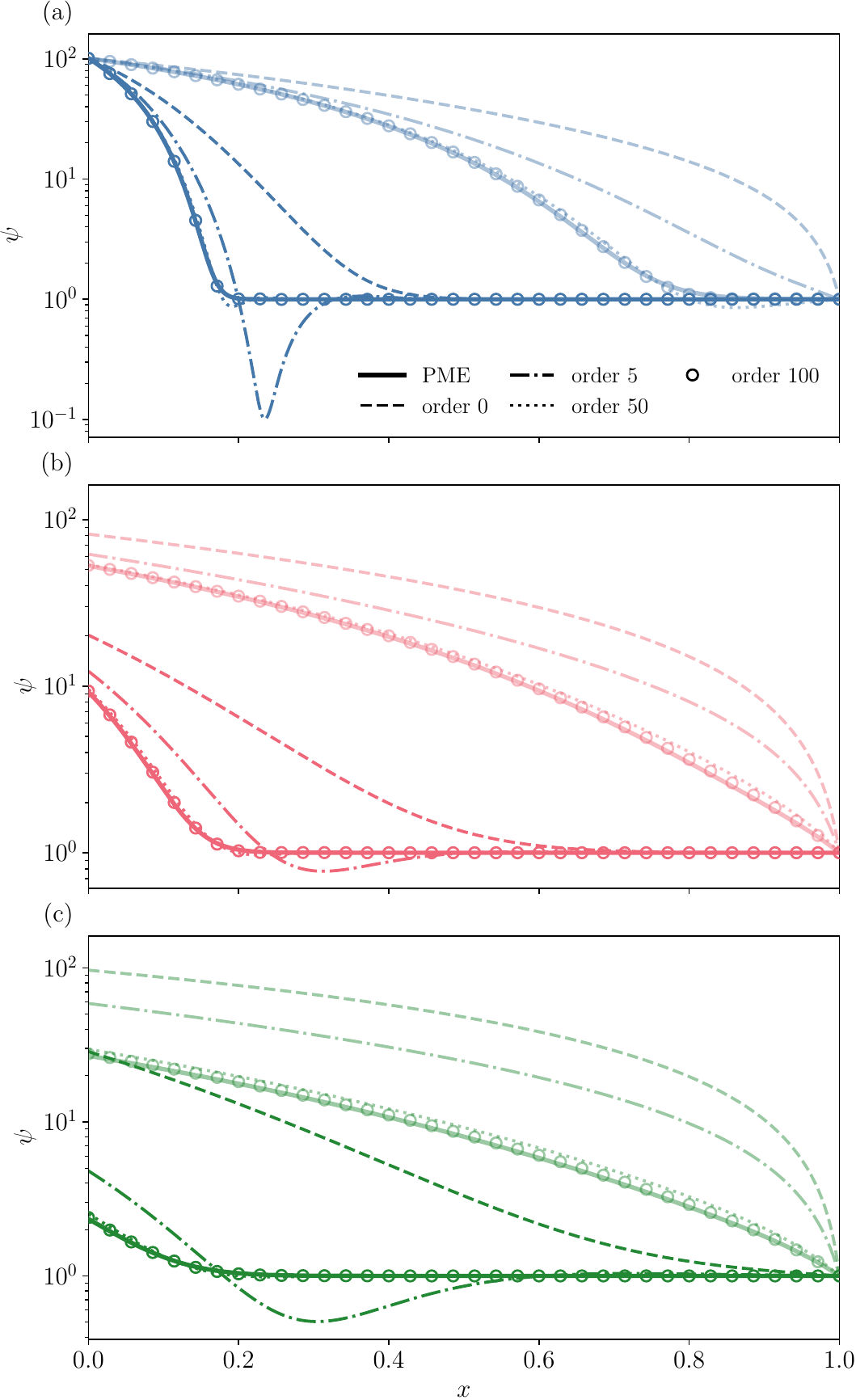}
	\caption{Pressure squared $\psi$ as a function of distance $x$ (logarithmic vertical axes) at $\Pi = 50$ ($\beta \approx 0.99$) for imposed inlet (a) pressure, (b) mass flux and (c) volumetric flux. Shown are (solid lines) the numerical PME solution and the perturbation series in $\beta = 2\Pi/(1+2\Pi)$ truncated at orders (dashed) 0, (dash-dotted) 5, (dotted) 50 and (circles) 100, at times (dark) $0.01t_\text{rel}$ and (light) $0.2t_\text{rel}$, with $t_\text{rel}$ calculated using Equation~\ref{eq:observed_diffusion_timescale}. At $t_\text{rel}$ all orders lie within about $1\%$ of the PME solution.} \label{fig:beta_series_convergence}
\end{figure}

Figure~\ref{fig:beta_series_convergence} illustrates the convergence of the series with truncation order at $\Pi=50$, which gives $\beta = 2\Pi/(1 + 2\Pi) \approx0.99$, for the three inlet conditions. 

Low-order truncations undershoot $\psi=1$ ahead of the pressure front, e.g.\ down to $\psi\approx0.1$ at order 5 and $0.01t_\text{rel}$ for imposed pressure (Fig.~\ref{fig:beta_series_convergence}a), and these undershoots vanish as the order increases.

\begin{figure}[H]
	\centering
	\includegraphics[width=0.6\textwidth]{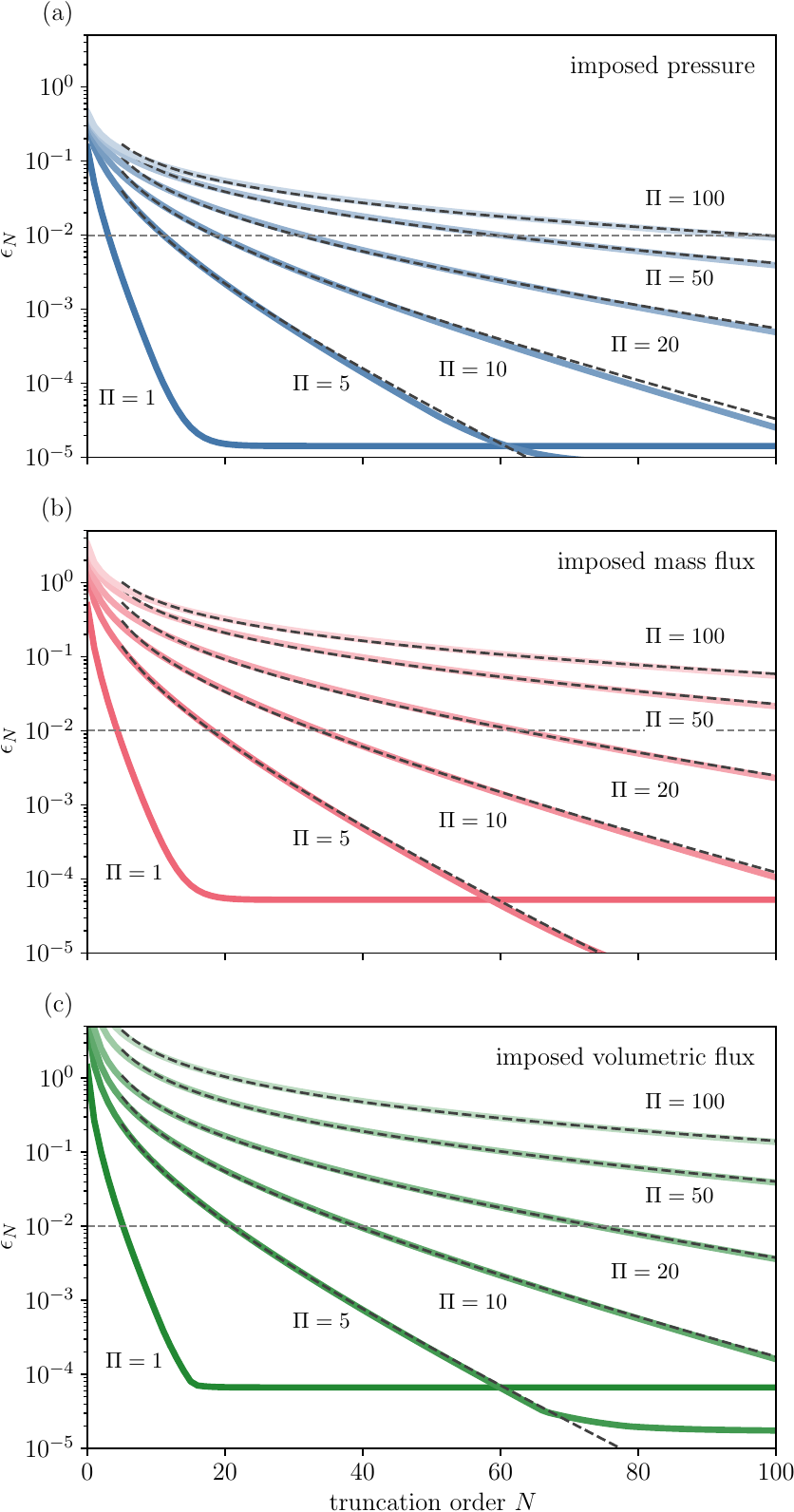}
	\caption{Maximum relative residual $\epsilon_N$ between the truncated perturbation series at order $N$ and the numerical PME solution over $0.01t_\text{rel}\le t\le t_\text{rel}$ (Eq.~\ref{eq:app_max_residual}) as a function of the truncation order $N$ for imposed inlet (a) pressure, (b) mass flux and (c) volumetric flux, at $\Pi=1$, $5$, $10$, $20$, $50$ and $100$ (dark to light), together with (dark grey dashed lines) Equation~\ref{eq:app_residual_law} fitted for $N\ge5$ and residuals above $10^{-3}$, for $\Pi\ge5$, since at $\Pi=1$ the residual reaches the numerical floor within a few orders. The grey dashed line marks a residual of $1\%$; vertical axes are clipped at 5.} \label{fig:beta_series_convergence_allPi}
\end{figure}

To quantify convergence over the transient and for all $\Pi$, we define the maximum relative residual of the series truncated at order $N$,
\begin{align}
	\epsilon_N=\max_{0.01t_\text{rel}\le t\le t_\text{rel}}\frac{\|\psi_N(\cdot,t)-\psi_\text{PME}(\cdot,t)\|}{\|\psi_\text{PME}(\cdot,t)-1\|},
	\label{eq:app_max_residual}
\end{align}
\noindent where $\psi_N$ is the series truncated at order $N$, $\psi_\text{PME}$ the numerical solution of the PME, and $\|f\|=(\int_0^1f^2\,dx)^{1/2}$ the $\ell_2$ norm in $x$; the maximum is taken over about 185 times for each $\Pi$. Figure~\ref{fig:beta_series_convergence_allPi} shows $\epsilon_N$ for $1\le\Pi\le100$. For the three inlet conditions, $\epsilon_N$ decreases monotonically with the truncation order and, at any given order, increases monotonically with $\Pi$.
At $\Pi=50$, $\epsilon_{100}=$ $0.4\%$, $2\%$ and $4\%$ for imposed pressure, mass flux and volumetric flux, respectively.

The maximum relative residual $\epsilon_N$ is well described by the product of geometric and algebraic factors, that is,
\begin{align}
	\epsilon_N \sim \beta^{N} N^{-a},
	\label{eq:app_residual_law}
\end{align}
\noindent (dark grey dashed lines, Fig.~\ref{fig:beta_series_convergence_allPi}a--c).
The geometric factor, $\beta^N$, is the residual of a geometric series, whose terms have constant amplitude, and decays slowly as $\beta\to1$. For example, at $\Pi=50$, $\beta^{100}\approx0.37$, that is, a residual of about $37\%$ at order 100. The additional algebraic factor, $N^{-a}$, reflects that, unlike in a geometric series, the corrections of our perturbation series decrease with their index $i$, as $\|V_i\|\sim i^{-3/2}$ for imposed pressure and mass flux and $\|V_i\|\sim i^{-7/4}$ for imposed volumetric flux. Thus, at small $\Pi$ the geometric factor decays fast, controlling the convergence, whereas at large $\Pi$, where $\beta^N$ decays slowly, the convergence is driven mainly by the algebraic factor. The exponent $a$ is close to one for $\Pi\lesssim20$ and decreases to $a\approx0.8$ at $\Pi=100$ for imposed pressure and mass flux. For imposed volumetric flux $a$ decreases from 1.2 to 1 at $\Pi=100$.  

\section{Eigenmodes of the PME linearized about the steady state} \label{app:eigenvalue_problem_degenerate_diffusion}

The solution to the linear diffusion Equation~\ref{eq:linearized_PME_around_ss}, repeated below, 
\begin{align}
	\frac{\partial \delta\psi}{\partial 		\theta'}= \sqrt{1-\beta x}\,\frac{\partial^2\delta\psi}{\partial x^2},
	\label{eq:app_linearized_PME_inlet_scaled}
\end{align}
\noindent is obtained by separation of variables. 
Making the substitution $z=1-\beta x$
and inserting the normal mode $\delta\psi=\Phi \,e^{-\mu'\theta'}$ gives the eigenvalue problem
\begin{align}
	\frac{d^2\Phi}{dz^2}+\frac{\mu'}{\beta^2}\,z^{-1/2}\,\Phi=0,\qquad \Phi(1-\beta)=0,
	\label{eq:app_limit_eigen}
\end{align}
\noindent with boundary data $\Phi=0$, $d\Phi/dz=0$ or $d\Phi/dz=\Phi/2$ at $z=1$ for imposed pressure, mass flux and volumetric flux, respectively. Its general solution is \citep[][Eq.~10.13.3, with $P=3/2$]{NIST:DLMF}
\begin{align}
	\Phi=\sqrt{z}\,[A\,J_{2/3}(\eta\,z^{3/4})+B\,Y_{2/3}(\eta\,z^{3/4})] 
	\label{eq:app_general_Bessel_solution}
\end{align}
\noindent with $A$ and $B$ constants, and $J_{2/3}$ and $Y_{2/3}$ the Bessel functions of first and second kind, respectively, of order $2/3$, and $\eta=4\sqrt{\mu'}/(3\beta)$.

Evaluating Equation~\ref{eq:app_general_Bessel_solution} at $z=1$ and imposing the boundary data gives a homogeneous linear system for $A$ and $B$, which define the eigenvalues $\mu'_n(\beta)$. 
As $\Pi \to 0$ and hence $\beta\to0$ (i.e., uniform diffusivity), the smallest eigenvalue $\mu'_1$ controlling the late-time decay tends to $\pi^2$ and $(\pi/2)^2$, for imposed pressure and fluxes, respectively.
As $\Pi\to\infty$ and $\beta\to1$, it is $\eta=\tfrac43\sqrt{\mu'}$, the outlet ($z=0$) becomes a singular point, and $\Phi(0)=0$ requires $B=0$ in Equation~\ref{eq:app_general_Bessel_solution}. Using the following identity between a Bessel function and its derivative \citep[][Eq.~10.6.2]{NIST:DLMF},
\begin{align}
	\frac{dJ_{2/3}(\xi)}{d\xi}=J_{-1/3}(\xi)-\frac{2}{3\xi}J_{2/3}(\xi),
\end{align}
\noindent the inlet conditions become
\begin{align}
	J_{2/3}(\eta)=0,\qquad J_{-1/3}(\eta)=0,\qquad \tfrac34\eta\,J_{-1/3}(\eta)=\tfrac12J_{2/3}(\eta),
	\label{eq:limit_roots}
\end{align}
\noindent for imposed pressure, mass flux and volumetric flux, respectively, where each root $\eta_n$ gives the  eigenvalue  $\mu'_n=(3\eta_n/4)^2$. The smallest roots are $\eta_1\approx3.376$, $1.866$ and $1.401$, which give $\mu'_1\approx6.41$, $1.96$ and $1.10$, respectively, in agreement within $1\%$ with the late-time decay of our simulations at $\Pi \geq 100$. The next roots give $\mu'_2\approx24.0$, $14.0$ and $13.2$, so that the higher modes decay at least about four times faster than the slowest one.

\end{document}